\pdfoutput=1
\documentclass[11pt,reqno,preprint]{article}
\usepackage{jheppub}
\usepackage{epsfig}
\usepackage{amssymb}
\usepackage{amsmath}
\usepackage{mathrsfs}
\usepackage{hyperref}
\usepackage{multirow}
\usepackage{textcomp}
\usepackage{subcaption}
\usepackage{pifont}
\usepackage{float}

\usepackage{tikz}
\usetikzlibrary{snakes}
\usetikzlibrary{decorations.pathreplacing,decorations.markings,decorations.pathmorphing}

\newfloat{movie}{tbp}{lop}
\floatname{movie}{Movie}
\def\be{\begin{equation}}
\def\ee{\end{equation}}
\def\ba{\begin{eqnarray}}
\def\ea{\end{eqnarray}}

\def\ep{\varepsilon}
\def\RAdS{R_{\rm AdS}}

\def\<{\langle}
\def\>{\rangle}
\def\.{{\cdot}}
\def\zb{\overline{z}}

\def\cO{\mathcal{O}}
\def\cP{\mathcal{P}}
\def\cG{\mathcal{G}}
\def\cC{{\mathcal{C}}}
\def\cU{{\mathcal{U}}}
\def\cM{{\mathcal{M}}}
\def\GN{{G_N^{(d+1)}}}

\def\e{\epsilon}
\def\Df{{\Delta_{\phi}}}
\def\nbar{\overline{n}}
\def\phiex{\<\phi\cO\phi\>}

\def\scrI{\mathcal{I}}
\def\cmark{\ding{51}}%
\def\xmark{\ding{55}}%
\DeclareMathOperator*{\Res}{Res}
\def\sopt{\sigma_{\rm o}}

\title{Holographic Cameras: An Eye For The Bulk}
\author{Simon Caron-Huot$^{a,b}$}
\affiliation{$^a$Department of Physics, McGill University, 3600 Rue University, Montr\'eal, H3A 2T8, QC Canada}
\affiliation{$^b$School of Natural Sciences, Institute for Advanced Study, Princeton, NJ 08540, USA}
\emailAdd{schuot@physics.mcgill.ca}
\abstract{
We consider four-point correlators in an excited quantum state of a field theory.
We show that, when the theory and state are holographic,
a judiciously applied Fourier transform produces high-quality images of point-like bulk particles, revealing the geometry in which they move. For translation-invariant states, the bulk Einstein's equations amount to
local differential equations on correlator data.
In theories or states that are not holographic, images are too blurry to extract a bulk geometry.
We verify this for gauge theories at various couplings and the 3D Ising model by adapting formulas from conformal Regge theory.
}

\begin{document}

\maketitle
\pagebreak

\section{Introduction}

The AdS/CFT or gauge-gravity correspondence shows that
the dynamics of certain strongly interacting quantum systems may be better understood
in terms of a dual ``bulk'' system, which enjoys one additional dimension of space and dynamical gravity \cite{Maldacena:1997re}.
Gravity is not part of the original system but only emerges in the bulk.
The new coordinate is loosely correlated with the size of an excitation, but, 
remarkably, small relative changes in it can represent sharply distinct bulk points: the bulk dynamics is local
down to scales much shorter than its curvature radius.
There has been much progress in understanding which quantum systems admit local duals
(building on the celebrated large-$N$ large-gap criterion \cite{Heemskerk:2009pn})
and on how to reconstruct various aspects of the bulk. Yet, the emergence of a bulk and its
associated entanglement structure from the boundary dynamics remain largely mysterious.
Most basically, if a system holographically encodes a bulk universe, why should it hide it?

The purpose of this note is to explore a decoding tool which comes with any holographic system: its own Hamiltonian.
A straightforward Fourier transform applied to four-point correlators will produce easily-interpretable images of
point particles moving in a non-trivial bulk geometry.  We dub the resulting devices \emph{holographic cameras}.

Since the cameras are relatively straightforward to describe (theoretically), let us introduce them immediately and give more context below. Half of the device is an ``holographic canon'' which shoots a particle into the bulk.
Starting from an arbitrary state $|\Psi\>$, we perturb it by acting with a local operator $\cO$ integrated against a wavepacket
\be
\mbox{cannon: }\quad |\Psi\> \mapsto \int_x\ \psi_{p,L}^*(x) \cO(x)|\Psi\>\quad\mbox{with}\quad  \psi_{p,L}(x) \sim e^{ip_\mu x^\mu-|\delta x|^2/(2L^2)}.  \label{excited state}
\ee
The wavepacket will be assumed to create high-energy excitations with relatively well-defined energy and momenta ($p^0L\gg 1$);
the position uncertainty $L$ should be small compared with intrinsic time and length scales of the state.
If the field theory admits long-lived quasi-particles with momentum near $p^\mu$, eq.~\eqref{excited state} would be a familiar way to excite them by driving them at resonance.
Rather, we will consider momenta $p^\mu$ that do \emph{not} correspond to boundary quasi-particles.
If the system admits a dual description in terms of bulk quasi-particles, eq.~\eqref{excited state} will instead excite them.

Concretely, we will mostly study relativistic conformal field theories, where $p^\mu$ is a timelike vector.
We will imagine that $\cO(x)$ is a simple operator such as the electromagnetic current $J^\mu(x)$ or the stress tensor $T^{\mu\nu}(x)$,
and treat it like a light scalar for simplicity.

To confirm that a projectile has indeed been fired, one could measure one-point functions in the state \eqref{excited state}.
Removing an inessential integration, this boils down to three-point functions
\be
\int_{\delta x}\ \psi_{p,L}(\delta x)\  \<\Psi|\ \cO(x{+}\tfrac12\delta x)\ \cO'(y)\ \cO(x{-}\tfrac12\delta x)\ |\Psi\>.
\label{excited 3pt}
\ee
The only integration here is over the separation $\delta x^\mu$ between the operators that create and absorb the excitation;
this suffices to control its energy and momentum.
Such correlators have been much studied and will be further discussed below.
If $\cO'$ is the stress tensor, for example, then \eqref{excited 3pt} measures the energy and momentum density at $y$.
In holographic theories in the vacuum, this reveals an essentially spherical expanding shell of energy \cite{Hofman:2008ar}.
One could also replace the state $|\Psi\>$ by a density matrix, ie.~a thermal state (to keep our notations light we won't indicate this).
In a thermal state, correlators like \eqref{excited 3pt} with boosted $p^\mu$
can probe jet-like excitations of a strongly interacting plasma (see \cite{Arnold:2011qi} and references therein).
In addition to the energy and charge spreading out with time,
among other features, one observes the jet center-of-energy slowing down and eventually stopping as it dissipates into the medium. 
These observations are all compatible with a dual picture of a point-like (or stringy) projectile 
reversibly moving in a bulk gravitational field. However, they are admittedly rather blunt probes of the bulk, and no smoking guns.

To better image the bulk itself, we propose two variants of \emph{holographic cameras}.
The first is inspired by a venerable technology: the radar. The second (``active'') camera
is more exotic and requires the ability to evolve small perturbations backward in time.

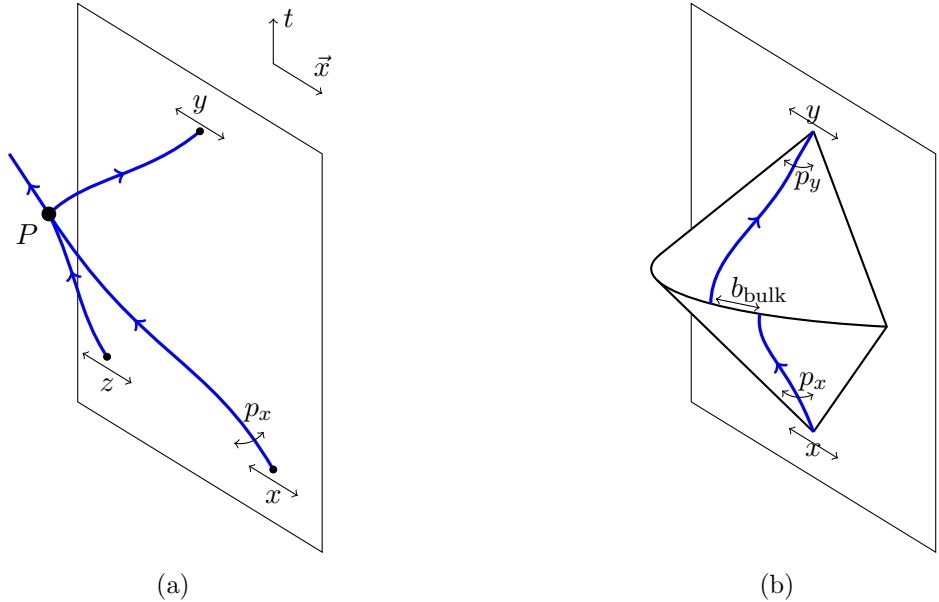
\begin{figure}[t]\centering
\begin{subfigure}[b]{0.40\textwidth}\centering\begin{tikzpicture}[xscale=1.3]
\draw (0,-0.1) -- (2.5,-2.1) -- (2.5,3.2) -- (0,5.2) -- (0,2.2) --cycle;
\draw[very thick,blue,decoration={markings,mark=at position 0.6 with \arrow{>}},postaction=decorate] (2.0,-1) .. controls (1.325,0.5) and (0.65,0.5) .. (-0.295,2.4);
\draw[very thick,blue,decoration={markings,mark=at position 0.5 with \arrow{>}},postaction=decorate] (-0.295,2.4) -- (-0.7,3.2);
\draw[very thick,blue,decoration={markings,mark=at position 0.6 with \arrow{>}},postaction=decorate] (0.3,0.5) .. controls (0,1.1) and (0,1.6) .. (-0.295,2.4);
\draw[very thick,blue,decoration={markings,mark=at position 0.5 with \arrow{>}},postaction=decorate] (-0.295,2.4) .. controls (0.1,2.9) and (0.7,2.9) .. (1.25,3.5);
\filldraw (-0.295,2.4) ellipse (2pt and 2.6pt);  \node[below left] at (-0.3,2.4) {$P$};
\draw[<->] (2,5) -- (2,4.4) -- (2.5,4.0);
\node [right] at (2,5) {$t$}; \node [above] at (2.5,4.1) {$\vec{x}$};
\filldraw (2,-1) ellipse (1pt and 1.3pt);
\draw[<->] (1.75,-0.95) -- (2.25,-1.35); \node[below] at (2,-1.15) {$x$}; 
\draw[<->] (1.6,-0.65) [out=-10,in=-120] to (1.9,-0.5); \node [above] at (1.85,-0.5) {$p_x$}; 
\filldraw (0.3,0.5) ellipse (1pt and 1.3pt);
\draw[<->] (0.05,0.55) -- (0.55,0.15);  \node [below] at (0.3,0.35) {$z$};
\filldraw (1.25,3.5) ellipse (1pt and 1.3pt);
\draw[<->] (1,3.8) -- (1.5,3.4); \node [above] at (1.25,3.6) {$y$};
\end{tikzpicture}\caption{\label{fig:radar}}\end{subfigure}
\hspace{0.1\textwidth}
\begin{subfigure}[b]{0.40\textwidth}\centering\begin{tikzpicture}[xscale=1.3]
\draw (0,1.08) -- (0,-0.1) -- (2.5,-2.1) -- (2.5,3.2) -- (0,5.2) -- (0,2.2);
\draw[thick] (1.25,-0.5) -- (2,0.9) -- (1.25,3.5); 
\draw[thick] (2,0.9) arc (255:166:3.25 and 0.8);  
\draw[thick] (1.25,-0.5) -- (-0.35,1.52);  \draw[thick] (-0.32,1.86) -- (1.25,3.5);
\draw[very thick,blue,decoration={markings,mark=at position 0.6 with \arrow{>}},postaction=decorate] (1.25,-0.5) .. controls (1,0.4) and (0.65,0.6) .. (0.7,1.07);
\draw[very thick,blue,decoration={markings,mark=at position 0.5 with \arrow{>}},postaction=decorate] (0.2,1.2) .. controls (0.2,1.9) and (0.8,2.3) .. (1.08,3.12) -- (1.25,3.5);
\draw[<->] (1,-0.45) -- (1.5,-0.85); \draw[<->] (1,3.8) -- (1.5,3.4);
\node[below,inner sep=0pt] at (1.25,-0.65) {$x$}; \node [above,inner sep=0pt] at (1.25,3.6) {$y$};
\draw[<->] (0.7,1.15) -- (0.25,1.27) ; \node [above right] at (0.3,1.1) {$b_{\rm bulk}$};
\draw[<->] (0.93,0.05) [out=-50,in=-150] to (1.25,0.0); \node [above] at (1.25,-0.08) {$p_x$}; 
\draw[<->] (0.95,3.12) [out=-50,in=-150] to (1.25,3.05); \node [below] at (1.2,3.1) {$p_y$};
\end{tikzpicture}\caption{\label{fig:camera}}\end{subfigure}
\caption{\label{fig:cameras}
Two models of holographic cameras.
In both, an energetic projectile is fired from a boundary point $x^\mu$ with initial momentum $p_x^\mu$.
The plane represents the $R^{1,d-1}$ boundary spacetime in which we can do measurements,
and the thick blue paths are null geodesics in the (arbitrary) bulk metric.
(a) In the radar camera \eqref{radar}, one sends a pulse from a point $z$
and records at $y$ its reflection off the projectile.
(b) In the active camera \eqref{camera}, one
records how the projectile interferes with the reverse time evolution of a second one reaching $y$ with momentum $p_y$.
In a theory with local bulk dynamics, this signal will be sharply peaked around vanishing bulk impact parameter $b_{\rm bulk}$.
}
\end{figure}

In the ``holographic radar'', the projectile's position is imaged by sending in a pulse from a point $z$,
whose reflection is then recorded at a later point $y$. This can be modelled by the correlator:
\be\label{radar}
\mbox{radar:}\quad \int_{\delta x} \psi_{p,L}(\delta x) \ \<\Psi| \cO(x{+}\tfrac12\delta x) \ \cO'(y)\ \cO'(z)\ \cO(x{-}\tfrac12\delta x) |\Psi\>\,.
\ee
We are imagining that $x$ and $z$ are spacelike from each other and $y$ is in their common future, see figure \ref{fig:radar}.
The operator ordering is that of a two-point function in the state $\cO|\Psi\>$.
The device is meant (ideally) to operate in the regime of bulk geometrical optics.
For suitable choice of $z$, the forward lightcone of $z$ intersects the projectile's trajectory at a point $P$,
and its reflection from there is visible at any point $y$ along the future lightcone of $P$.
The signature of a bulk point-like projectile is a sharp singularity or peak as $y$ approaches that lightcone.

The holographic ``active camera'' instead uses two pairs of nearly coincident points,
arranged in an out-of-time-order fashion that combines a cannon and its time-reversed version.
It is defined mathematically by the following correlator:
\be
\begin{aligned}
\mbox{camera:}\quad   G(x,p_x,L_x; y,p_y,L_y) &\equiv \int_{\delta x,\delta y}\psi_{p_x,L_x}(\delta x)\psi_{p_y,L_y}(\delta y)
\\ & \hspace{-10mm}\times
\<\Psi|\ \cO'(y{+}\tfrac12\delta y)\ \cO(x{+}\tfrac12\delta x)\ \cO'(y{-}\tfrac12\delta y)\ \cO(x{-}\tfrac12\delta x)\ |\Psi\>\,.
\label{camera}
\end{aligned}\ee
We imagine that $y$ is in the future of $x$ ($y{\succ} x$), as depicted in figure \ref{fig:camera}.
The out-of-time-ordering of operators (``$yxyx$") in \eqref{camera} is unusual but crucial to this device.
It means that an excitation created near $x$ by the rightmost $\cO$ is first time-evolved to $y$,
where a second excitation is produced with $\cO'$. However, the two excitations are then evolved \emph{backward}
to the original time $x^0$!
This yields a ket $\cO'\cO|\Psi\>$ at $x^0$ which is then compared with a second ket $(\cO\cO'|\Psi\>)^\dagger$, in which
$\cO'$ was applied first.
In words, eq.~\eqref{camera} measures how creating $\cO$ interferes with one's ability to evolve $\cO'$ backward in time.

The smoking gun that $\cO$ and $\cO'$ produce point-like bulk objects is a sharp peak around shooting angles where their trajectories intersect ($b_{\rm bulk}\to 0$ in figure \ref{fig:camera}).
Tracking this peak as a function of $y$ and $p_y$ produces a movie of the bulk trajectory of $\cO$.
If the peak is sharp, a bulk's geometry may be inferred,
somewhat analogously to how the gravity fields of Earth and Sun are inferred from trajectories of falling objects and moving planets.
The radar-like system can produce similar movies by varying the time $z$ at which the pulse is sent.

The Fourier transforms near $x$ and $y$ will resolve bulk angles rather similarly to how
a lens (or biological eye) resolves the arrival angle of photons.
Angular resolution will be essentially the ratio of wavelength to detector size: $\delta\theta_i \sim 1/(p_iL_i)$ where $i=x,y$.
Fourier transforms with respect to $y$ and $z$ could also be incorporated in the radar-like system \eqref{radar} to control
the energy and angles of the pulse. Both cameras resolve the bulk more finely than the 3-point function in \eqref{excited 3pt} because
high probe energies grant access to short wavelengths.\footnote{
Integrating $\cO'(z)$ against a high-frequency Gaussian can make the pulse obey geometrical optics.
Two-point functions transformed in this way can image Einstein rings and other features of AdS black holes
\cite{Hashimoto:2018okj,Hashimoto:2019jmw,Kaku:2021xqp}.
}

The active camera differs from any mass-market camera in a fundamental way: instead of
passively recording photons reaching it, \eqref{camera} actively gathers information about the past.
This counter-intuitive property is related to the out-of-time-ordering of the operators.
This correlator can only be ``measured'' without violating causality
if one has the ability to use a system's Hamiltonian to evolve small perturbations backward in time.
This is somewhat related to the precursor concept \cite{Polchinski:1999yd}.
The radar camera, on the other hand, is given by a straightforward expectation value
that is measurable without such a super-power.

Both cameras come with a few knobs: shooting positions $x^\mu$, $y^\mu$ (and $z^\mu$), energy-momenta $p_x^\mu$ and possibly $p_y^\mu$, and optical sizes $L_i$ for each wavepacket.
These parameters can be tuned to reveal narrow peaks as a function of angles or arrival times, thus bringing the bulk into focus.
Intuitively, it is desirable that the probes have large enough energies to satisfy geometrical optics,
yet small enough that we can ignore backreaction on the background.
Success is not guaranteed: we will describe situations where crisp movies can be produced, and others that only lead to inconclusive blurs.

In this paper, we will mostly discuss the active camera.
This is for a superficial reason: the author finds two null geodesics easier to visualize than three. The out-of-time-order correlator in \eqref{camera} also seems to enjoy interesting theoretical properties.

Out-of-time-ordered correlators have been much discussed theoretically in recent years.
In certain regimes, they display exponential growth with time:
$\< \Psi| [\cO_{t_1}, \cO'_{t_2}]^2 |\Psi\> \sim e^{\lambda_* |t_2-t_1|}$, with $\lambda_*$
interpreted as a quantum analog of Lyapunov exponents in chaotic systems \cite{Shenker:2013pqa,Maldacena:2015waa}.
In thermal equilibrium, the exponent satisfies the bound on chaos $\lambda_* \leq \frac{T}{2\pi}$, which
is famously saturated by systems holographically dual to black holes.
The focus of this paper is on ``pre-chaos'' times $|x-y|$ that are not necessarily large compared to intrinsic timescales of the system,
only large compared with the energies $p_x^0$, $p_y^0$ used to probe it.
Nonetheless, the bulk description will be relatively similar:
we will find for example tree-level graviton exchange between the geodesics in figure \ref{fig:camera}.

An extensive literature is devoted to reconstructing aspects of bulk physics from boundary observables.
We will not attempt to review it here (see for example \cite{Kajuri:2020vxf}).
Perhaps the earliest ideas exploit the so-called Fefferman-Graham expansion:
given one-point functions of the stress tensor $\<T^{\mu\nu}(x)\>_\Psi$ in some state,
Einstein's equations can be radially integrated to reconstruct a bulk metric $g_{ab}(x,r)$, at least up to some depth.
Relatedly, the HKLL procedure inverts bulk equations of motion to express (fluctuations of) bulk fields as specific, background-dependant, integrals over boundary operators \cite{Hamilton:2006az}.
Various geometric data, such as lengths of bulk geodesics or areas of minimal surfaces,
are also related by the bulk dynamics to boundary correlators or entanglement entropy, see \cite{Hubeny:2012ry}
for a general discussion.
In such ways, bulk physics can be reconstructed from boundary measurements by solving various inverse problems.

A key facet of the ``cameras'' is that they can image local bulk excitations with no prior knowledge of bulk equations of motion or of its geometry. Little to no data analysis required, for that matter: the system's own Hamiltonian does the hard work!
While the images may sometimes be used to reconstruct a bulk geometry, the true purpose of the cameras
seems rather complementary: we want to \emph{observe} bulk physics.

An eventual application we have in mind is to theoretically study strongly coupled quantum systems.
From this perspective, we find it appealing to conceptualize the bulk geometry in terms of four-point correlators
that satisfy self-consistency conditions like OPE associativity. 
Not having to refer to bulk equations of motion could also be valuable in situations where those are not precisely known.
Finally, we find it encouraging that a bulk geometry is only inferred \emph{when one exists}.%
\footnote{
Consider for example an excited state in the 3D Ising model.
This model is not generally believed to be dual to Einstein's gravity, but an enthusiastic student could still plug
the one-point data $\<T^{\mu\nu}(x)\>_\Psi$ of some state into Einstein's equations and calculate a bulk metric.
Unfortunately, this procedure gives no insight as to whether this metric holds any significance for the 3D Ising model:
this can only be decided by further calculations and tests.
In contrast, an experimentalist who could measure \eqref{radar} or \eqref{camera} could
readily judge from their data (from the sharpness and shape of peaks) whether inferring a bulk metric is sensible.}

Our approach bears similarity with the bulk-point limit of correlators and light-cone cuts \cite{Maldacena:2015iua,Engelhardt:2016wgb}, where one attempts to fire energetic excitations from the boundary onto a common bulk point.
At a technical level, we will add little new: the correlators \eqref{radar}-\eqref{camera} differ from earlier ones by tweaks on
operator ordering and Fourier wavepackets.
Qualitatively, these tweaks are significant.
They enable to focus on a given bulk point $P$ using fewer boundary points and a much smaller boundary footprint,
just using two (or three) small open sets that intersect the past and future lightcones of $P$.
The images also seem easier to interpret, since only two (or three) bulk geodesics need to be visualized, in any spacetime dimension.
Finally, they show that simple correlators can simultaneously reveal the local dynamics of bulk excitations and the geometry they evolve in.

As mentioned, our key formulas will be adaptations of existing ones (from eikonal approximations and conformal Regge theory).
However, since their derivations may not be widely known and have many subtle points,
we felt that reviewing them before adapting them might not do proper justice to the original papers.
Instead, we opt to give a self-contained presentation from a hopefully uniform perspective,
referring to original sources along the way.

This paper is organized as follows.
Section \ref{sec:eikonal} is devoted to the eikonal approximation \eqref{G naive} for the out-of-time-ordered correlator in
a general background, including useful limits in subsection \ref{ssec:limits}.
Most of the discussion is based on technical examples, while the role
of symmetries and large-$N$ expansions is discussed more qualitatively in subsection \ref{ssec:abstract}.
Section \ref{sec:CRT} explains how the camera correlators, for conformal theories in the vacuum states,
is captured by conformal Regge theory.
In particular, relations between various perspectives are explained in \ref{ssec:relations}.
The extensive literature on Regge theory enables us to plot the correlators in theories that go beyond gravity,
including supersymmetric Yang-Mills theory at weak and strong coupling and the critical 3D Ising model.
The importance of the large-N large-(higher-spin)-gap criterion to obtain sharp images is emphasized.

Section \ref{sec:BTZ} shows the images obtained in a thermal state of a (strongly coupled)
CFT${}_2$, and how these images straightforwardly
reveal a bulk metric. In particular, for any translation-invariant state in CFT$_d$, the bulk's Einstein equations
translates into a simple differential equation \eqref{Einstein} on correlator data.
Section \ref{sec:conclusions} summarizes our results and discusses some open questions.
Three appendices record useful formulas on: Fourier transforms and harmonic analysis, shockwaves in AdS space,
and conformal Regge theory.

\section{Review and extension of eikonal approximation} \label{sec:eikonal}

When the theory and state $|\Psi\>$ admit a holographic description,
we expect to find a regime where the correlator \eqref{camera} is dominated by tree-level graviton exchange
between the null geodesics in \ref{fig:camera}.
The purpose of this section is to elaborate on eq.~\eqref{G naive} below, which expresses this.
The reader satisfied by this formula could in principle skip to section \ref{sec:CRT},
since this section and the next one are largely complementary to each other.

Let us state directly the key formula which expresses the correlator \eqref{camera} in the bulk tree-level
approximation, at the same time setting up notation:
\be
 G(x,p_x; y,p_y) \approx 1- 8\pi i \GN \int_{-\infty}^{u_0} du \int_{v_0}^{+\infty} dv \ \< h_{uu}(X) h_{vv}(Y) \>_{\Psi,\rm ret}\,. 
\label{G naive}
\ee
The integrand is a bulk-to-bulk propagator for metric perturbations;
$X(u)$ and $Y(v)$ are coordinates along null geodesics with initial conditions $(x,p_x)$ and $(y,p_y)$,
ie. the blue curves in figure \ref{fig:camera},
and the times $u$ and $v$ are suitably normalized affine times (see \eqref{p geodesic} and \eqref{p' geodesic} below).
Intuitively, \eqref{G naive} arises from the classical coupling between a point particle and a metric perturbation,
\be
 S= -m\int d\tau \sqrt{-g_{\mu\nu}\dot{x}^\mu \dot{x}^\nu} \quad\Rightarrow\quad
 \delta S =\frac{\kappa}2 \int P^\mu  h_{\mu\nu} dx^\nu \quad\mbox{with}\quad
 P^\mu \equiv \frac{m \dot{x}^\mu}{\sqrt{-g_{\rho\sigma}\dot{x}^\rho\dot{x}^\sigma}}\,, \label{naive S}
\ee
where we write perturbations as $g_{\mu\nu}=g_{\mu\nu}^\Psi + \kappa h_{\mu\nu}$ with $g_{\mu\nu}^\Psi$ is the background.
The bulk gravitational constant is
\be
 \kappa^2 = 32\pi \GN\,.
\ee
We work in mostly-plus metric and sometimes use the following abbreviation for normalized timelike vectors, and reflection tensor:
\be
 \hat{p}^\mu = p^\mu/\sqrt{-p^2},\qquad \mathcal{I}^{\mu\nu}_x=\eta^{\mu\nu}-2x^\mu x^\nu/x^2.
\ee
Finally, we will sometimes omit the argument $L_x$, $L_y$ in formulas like \eqref{G naive}
to indicate that the finite width of wavepackets was ignored.

Conceptually, we propose to
view \eqref{G naive} as a kind of Operator Produced Expansion expressing
the limit $\delta x\to 0$ of products $\cO(x{+}\frac12\delta x)(\cdots)\cO(x{-}\frac12\delta x)$ in terms of simpler objects like $\int du h_{uu}$. This expansion is conceptually similar to how shockwaves appear in the Regge limit of correlators.
Our presentation will thus overlap greatly with \cite{Cornalba:2008qf,Cornalba:2006xk,Balitsky:2001gj,Kovchegov:2012mbw,Afkhami-Jeddi:2017rmx}, adapted to a less familiar operator ordering.
We introduce the OPE viewpoint starting with three-point functions in perturbative $\phi^4$ theory,
and gradually move on to gravity in AdS space and to four-point functions.

\subsection{Folded OPE in $\phi^4$ theory} \label{ssec:folded}

Consider massless $\lambda\phi^4$ theory in $d\approx 4$ dimensions, with action:
\be
 S=\ -\!\int d^dx\left(\tfrac12(\partial \phi)^2+\tfrac{1}{4!}\lambda\phi^4\right).
\ee
We will work perturbatively in small $\lambda$; a mass term could be added and would not modify the short-distance effects to be discussed.

Let us begin by studying a vacuum three-point function analogous to \eqref{excited 3pt}:
\be
 \phiex \equiv \<\Omega|\ \phi(\delta x)\ \cO(y)\ \phi(0)\ |\Omega\>. \label{phi4 3pt}
\ee
We would like to understand the coincidence limit $\delta x\to 0$ when $y$ is timelike from the origin.
Crucially, the operators are not time-ordered and the two $\phi$'s are not next to each other.
If that were the case,
we could simply apply the standard Operator Product Expansion to approximate $\phi(\delta x)\phi(0)$ as a series in local operators.
Instead, we will find a more interesting picture where energetic
$\phi$ quanta propagate a finite distance before being measured by $\cO(y)$.
For lack of established terminology, we refer to this as the \emph{folded OPE limit}.

To evaluate \eqref{phi4 3pt}, it is helpful to temporarily revert to a Schr\"odinger picture:
\be
 \phiex =
 \<\Omega|\ \phi(\vec{\delta x})\ (U_{y^0\leftarrow\delta x^0})^\dagger\ \cO(\vec{y})\ (U_{y^0\leftarrow0})\ \phi(\vec{0})\ |\Omega\> \label{phi ex}
\ee
where $U_{t_2\leftarrow t_1}=e^{-i H(t_2-t_1)}$ is the time-evolution operator.
Standard arguments allow to rewrite $U$ as a path integral over field configurations $\phi(x,t)$.
The same can be done for $U^\dagger$, of course.
The quirk is that the path integral now requires a doubled set of fields:
\be
 \phiex = \<\ \phi_2(\delta x)\ \cO_1(y)\  \phi_1(0)\ \>_\cC,\qquad
\<\cdots\>_\cC\equiv \frac{1}{Z_\cC} \int [d\phi]_\cC\ e^{i S[\phi_1] -i S[\phi_2]} (\cdots) \label{SK int}
\ee
where $Z_\cC$ is the result of the path integral without insertion.
The time contour $\cC$, usually called Schwinger-Keldysh contour, can be visualized as:
\be\cC:\quad
\vcenter{\hbox{
\begin{tikzpicture}
\draw (-2,0.5) -- (-2,0) -- (5,0) -- (5,-0.3) -- (-2,-0.3) -- (-2,-0.8);
\draw (2,0.1) -- (2.15,0) -- (2,-0.1);
\filldraw (0,0) circle (2pt); \filldraw (4,0) circle (2pt); \filldraw (0.5,-0.3) circle (2pt);
\node [above] at (0,0) {$\phi_1(0)$};
\node [above] at (4,0) {$\cO_1(y)$};
\node [below] at (0.5,-0.3) {$\phi_2(\delta x)$};
\draw (7,0.35) -- (7,-0.15) -- (7.5,-0.15); \node [above] at (7.25,-0.15) {$t$}; 
\node [right] at (5,0.1) {$1$}; \node [right] at (5,-0.4) {$2$};
\end{tikzpicture}}}
\label{SK contour}
\ee
Subscripts in \eqref{SK int} indicate which branch a given field lives on.\footnote{The measurement $\cO(y)$ could be inserted on either branch since the branches cancel each other outside the past lightcone of $y$.
This is the famous ``largest time equation'' \cite{tHooft:1973wag}.
}

Projection onto the vacuum state $|\Omega\>$ can be implemented by the usual $i0$ prescriptions:
the contours are slightly tilted so that ${\rm Im}(t)\to \pm\infty$ at the start and end, respectively.
More generally, an arbitrary initial state or density matrix could be inserted as boundary conditions at the left ends of the contour.

We briefly review the essential rules for perturbation theory, which can be deduced from the path integral \eqref{SK int},
or by analytically continuing the usual rules for Euclidean quantum field theory
(see \cite{Haehl:2017qfl} for a more detailed discussion and further references).
Each interaction vertex can now be inserted on either branch, and there is an overall sign change on the second branch.
Propagators between fields on the first branch are the usual time-ordered ones, and those on the second branch are anti-time-ordered.
Those connecting different branches are Wightman functions. In position space:
\be\begin{aligned} \label{propagators}
\< \phi_1(x) \phi_1(0) \>_\cC&=\frac{P_{\Df}}{(x^2+i0)^\Df}\,, \hspace{20mm}
\< \phi_2(x) \phi_2(0) \>_\cC = \frac{P_{\Df}}{(x^2-i0)^\Df}\,, \\
\< \phi_2(x) \phi_1(0) \>_\cC &= \frac{P_{\Df}}{(-(x^0-i0)^2 + \vec{x}^2)^\Df} \equiv \frac{P_{\Df}}{(x_-^2)^\Df}\,.
\end{aligned}\ee
The $i0$ prescription in the last case is easy to guess from the contour \eqref{SK contour}:
the second branch is shifted by a small negative imaginary time.
In general we denote by $x_\pm$ a vector with a small imaginary time shift: $x_\pm^0=x^0\pm i0$.
The exponent is $\Df=\frac{d-2}{2}$; the normalization $P_{\Df}$, in \eqref{PDelta}, won't be too important here.

The formalism manifests the fact that points on different branches of \eqref{SK contour}  are not ``close'': a particle
cannot connect them without going around the fold.

Nonetheless, the correlator \eqref{phi4 3pt} is singular as $\delta x\to 0$. Let us see this explicitly for
the connected diagram at order $\lambda$.
Accounting for the contributions from each branch, we have
\be \label{phi4 lambda}
 \phiex\big|_{\lambda} =
-i\lambda P_\Df^2\int d^dx \left[ \frac{\< \frac12\phi^2_1(x) \cO_1(y)\>}{(x^2+i0)^\Df [(x{-}\delta x_-)^2]^\Df}
-\frac{\< \frac12\phi^2_2(x) \cO_1(y)\>}{(x_-^2)^\Df [(x{-}\delta x)^2-i0]^\Df}
\right].
\ee
We kept the numerator unevaluated, anticipating a passive spectator role for $\cO_1(y)$.

A standard Landau-(Coleman-Norton) analysis indicates that singularities in position space come from integration regions
representing classical paths of particles with arbitrary high energy (see for example \cite{Alday:2010zy} or
section 2 of  \cite{Maldacena:2015iua}).
The defining feature of the folded OPE limit is that \emph{infinitely} many such paths exist:
any lightray leaving the origin on the first fold will return to it on the second fold.\footnote{
On the second branch, positive-frequency excitations propagate backward in time.}
The $\delta x\to 0$ singularity comes from the full lightcone!

\begin{figure}[t]\centering
\begin{subfigure}[b]{0.4\textwidth}\raisebox{12mm}{\begin{tikzpicture}[xscale=1.3]
\draw[thick] (-2.07,2.07)--(0,0)--(2.07,2.07);
\draw[thick,shift=(-5:2.128 and 0.6)] (0,2.21) arc (-5:-123:2.128 and 0.6);
\draw[thick,shift=(-175:2.128 and 0.6)] (0,2.21) arc (-175:-128:2.128 and 0.6);
\draw[thick,dashed,shift=(-5:2.128 and 0.6)] (0,2.21) arc (-5:185:2.128 and 0.6);
\draw[very thick,blue,decoration={markings,mark=at position 0.55 with \arrow{<},mark=at position 0.95 with \arrow{<}},postaction=decorate] (0,0)--(-1.5,2.1) arc (215.5:31.8:0.1) -- (-1.11,1.79);
\draw[very thick,blue] (0,0.1)--(-0.98,1.58);
\filldraw (0,0) ellipse (1pt and 1.3pt);  \node [below] at (0,0) {$x{+}\delta x$};
\filldraw (0,0.1) ellipse (1pt and 1.3pt);  \node [above] at (0,0.2) {$x$};
\draw[<->] (-0.53,0.6) [out=-50,in=-140] to (-0.1,0.6); \node [above,inner sep=2pt] at (-0.2,0.6) {$n$}; 
\begin{scope}[shift={(0.15,0.05)}]
	\draw[thick,<->] (-1.94, 3)--(-1.576, 2.4)--(-1.1, 3);
	\node[above left,inner sep=1pt] at (-1.94,3) {$u$}; \node[above right,inner sep=1pt] at (-1.1,3) {$v$};
\end{scope}
\filldraw (0,4) ellipse (1pt and 1.3pt); \node [right] at (0,4) {$y$};
\filldraw[blue] (-0.7, 1.14) ellipse (1pt and 1.3pt);
\draw[thick,dash pattern=on 10pt off 8pt on 135pt off 9pt] (-0.7, 1.14) [out=85,in=-180+65] to (0,4) [out=-180+85,in=65] to (-0.7, 1.14);
\end{tikzpicture}}
\caption{$\lambda\phi^4$ interaction in field theory \label{fig:folded phi4}}
\end{subfigure}
\hspace{0.1\textwidth}
\begin{subfigure}[b]{0.4\textwidth}
\begin{tikzpicture}[xscale=1.3]
\draw (0,1.08) -- (0,-0.1) -- (2.5,-2.1) -- (2.5,3.2) -- (0,5.2) -- (0,2.2);
\draw[thick] (1.25,-0.5) -- (2,0.9) -- (1.25,3.5);
\draw[thick] (2,0.9) arc (255:221:3.25 and 0.8); \draw[thick] (-0.31,1.87) arc (166:215:3.25 and 0.8);
\draw[thick] (1.25,-0.5) -- (-0.35,1.52);  \draw[thick] (-0.32,1.86) -- (1.25,3.5);
\draw[<->] (2,5) -- (2,4.4) -- (2.5,4.0);
\node [right] at (2,5) {$t$}; \node [above] at (2.5,4.1) {$\vec{x}$};
\begin{scope}[shift={(1.75,-0.4)}]
	\draw[thick,<->] (-1.9, 3)--(-1.54, 2.245)--(-1.14, 2.85);
	\node[above left,inner sep=1pt] at (-1.9,3) {$u$}; \node[above right,inner sep=1pt] at (-1.14, 2.85) {$v$};
\end{scope}
\draw[<->] (0.7,0.24) [out=-50,in=-150] to (1.1,0.2); \node [above] at (1.05,0.1) {$\theta$}; 
\draw[very thick,blue,decoration={markings,mark=at position 0.55 with \arrow{<},mark=at position 0.95 with \arrow{<}},postaction=decorate] (1.25,-0.5) -- (0.1,1.5) arc (205:25:0.1) -- (0.44,1.22);
\draw[very thick,blue] (1.25,-0.4) -- (0.53,1.01);
\filldraw (1.25,-0.4) ellipse (1pt and 1.3pt); \node[above,inner sep=2pt] at (1.25,-0.25) {$x$}; 
\filldraw (1.25,-0.5) ellipse (1pt and 1.3pt); \node[below,inner sep=0pt] at (1.25,-0.5) {$x{+}\delta x$}; 
\filldraw[blue] (0.68, 0.7) ellipse (1pt and 1.3pt);
\draw[decorate, decoration={snake,amplitude=1.6,segment length=7}] (0.68,0.7)--(1.25,3.5);
\node [above,inner sep=0pt] at (1.25,3.6) {$y$};
\end{tikzpicture}
\caption{$h\partial\phi\partial\phi$ interaction in bulk gravity \label{fig:folded gravity}}
\end{subfigure}
\caption{
Folded OPE limit of three-point functions, in perturbative field theory and in perturbative gravity.
Singularities as $\delta x\to 0$ come from paths which leave $x$ on the first timefold and return nearby on the second timefold,
approximately following null geodesics (thick blue lines).
The two panels need not be continuously connected: boundary and bulk perturbation theory don't normally overlap.
They look technically similar because the
Landau analysis applies to any perturbative description, whether it exists in the original spacetime or in a dual one.
}
\end{figure}
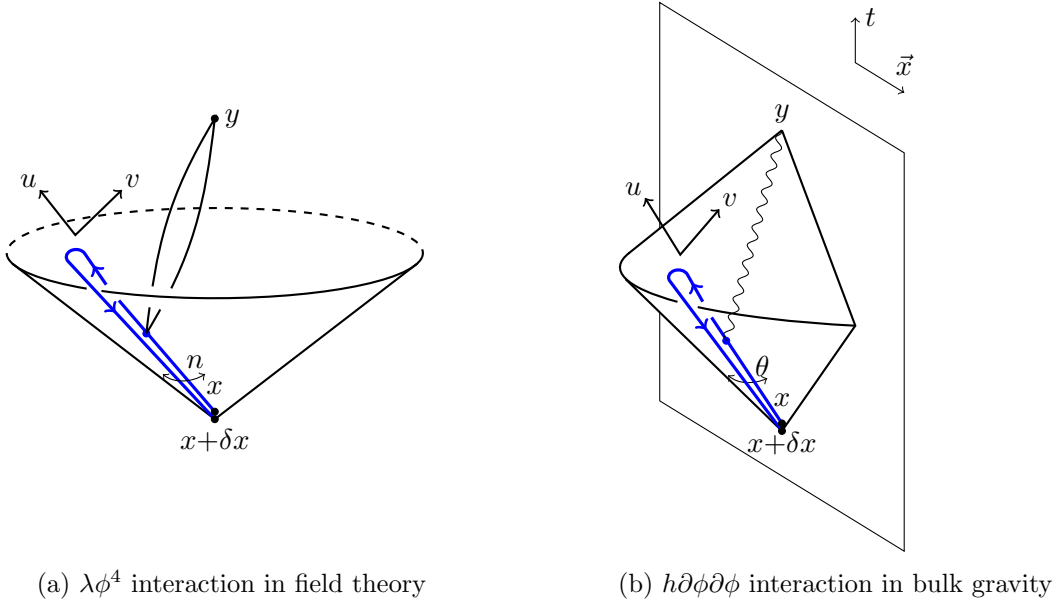

This can be confirmed by expanding \eqref{phi4 lambda} near the cone,
as depicted in figure \ref{fig:folded phi4}.
We parametrize it in terms of a null vector $n^\mu$
\be
 x^\mu = u n^\mu + v \nbar^\mu \qquad\mbox{with}\qquad n^\mu = (1,\hat{n}), \qquad \nbar^\mu = \tfrac12(1,-\hat{n})
\ee
where $v\to 0$ is a retarded time near the cone,
$u$ is an affine time along lightrays, and $\hat{n}\in S^{d-2}$ is a unit vector in $d-1$ dimensions.
(Normalized so that $n{\cdot}\nbar=-1$.)
The integral \eqref{phi4 lambda} becomes
\be
 \phiex\big|_{\lambda} =
-i\lambda P_\Df^2 \int \frac{(u-\tfrac{v}{2})^{d-2}d^{d-2}n\ du\ dv\ \< \frac12\phi^2_1(un+v\nbar)_1 \cO_1(y) \>_\cC}
{(-2uv+i0)^{\Df}\big[{-}2u(v+n\.\delta x+i0) +\ldots \big]^{\Df}}
+\mbox{($\phi_2^2$ term)}
\label{scalar integral near cone}
\ee
where we dropped subdominant terms as $\delta x\to 0$. (Explicitly, ``$\ldots$"$=+\delta x\.(\delta x-2v\nbar)$.)
The salient feature is that as $\delta x\to 0$, the $v$ contour gets pinched between branch points:
\be
\vcenter{\hbox{
\begin{tikzpicture}
\draw (-3,0) -- (-0.4,0) .. controls (0,-0.3) .. (0.4,0) -- (5,0);
\draw (-1.15,0.1) -- (-1,0) -- (-1.15,-0.1);
\draw [decorate,decoration={snake,zigzag}] (-0.1,0.0) .. controls (0.58,0.24) .. (5,0.2);
\draw [decorate,decoration={snake,zigzag}] (0.58,-0.2) -- (5,-0.2);
\draw (7,0.35) -- (7,-0.15) -- (7.5,-0.15);
\node [above] at (7.25,-0.15) {$v$}; 
\filldraw (-0.1,0) circle (2pt); \filldraw (0.58,-0.2) circle (2pt);
\node [above] at (-0.1,0.15) {$0{+}i0$};
\node [below] at (0.58,-0.3) {$-n\.\delta x{-}i0$};
\end{tikzpicture}}} \nonumber
\ee
We drew the picture for $u>0$.  For $u<0$, the two cuts are on the same side and there is no pinch.
Thus only the future lightcone contributes a singularity.
Assuming that the numerator of \eqref{scalar integral near cone} is regular as $v\to 0$, we can neglect
its $v$-dependence and analytically integrate over $v$, see $I_{\Df,\Df}$ in \eqref{I12 result}.
The contribution from the second timefold gives the same up to sign, so the result is\footnote{
For reference, the overall constant is $\tilde{P}_d= P_{\Df}I_{\Df,\Df} = 
\frac{\Gamma(2\Df-1)}{16\pi^{d-1}}e^{-\frac{i\pi}{2}(2\Df-1)}$ using $\Df=\frac{d-2}{2}$.
}
\be
\phiex\big|_{\lambda} \to -i\lambda  \int \frac{\tilde{P}_d\ d^{d-2}n}{(-2n\.\delta x_-)^{2\Df-1}}
 \int_0^\infty du \<\tfrac12(\phi_1^2-\phi_2^2)\cO_1(y)\>_\cC + \mbox{(less singular as $\delta x\to 0$)}. \label{phi product}
\ee
Let us interpret this result.  Crucially, $\cO_1(y)$ plays a passive role, as anticipated.
The result is thus really a statement about products of $\phi$'s,
\be
 \phi_2(\delta x)\ \phi_1(0) \supset 
 -i\lambda \int \frac{\tilde{P}_d\ d^{d-2}n}{(-2n\.\delta x_-)^{2\Df-1}} L_{n,\cC}[\tfrac12\phi^2] \label{phi4 OPE}
\ee
in the sense that inserting either side in a path integral gives the same result as $\delta x\to 0$.
We defined the ``light transform'' along the Schwinger-Keldysh contour $\cC$:
\be
 L_{n,\cC}[O]= \int_0^{\infty} du \left[ O_1(un) - O_2(un) \right] \equiv \int_{\cC} du\ O(un)
\ee
where the integral starts from the origin $u=0$ on the first branch and ends at the origin on the second branch.

The inclusion $\supset$ in \eqref{phi4 OPE} indicates that the full OPE contains many other terms,
notably the identity (proportional to $\< \phi_2(\delta x)\ \phi_1(0)\>$) and  bilinears in $\phi$ (so-called double-twists),
which exist already in the non-interacting theory.
The matching calculations necessary to disentangle all those are in fact particularly subtle in $\phi^4$ theory
($\phi^2$ being itself a double-twist). This will be discussed below
and needs not concern us yet: here we are simply describing a way to rewrite the $\delta x\to 0$ limit of
one diagram in \emph{some} simpler way.

Eq.~\eqref{phi4 OPE} captures the singular part of correlators like \eqref{phi ex}
as $\delta x\to 0$.  It represents physically a $\phi$ excitation moving along the lightlike direction $n^\mu$
and interacting with the external world through
a coupling $\int du \frac{-i\lambda\phi^2}{2}$ along its worldline. The angular distribution of the excitation
depends on how the limit $\delta x\to 0$ is approached. This is described by the measure $\frac{\tilde{P}_d\ d^{d-2}n}{(-2n\.\delta x_-)^{2\Df-1}}$.
For example, if $\delta x$ is purely in the time direction, then the distribution is spherically symmetric.

The integrand in \eqref{phi product} is a retarded product supported within the past lightcone of $y$.
This makes the $u$ range effectively finite: whatever happens after the measurement can't affect the measurement.

It is easy to see that \eqref{phi4 OPE} holds even if the quantum state of the theory is not the vacuum.
The essential requirement was that neither the state $|\Psi\>$ nor $\cO(y)$ create
sharp features within a distance $v\sim n\.\delta x$ of the lightcone,
so as to not invalidate the expansion leading to \eqref{phi product}.\footnote{In a conformal field theory, the OPE of any two operators acting on the vacuum is convergent for any $\delta x^\mu$ \cite{Mack:1976pa}.
We do not know whether the OPE that starts with \eqref{phi4 OPE} is convergent or only asymptotic.}

It is instructive to evaluate \eqref{phi4 OPE} for a concrete three-point function in the vacuum,
say with $\cO=\phi^2$.  The diagram shown in figure \ref{fig:folded phi4}, for one value of $n$, gives
the integral
\be
 \< L_{n,\cC}[\tfrac12\phi^2] \ \phi_1^2(y) \>_\cC = \int_\cC \frac{du\ P_{\Df}^2}{(y^2-2u n\.y)^{d-2}} = \frac{-2i\sin(\pi(d{-}3))}{d-3}
 \frac{P_{\Df}^2}{(-y^2)^{d-3}(-2n\.y)}\,. \label{vanishing u integral}
\ee
The final integration over $n$ is generally nontrivial, but notice that the above vanishes in $d=4$,
where the theory is conformal (at this order in perturbation theory).
This is related to the statement that light transforms annihilate the vacuum \cite{Kravchuk:2018htv}: in a conformal theory
we only expect the expectation value of $L_{n,\cC}[\cdots]$ to be nonvanishing in the presence of other operators
both inside \emph{and} outside the cone.  Alternatively, the contribution \eqref{phi4 OPE} had to vanish
in a conformal theory because the shape of vacuum three-point functions is fixed by symmetries, and one can check
that it does not contain any term with exponent $\sim 1/\delta x^{2\Df-1}$.
Regardless, eq.~\eqref{phi4 OPE} remains valid, and nontrivial, in excited states.

We conclude this $\phi^4$ discussion with a technical comment.
Calling \eqref{phi4 OPE} an OPE is a slight abuse of language:
$L_{n,\cC}[O]$ is not an ``operator'' acting on the standard Hilbert space of the theory.
It is only an operator in the sense that its insertion makes sense inside the (Schwinger-Keldysh) path integral.
This subtlety makes a nonperturbative classification of ``operators which can be inserted on a null cone''
more challenging than that of local operators, which are simply states of the theory on a small Euclidean sphere in radial quantization.

\subsection{More loops, more fields, more theories}\label{ssec:abstract}

Pinching contour arguments of the sort just reviewed give a clear picture of perturbative singularities 
in terms of classical particles. We can use it to anticipate how the OPE gets modified
for other operators,  at higher orders in $\lambda$ or $\delta x$, or in other theories.

A simple generalization is to replace each $\phi$ in \eqref{phi4 3pt} by $\phi^m$.
A contribution with $m'$ copies of \eqref{phi4 OPE} and $(m{-}m')$ direct Wick contractions between the $\phi$'s
then yield an integral over $m'$ distinct lightrays:
\be
 \phi^m(\delta x)_2\ \phi^n(0)_1 \supset \< \phi_2\phi_1\>_\cC^{m-m'} 
 \int \prod_{i=1}^{m'} \frac{-i\lambda \tilde{P}_d\ d^{d-2}n_i}{(-2n_i\.\delta x)^{2\Df-1}} L^\cC_{n_i}[\tfrac12\phi^2]\,.
\label{phi4 m example}
\ee
This demonstrates a basic fact: configurations with multiple particles on the null cone do contribute to folded OPEs.
Single-particle excitations cannot always give the full answer.
Of course, we expect the same phenomenon when \eqref{phi4 OPE} itself is expanded to sufficiently high order in perturbation theory,
since a generic diagrams contains many multi-particle intermediate state.
These may however be suppressed by powers of $\delta x$, as happens in the example of \eqref{phi4 m example}:
the $m'$th term is suppressed by a relative $|\delta x|^{m'}$.

It will be useful to understand and classify these factors by symmetry arguments.
As explained in the next section, in conformal theories in the vacuum state, the limit under consideration is equivalent to the Regge limit.
The present discussion greatly benefits from hindsight from the extensive literature on this topic.

Consider another system: scalar QED. The Lagrangian density is
\be
 L_{\rm scalar\ QED} = -\frac{F_{\mu\nu}F^{\mu\nu}}{4e^2}- (\partial^\mu{+}iA^\mu) \Phi^*(\partial_\mu{-}iA_\mu)\Phi\,.
\ee
The OPE of $\Phi_2 \Phi^*_1$ can now contain line integrals of the gauge field $\int A_u du$.
Its calculation is essentially identical to \eqref{phi4 lambda} above, with the crucial
difference that the $\Phi\Phi^* A$ vertex contains a derivative so we end up with a more singular denominator.
To be precise, we should only record results for gauge-invariant local operators. For $\cO\equiv (\Phi^*\Phi)$ we get terms with
two integrals $\int A_u du$:
\be
\cO_2(\delta x)\ \cO_1(0)
\supset P_\Df^2e^{-2\pi i\Df} \int
\frac{d^{d-2}n_1/\Omega_{d-1}}{(-n_1\.\delta x)^{d-2}}
\frac{d^{d-2}n_2/\Omega_{d-1}}{(-n_2\.\delta x)^{d-2}} L_{n_1,\cC}[-iA_u] L_{n_2,\cC}[iA_u]. \label{QED stupid}
\ee
Before discussing the dependence on $\delta x$, let us discuss a simple improvement.
The above is only the first term in an expansion in small $A$, and one may worry that the series breaks down
if $\< L_{n,\cC}[iA_u]\>$ is not small in the state under consideration.
Indeed, diagrams with multiple insertions of $A_u$ along each lightray also contribute at the same order in $\delta x$.
However, it is relatively easy to show that these simply sum up to Wilson lines.
Thus, it is better to rewrite \eqref{QED stupid} in terms of Wilson lines:
\be
\cO_2(\delta x)\ \cO_1(0)
 \sim P_\Df^2e^{-2\pi i\Df} \int
\frac{d^{d-2}n_1/\Omega_{d-1}}{(-n_1\.\delta x)^{d-2}}
\frac{d^{d-2}n_2/\Omega_{d-1}}{(-n_2\.\delta x)^{d-2}}\ \cU_{n_1}^\dagger\cU_{n_2}  + \ldots,
\quad \cU_n \equiv e^{i \int_\cC du A_\mu n^\mu(un)} .\label{QED ok}
\ee
Since $\cU_n$ is a unitary, its expectation values are bounded. The advantage of \eqref{QED ok} over \eqref{QED stupid} is
that the latter remains valid even the $\cU$'s are not parametrically close to identity.
The omitted terms have coefficients  suppressed by $e^2$ or higher powers of $\delta x$; they can contain more Wilson lines.
These statements naturally generalize to non-abelian gauge theories.  

No terms involving direct Wick contraction should to be added to \eqref{QED ok}. This is because $\cO$'s are made of charged excitations, which cannot connect the two local operators without dragging a Wilson line along the contour $\cC$.
Rather, the vacuum expectation value of $\< \cO_2 \cO_1\>$ is automatically accounted for by \eqref{QED ok} by taking the weak-field limit $\<\cU_n\> \mapsto 1$ and integrating over angles.\footnote{
If the gauge group is $G$, operators on the right-hand-side of \eqref{QED ok} must be
invariant under independent $G\times G$ gauge transformations at the two endpoints.
``Weak field'' really means that $\< \cU_n \>$ is $n$-independent,
thus spontaneously breaking this group to its diagonal: $G\times G \to G$.
Small perturbations can be parametrized as $\mathcal{U}_n=e^{igW_n^a T_a}$ where $W_n$ is called the Reggeized gluon, see \cite{Caron-Huot:2013fea} and references therein.
}

Extrapolating these examples, one might guess that exchange of a field of spin $J$
comes with a factor $1/(-2n\.\delta x)^{2\Df+J-1}$.
This is nicely formalized by the \emph{light transform} of \cite{Kravchuk:2018htv}. It shows that in conformal theories,
if $H$ is an operator of dimension $\Delta$ and spin $J$ (represented as a $J$-index symmetrical traceless tensor $H_{\mu_1\cdots \mu_J}$),
then the integral $L_{n,\cC}[H_{u\cdots u}]$ transforms like an operator with transformed quantum numbers:
\be
 \Delta_L=1-J, \qquad J_L = 1-\Delta\,. \label{light transform}
\ee
Applied to the operator $\phi^2$ which has $(\Delta,J)=(2\Df,0)$, this gives $(\Delta_L,J_L)=(1,1-2\Df)$.
This explains precisely the $\delta x$ dependence in \eqref{phi4 OPE}.
(The leading coefficient of $L$ in the OPE of two scalars of dimensions $\Delta_1$ and $\Delta_2$
by definition takes the form $\cO_1(\delta x)\cO_2(0)\sim (n\.\delta x)^{J_L}/(\delta x^2)^{\frac{\Delta_1+\Delta_2-\Delta_L+J_L}{2}}$.)
This also explains why multi-particle contributions in scalar field theory must be power-suppressed, as we saw in \eqref{phi4 m example}.

Conformal symmetry does not play a role here:
$J_L$ is the Lorentz symmetry of the tangent space near $x^\mu$ (of which $\delta x^\mu$ is an element)
while $\Delta_L$ acts by rescaling $\delta x$.

As mentioned below \eqref{phi4 OPE}, the full OPE in $\phi^4$ theory starts with double-twist operators at tree-level.
One might be tempted to do a derivative expansion $(\phi_2(0)+\delta x^\mu\partial_\mu \phi_2(0)+\ldots)\phi_1(0)$,
however these products are not really well-defined and
the more robust way to write the free-theory OPE is to integrate over
nonlocal lightray operators which analytically continue $\phi\partial^J\phi$, as discussed recently in \cite{Caron-Huot:2022eqs}.
A peculiarity of this theory is that the leading trajectory with $\Delta(J)\approx d-2+J$ mixes with its shadow with $\Delta\mapsto d-\Delta$ already in the free theory. This makes the renormalization of $L[\phi^2]$, which lives precisely at their intersection, particularly subtle.
The upshot, according to \cite{Caron-Huot:2022eqs},
is that at small but finite coupling, the contribution \eqref{phi4 OPE} can be subsumed into an $O(\lambda)$ correction to double-twist OPE coefficients. This will be seen quite explicitly in eq.~\eqref{CRT Ising} below. Since operators like \eqref{QED ok} do not lie near tree-level intersections, we do not expect them to be affected by such subtleties (nor the operators we study in gravity below).

For gauge theories, the relation \eqref{light transform} gives $\Delta_L[\cU]=0$ which
explains why operators with arbitrary many Wilson lines
in \eqref{QED ok} are \emph{not} suppressed by $\delta x$: Wilson lines are marginal operators in the folded OPE.
Quantum corrections however change $\Delta_L$. This is captured by the famous BFKL evolution equation,
which has been recast in terms of Wilson lines in many references, see for example \cite{Mueller:1994jq,Balitsky:1995ub}.
Operators with different numbers of Wilson lines mix at the cost of a loop factor $g^2$.
(In the t' Hooft planar limit, the parameter is really $g^2\sim 1/N_c$ \emph{not} $g^2N_c$:
Wilson line number is effectively conserved at large-$N_c$, see \cite{Caron-Huot:2013fea}.)

Crucially, the quantum corrections increase $J$ and make the $\delta x\to 0$ limit \emph{more} singular.
This is the famous fact that the BFKL Pomeron has intercept $J_* =1+O(\alpha_s C_A) >1$.
This means that in non-abelian gauge theories, contributions with more Wilson lines,
although formally suppressed by powers of $g^2$, grow faster with energy.
This ultimately causes the structure in \eqref{QED ok} to break down at exponentially small $\delta x$.
(In some situations, this breakdown is necessary to satisfy
nonperturbative bounds on correlators, like the Froissart-Martin bound.)

In other words, the form \eqref{QED ok} may not give the true asymptotic limit $\delta x\to 0$,
but only a transient ``pre-saturation'' regime at intermediate $\delta x$. This is still an interesting regime and we will try to stay within it.
Large-$N_c$ helps by parametrically delaying saturation (sometimes called scrambling), to $|\delta x|^{J_*-1}\sim 1/N_c^2$.
A similar comment will apply to tree-level graviton exchange below.

\subsection{Scalar coupled to gravity in AdS}

We now move on to our main interest: theories with a perturbative description in a dual,
asymptotically AdS${}_{d+1}$, spacetime. Consider a scalar field coupled to gravity
\be
 S = \int d^{d+1}x \sqrt{-g}\left(\frac{R-\Lambda}{16\pi G} - \tfrac12 (\partial \phi)^2 -\tfrac12m^2 \phi^2\right)
\ee
with $\Lambda=-d(d{-}1)/\RAdS^2$ the cosmological constant.
The bulk field $\phi$ is dual to a scalar operator $\cO$
of scaling dimension $\Delta$ with $\Delta(\Delta-d)=m^2\RAdS^2$.

\def\gravex{\<\phi T \phi\>}

We would like to understand the OPE of boundary operators $\cO$ in the presence of other operator insertions.
From the preceding discussion, 
we anticipate graviton exchange to dominate whenever it is nonvanishing,
so let us study three-point functions like
\be
 \gravex \equiv \<\Psi| \cO(\delta x)\ T^{\mu\nu}(y)\ \cO(0)\ |\Psi\> \label{grav ex}
\ee
at tree level, where there is a single diagram shown in figure \ref{fig:folded gravity}.
We start with a bulk geometry that is pure AdS and generalize shortly.
The AdS metric in Poincar\'e coordinates is
\be
 ds^2_{\rm AdS} = \RAdS^2 \frac{dr^2+\eta_{\mu\nu}dx^\mu dx^\nu}{r^2} \,, \label{AdS}
\ee
where the field theory is at the $r\to 0$ boundary and $\eta_{\mu\nu}$ is the flat Minkowski metric.

We are interested in the singular part of the diagram as $\delta x\to 0$.
From a Landau analysis similar to \eqref{phi4 OPE} we anticipate that the region near the future lightcone of the origin will dominate
as $\delta x\to 0$, so we directly zoom onto this region.
The base of the cone is now a $(d-1)$-dimensional hyperbolic space $H_{d-1}$.
It can be naturally parametrized by a timelike unit vector $\theta^\mu$ ($\theta^2=-1$),
which gives the bulk null vector $\theta^\mu \partial_\mu + \partial_r$.
We parametrize a neighborhood of the cone by variables $(\theta,u,v)$ (see figure \ref{fig:folded gravity})
\be
 (x^\mu,r) = \tfrac{1}{-u}(\theta,1) + \tfrac{v}{2} (\theta,-1),
\ee
with $u$ is an affine coordinate and $v$ is a retarded time away from the null cone ($v=0$).
Note that null geodesics take an infinite amount of affine time but a finite coordinate time
to reach the boundary. This is why the boundary is reached at $u=-\infty$.
Using affine time makes the parametrization slightly awkward but it makes the calculation
more similar to \eqref{phi4 OPE} and will also simplify things below.
Finally, we need the scalar bulk-to-boundary propagator (with canonical normalization in the field theory $\< \cO \cO\>= 1/(x^2)^\Delta$)
\be \label{bulktobdy}
\< \phi(r,y) \cO(x)\> = C_\Delta^{1/2}\left(\frac{r}{r^2+(x-y)^2}\right)^\Delta,\quad
C_\Delta = \frac{\Gamma(\Delta)}{2\pi^{\frac{d}{2}}\Gamma(\Delta-\tfrac{d}{2}+1)}.
\ee

Accounting for the two derivatives at the vertex, the diagram to leading order in $v\sim \delta x\to 0$ is given by
the bulk analog of \eqref{scalar integral near cone}%
\footnote{Here we are simply extending the boundary Schwinger-Keldysh timefolds into the bulk, following \cite{Son:2002sd,Skenderis:2008dg}. Special prescriptions near black hole horizons have been discussed \cite{Glorioso:2018mmw,Chakrabarty:2019aeu}. They are not needed here
since our observables only probe the bulk's perturbative density matrix at the past edge of
the causal diamond between $x$ and $y$.}
\be
 \gravex \approx
 4i\Delta^2 C_\Delta \int \frac{du\ dv\ d^{d-1}\theta \ \< T_1^{\mu\nu}(y) \big(\delta g_{uu,1}- \delta g_{uu,2}\big) \>_\cC}
 {(-2v+i0)^{\Delta+1}(-2v-2\theta\.\delta x-i0)^{\Delta+1} }.
\ee
Neglecting as before the $v$-dependence of the correlator, we can integrate over $v$ using \eqref{I12 result} to get the OPE
\be
 \cO_2(\delta x)\ \cO_1(0) \sim \< \cO_2(\delta x)\ \cO_1(0)\> + \frac{\tilde{C}_\Delta}{2}\int_{H_{d-1}} \frac{d^{d-1}\theta}{(-\theta\.\delta x_-)^{2\Delta+1}} L_{0,\theta}[\delta g_{uu}] +\ldots
 \label{grav OPE}
\ee
The exponent in the denominator agrees with the formula $2\Delta+J-1$ with $J=2$.
For reference, the normalization is
$\tilde{C}_\Delta= \frac{e^{-i\pi \Delta}\Gamma(2\Delta+1)}{(2\pi)^d P^>_\Delta}$ with $P^>_\Delta$ in \eqref{Fourier Wightman};
$d^{d-1}\theta$ is the canonical measure on the unit hyperboloid and $L$ is further discussed below.

Eq.~\eqref{grav OPE} will be our main tool to study the OTOC and its Fourier transform.

It will be useful to write the bulk light-transform covariantly.
Consider a metric which asymptotically approaches the AdS form \eqref{AdS} as $r\to 0$.
Sufficiently close to the boundary, the (boundary) momentum $g_{\mu a}\frac{dX^a}{du}$ becomes independent of affine time.
We normalize the affine time $u$ by equating this momentum to a timelike vector $p_\mu$.
The light transform $L_{x,p}$ is then defined as follows: first, we find
the null geodesic in the background geometry with affine parameter $u$ and appropriate initial momentum;
then we integrating the pullback of metric perturbations $\int h_{uu} du$ along this geodesic:
\be 
 \lim\limits_{u\to-\infty} (x,r) = (x,0) + \frac{\left(p^\mu,\sqrt{-p^2}\right)}{up^2 \RAdS^2}  + O(u^{-2})
  \quad\Rightarrow\quad L_{x,p}[h_{uu}]\equiv \int_{-\infty}^{u_0} du (h_{uu,1}-h_{uu,2}).
\label{p geodesic}
\ee
This definition is homogeneous in $p$:  $L_{x,\beta p}[h_{uu}] = \beta L_{x,p}[h_{uu}]$.
(We will use the same notation $L_{x,\theta}$ when $\theta^\mu$ is a unit vector instead of a ``momentum''.)
The turning point $u_0$ is any value sufficiently large to cover all regions with nonvanishing integrand,
for example to cover the past lightcone of $y$ in the current example.

With these definition, $p_\mu$ transforms like a cotangent vector at the boundary point $x$.
It parametrizes the direction in which a null geodesic is fired into the bulk, together with its energy.
From the boundary perspective, we do not control what the geodesic finds along its route, but we can control how it starts.

We claim that with this definition, the OPE \eqref{grav OPE} holds in any background geometry.
This can be established by studying the propagator in a neighborhood of the null cone from $x$.
In general this can be conveniently parameterized as (eq.~(5.5) of the lecture notes \cite{Witten:2019qhl}):
\be
 ds^2_{\rm cone} = -2Qdu dv + \tilde{g}_{ij} (d\theta^i +\xi^i dv)(d\theta^j +\xi^j dv), \label{ds2 cone}
\ee
where $Q$, $g_{ij}$ and $\xi^i$ can depend on all coordinates $(u,v,\theta)$ but $Q$ is independent of $u$ when $v=0$.
Null geodesics on the cone are paths with $v=0$ and constant $\theta$.
The scalar field's equations of motion near the cone take the form
\be \label{scalar EOM cone}
\left[ \partial_u \sqrt{\tilde{g}}\mathcal{D}_v +  \mathcal{D}_v \sqrt{\tilde{g}}\partial_u - \partial_i Q\sqrt{\tilde{g}}\tilde{g}^{ij} \partial_j + m^2 Q\sqrt{\tilde{g}}\right]\phi=0
\ee
with $\mathcal{D}_v=\partial_v-\xi^i\partial_i$.
The idea is to expand near the null cone and look at the most singular term $\phi \sim \phi_0(u,\theta) v^{-\Delta}$;
we see that the equation of motion sets $\tilde{g}^{1/4}\phi_0$ to be $u$-independent.
This means that the most singular part of the propagator simply follows geodesics in such a way that the graviton source term $\sim d^{d-1}\theta \sqrt{\tilde{g}}(\partial_v \phi^0)^2$ does not deform.\footnote{
If the bulk theory contains a dilaton-like scalar $\varphi$, such that scalar kinetic term is multiplied by $e^{-\varphi}$,
the propagator will be such that the source term $d^{d-1}\theta \sqrt{\tilde{g}}e^{-\varphi}(\partial_v \phi^0)^2$ is unaffected by the dilaton.
The results of this section thus apply in the Einstein frame metric.
}

The effects of spatial derivatives (or of mixing between the scalar and graviton, neglected in \eqref{scalar EOM cone}) are suppressed by powers of $v$.
Note that the $v\to 0$ expansion is subject to similar limitations as any WKB-like expansions; it may break down for example
if the effective energy $\sim 1/v$ is not sufficiently large compared with variations of the background geometry.
Note that at early times when the excitation is close to the boundary, the locally measured energy is not large and a WKB approximation is not justified; the small-$v$ expansion of \eqref{scalar EOM cone} nonetheless makes sense.

In the vacuum state, the three-point function $\gravex$ obtained from \eqref{grav OPE}
vanishes for timelike $y$ by a mechanism similar to \eqref{vanishing u integral}.
Yet the three-point function is nontrivial: there is a $\delta$-function at lightlike $y$,
given by the shockwave discussed in appendix \ref{app:shockwave}. 
This is the expanding shell of energy described in conformal collider physics \cite{Hofman:2008ar}. 
The shell has a small thickness $\delta x$ in retarded time which becomes a $\delta$-function in the OPE under consideration.

\subsection{Focusing using Fourier transforms and wavepackets}

We can now see how the Fourier transform described in introduction localizes around a single bulk lightray.
Let us first ignore the Gaussian envelope, and consider a momentum $p^\mu=(\omega,\vec{0})$ purely along the time direction.
The time integral against \eqref{grav OPE} gives
\be
\int \frac{dt \ e^{i\omega t}}{(\theta^0 t -\vec{\theta}\.\vec{x}-i0)^{2\Delta+1}} \propto \omega^{2\Delta} e^{i\omega\frac{\vec{\theta}\.\vec{x}}{\theta^0}}\,. 
\label{shooting fourrier 0}
\ee
Integrating this over $\vec{\delta x}$ would give a $\delta$-function that sets $\vec{\theta}=0$,
thus localizing the OPE \eqref{grav OPE} to a single lightray fired at a square angle into the bulk.

More generally, one could include a Gaussian envelope of width $L$ for $\vec{x}$ and get:
\be
\int \frac{d^dx \ e^{i\omega t-\frac{\vec{x}^2}{2L^2}}}{(-\theta\.x_-)^{2\Delta+1}} \propto \omega^{2\Delta} e^{-\frac{|\vec{\theta}|^2}{2\theta_0^2}(\omega L)^2}\,. 
\label{shooting angle uncertainty}
\ee
We see that the shooting angle $\vec{\theta}$ is Gaussian distributed with width $1/(\omega L)$.
This is the standard diffraction limit for the angular resolution of an optical lens of size $L$.
We conclude that the optical aspects of looking into the bulk are very similar to conventional optics.

Let us define a standard set of wavepackets.
For $p$ an arbitrary positive-frequency timelike vector, we can covariantly boosting the above.
Additionally, we can include independent widths $L$ and $L_0$ for the space and time components in the rest frame.
A natural normalization is provided by the vacuum two-point function:
\be \label{psi}
\psi_{p,L}(\delta x) \equiv \frac{P^>_\Delta}{(-p^2)^{\Delta-\frac{d}{2}}} e^{-ip\.\delta x-\frac{(x+\hat{p} \hat{p}\.x)^2}{2L^2} - \frac{(x\. \hat{p})^2}{2L_{0}^2}}
\quad \Rightarrow\quad
 \int d^d\delta x\  \psi_{p,L}(\delta x) \<\Omega| \cO(\delta x)\ \cO(0) | \Omega\> =1,
\ee
where $P^>_\Delta$ is in \eqref{Fourier Wightman}.
We will be interested in the regime $L|p|\gg 1$ where the shooting angle is well-defined, so we ignore here $1/(L|p|)$ corrections to the normalization.
In the limit that $L\to \infty$, we find that the omitted proportionality constants in \eqref{shooting fourrier 0} nicely cancel the factors in \eqref{grav OPE}, so its transform gives simply
\be
\int d^d\delta x\  \psi_{p,L=\infty}(\delta x)\ \cO_2(x{+}\tfrac12\delta x)\ \cO_1(x{-}\tfrac12\delta x)
\sim 1 + \frac{i}{2} L_{x,p}[\delta g_{uu}] + \ldots\,, \label{AdS OPE p}
\ee
as anticipated in eq.~\eqref{naive S}.
The omitted terms, which are the same as in \eqref{grav OPE}, give insight into the limits of this approximation.

To account for the Gaussian envelope, we can now use that multiplication by a Gaussian in $x$ is equivalent to convolution by a Gaussian in $p$. Thus \eqref{AdS OPE p} needs to be convolved with a Gaussian of width $1/(L|p|)$ in the orientation of $p$,
and with a Gaussian of width $1/L_0$ in its magnitude.
We will typically be interested in situations where the energy dependence of correlators is mild, so the latter is a negligible effect.
We thus omit $L_0$ from our notations. The angular uncertainty $1/(L|p|)$ is important however.

The reader should be aware that if $p^\mu$ is highly boosted,
then the footprint of our covariant wavepackets can become much larger than $L$ along the longitudinal direction.
A highly boosted source with a finite footprint would create waves that rapidly disperse, and do not follow a unique
bulk geodesic \cite{Arnold:2011qi}.  In this paper we will only consider momenta $p^\mu$ that are not parametrically boosted.

\subsection{Four-point correlators}

The out-of-time-order correlator featured in \eqref{camera} can now be obtained using
two copies of the OPE. Technically, it involves a four-sheeted timefold
\be
\vcenter{\hbox{
\begin{tikzpicture}
\draw (-1,0.3) -- (-1,0) -- (5,0) -- (5,-0.3) -- (-1,-0.3) -- (-1,-0.6) -- (5,-0.6) -- (5,-0.9) -- (-1,-0.9) -- (-1,-1.2);
\node [above] at (0,0) {$\cO(x{-}\tfrac12\delta x)$}; 
\node [above] at (4,0) {$\cO'(y{-}\tfrac12\delta y)$};
\node [below] at (0.5,-0.9)  {$\cO(x{+}\tfrac12\delta x)$};
\node [below] at (4.5,-0.9)  {$\cO'(y{+}\tfrac12\delta y)$};
\node [right] at (5,0) {\small $1$}; \node [right] at (5,-0.3) {\small $2$}; \node [right] at (5,-0.6) {\small $3$}; \node [right] at (5,-0.9) {\small $4$};
\filldraw (0,0) circle (2pt); \filldraw (4,-0.3) circle (2pt); \filldraw (0.5,-0.3) circle (2pt); \filldraw (4.5,-0.6) circle (2pt);
\draw (7,0.35) -- (7,-0.15) -- (7.5,-0.15);
\node [above] at (7.25,-0.15) {$t$}; 
\end{tikzpicture}}} \label{SK fourfold 1}
\ee
For clarity we have exaggerated the infinitesimal separation between the folds, 
but it will be important that they are actually all atop each other (this is unlike the late-time limit of \cite{Maldacena:2015waa}).

The OPE between the $\cO$ is still given by \eqref{grav OPE}, while that between the $\cO'$ is a time-reversed version of it:
\be
\cO'_3(y{+}\tfrac12\delta y)\ \cO'_2(y{-}\tfrac12\delta y) \sim \< \cO_3'(\delta y)\ \cO_2'(0)\> +
\frac{\tilde{C}_{\Delta'}\kappa}{2}\int_{H_{d-1}} \frac{d^{d-1}\theta'}{(-\theta'\.\delta y_-)^{2\Delta'+1}} L'_{y,\theta'}[h_{vv}]
\label{grav OPE y}
\ee
where the light transform is defined analogously to \eqref{p geodesic}, for a geodesic that now reaches the boundary at $v={+}\infty$:
\be 
 \lim\limits_{v\to+\infty} (y,r) = (y,0) + \frac{\left(p^\mu,-\sqrt{-p^2}\right)}{vp^2\RAdS^2}  + O(v^{-2})
  \quad\Rightarrow\quad L'_{y,p}[h_{vv}]\equiv \int_{v_0}^{\infty} dv (-h_{vv,2}+h_{vv,3}).
\label{p' geodesic}
\ee
Note that the excitation has positive energy ($\theta'^0>0$ in \eqref{grav OPE y})
but it propagates backward in time, since it starts on the second branch of the contour \eqref{SK fourfold 1}, whence the sign in $-h_{vv,2}$.

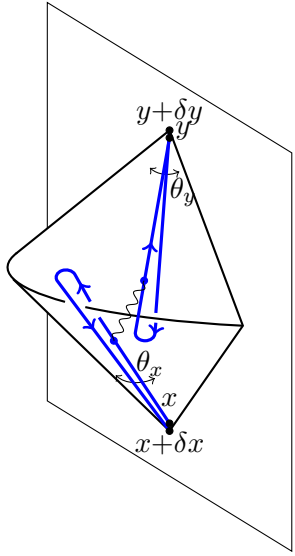
\begin{figure}[t]\centering
\begin{tikzpicture}[xscale=1.3]
\draw (0,1.08) -- (0,-0.1) -- (2.5,-2.1) -- (2.5,3.2) -- (0,5.2) -- (0,2.2);
\draw[thick] (1.25,-0.5) -- (2,0.9) -- (1.25,3.5);
\draw[thick] (2,0.9) arc (255:221:3.25 and 0.8); \draw[thick] (-0.31,1.87) arc (166:215:3.25 and 0.8);
\draw[thick] (1.25,-0.5) -- (-0.35,1.52);  \draw[thick] (-0.32,1.86) -- (1.25,3.5);
\draw[<->] (0.7,0.24) [out=-50,in=-150] to (1.1,0.2); \node [above] at (1.05,0.1) {$\theta_x$};
\draw[<->] (1.05,3) [out=-50,in=-150] to (1.35,2.95); \node [below right] at (1.15,3) {$\theta_y$}; 
\draw[very thick,blue,decoration={markings,mark=at position 0.55 with \arrow{<},mark=at position 0.95 with \arrow{<}},postaction=decorate] (1.25,-0.5) -- (0.1,1.5) arc (205:25:0.1) -- (0.44,1.22);
\draw[very thick,blue] (1.25,-0.4) -- (0.53,1.01);
\filldraw (1.25,-0.4) ellipse (1pt and 1.3pt); \node[above,inner sep=2pt] at (1.25,-0.25) {$x$}; 
\filldraw (1.25,-0.5) ellipse (1pt and 1.3pt); \node[below,inner sep=0pt] at (1.25,-0.5) {$x{+}\delta x$}; 
\filldraw[blue] (0.68, 0.7) ellipse (1pt and 1.3pt);
\draw[very thick,blue] (1.25,3.5) -- (1.11,1.07);
\draw[very thick,blue,decoration={markings,mark=at position 0.05 with \arrow{>},mark=at position 0.55 with \arrow{>}},postaction=decorate] (1.11,0.94) -- (1.1,0.8) arc (0:-190:0.1) --(1.25,3.4);
\filldraw (1.25,3.5) ellipse (1pt and 1.3pt); \node[above,inner sep=2pt] at (1.25,3.5) {$y{+}\delta y$};
\filldraw (1.25,3.4) ellipse (1pt and 1.3pt); \node[right,inner sep=2pt] at (1.25,3.5) {$y$}; 
\filldraw[blue] (0.99, 1.5) ellipse (1pt and 1.3pt);
\draw[decorate, decoration={snake,amplitude=1.6,segment length=6}] (0.68,0.7)--(0.99,1.5);
\end{tikzpicture}
\caption{
Graviton exchange contribution to the out-of-time-order correlator in the eikonal limit $\delta x$, $\delta y\to 0$. \label{fig:4pt}
}
\end{figure}

The four-point OTOC is now reduced to a graviton two-point function:
$\< L[h_{uu}]L[h_{vv}]\>$, as depicted in figure \ref{fig:4pt}.
The relevant combination turns out to be a retarded propagator:
\be\begin{aligned}
 \< (h_{uu,1}-h_{uu,2})(X)\ (-h_{vv,2}+h_{vv,3})(Y) \>_\cC &= \<\Psi| [h_{uu}(X),h_{vv}(Y)] |\Psi\> \theta(Y{\succ}X)
\\
&\equiv i \< h_{uu}(X) h_{vv}(Y) \>_{\Psi,\rm ret} \,.
\end{aligned}\ee
The retarded propagator depends only on the bulk geometry dual to the state $\Psi$ and
is insensitive to the presence of bulk gravitons (which could come, for example, from Hawking radiation if there is a black hole in the bulk).
It is useful to define a shorthand for its integral against null geodesics:
\be \label{L}
 \Pi_\Psi(x,p_x; y,p_y) \equiv -2i\< L_{x,p}[h_{uu}] L'_{y,p'}[h_{vv}] \>_\cC =
 2\int_{-\infty}^{u_0} du \int_{v_0}^\infty dv \< h_{uu}(X) h_{vv}(Y) \>_{\Psi,\rm ret}\,.
\ee
The normalization choice will become clear below.
The OPE \eqref{grav OPE} thus gives us a formula for the OTOC at tree-level in a general geometry
\be\begin{aligned} \label{OTOC position}
 &\<\Psi|\ \cO'(y{+}\tfrac12\delta y)\ \cO(x{+}\tfrac12\delta x)\ \cO'(y{-}\tfrac12\delta y)\ \cO(x{-}\tfrac12\delta x)\ |\Psi\>
\\ &\quad \approx \frac{1}{(\delta x_-)^{2\Delta}(\delta y_-)^{2\Delta'}}
-4\pi i G \int \frac{d^{d-1}\theta\ \tilde{C}_\Delta}{(-\theta\.\delta x_-)^{2\Delta+1}}  \int \frac{d^{d-1}\theta'\ \tilde{C}_{\Delta'}}{(-\theta'\.\delta y_-)^{2\Delta'+1}} \Pi_\Psi(x,\theta_x; y,\theta_y).
\end{aligned}\ee
Equivalently, its Fourier transform \eqref{camera} using the wavepacket normalized precisely in \eqref{psi} gives,
in the geometrical optics regime,
\be
 G_\Psi(x,p_x; y,p_y) -1 = -4\pi i \GN \Pi_\Psi(x,p_x; y,p_y) + \ldots \label{Gp} \qquad (|p_i|L_i \ll 1),
\ee
as quoted in \eqref{G naive}.
Optical limitations at finite $|p|L$ can be accounted for by a convolution as discussed below \eqref{AdS OPE p}.

Eq.~\eqref{OTOC position} and \eqref{Gp} are the main results of this section.
For conformal field theories in the vacuum state, \eqref{OTOC position} is essentially equivalent to eq.~(1.5) of \cite{Cornalba:2006xk}
and \eqref{Gp} to the impact parameter representation of \cite{Cornalba:2006xm} (see also \cite{Meltzer:2017rtf,Kulaxizi:2017ixa}).
The propagator $\Pi_\Psi$ entering \eqref{OTOC position} was normalized so it reduces to the transverse propagator,
also called $\Pi$ in that reference, up to a factor of energy, as discussed shortly.
They are meant to capture the contribution to the correlator with the fastest energy growth 
while the right-hand-side of \eqref{Gp} is still smaller than unity.

The eikonal approximation \eqref{grav OPE} also controls the radar-like correlator \eqref{radar}
in the regime where the projectile has a much higher energy than the probe. This is further discussed in section \ref{ssec:flat} below.

\subsection{Useful limits: near coincidence and vacuum AdS} \label{ssec:limits}

Let us finish this section with an important special case of \eqref{L}: when two geodesics nearly intersect at a point.
The singular behavior in this limit is controlled by the graviton retarded propagator in a small neighborhood of this point.
It takes the flat space form
\be \label{prop flat}
 \< h^{ab}(X) h^{cd}(Y) \>_{\rm ret} = \int \frac{d^{d+1}p}{(2\pi)^{d+1}} \frac{e^{i p{\cdot}(Y-X)}}{p_+^2}
 \left[ \frac12(P^{ac}_p P^{bd}_p+P^{ad}_p P^{bc}_p) - \frac{1}{d-1}P^{ab}_pP^{cd}_p\right]
\ee
where $P^{ab}_p = g^{ab} - \frac{p^ap^b}{p^2_+}$ is a projector, and
the $+$ subscript indicates the small imaginary shift $p^0\mapsto p^0+i0$ appropriate to the retarded boundary condition.
Suppose the metric is given locally in the form \eqref{ds2 cone}.
The integrals over null times $u$ and $v$ in \eqref{L} set two momenta to zero,
and the remaining $(d-1)$ integrals give the Green's function of the transverse Laplacian.
Reinstating factors of the local metric $g_{ab}^*$ near the collision, the singular behavior is thus:
\be
 \Pi_\Psi(x,p_x;y,p_y) \approx \frac{P^{(d-1)}_{(d-3)/2}}{(g_{ij}^* \delta\theta^i \delta \theta^j)^{\frac{d-3}{2}}} \left[-2g_{ab}^* \frac{dX^a}{du}\frac{dY^b}{dv}\right]\qquad (\delta\theta\to 0)
 \label{L flat}
\ee
with $P^{(d-1)}_{(d-3)/2} = \frac{\Gamma(\tfrac{d-3}{2})}{4\pi^{\frac{d-1}{2}}}$ (see eq.~\eqref{PDelta}).
The various factors have simple interpretation in terms of the center-of-mass energy and
impact parameter in the bulk.  For example, in vacuum AdS, a short calculation yields
\be\begin{aligned} \label{sbulk}
s_{\rm bulk} &\equiv -2g_{ab}^* \frac{dX^a}{du}\frac{dY^b}{dv} = \frac{2r^2}{\RAdS^2} (|p_x||p_y|-p_x{\cdot}p_y)\Big|_{r^2=\frac{|x-y|^2}2\frac{|p_x||p_y|}{|p_x||p_y|-p_x\.p_y}} = \frac{|p_x||p_y| |x-y|^2}{\RAdS^2}\,,
\\
b_{\rm bulk} &\equiv \left(g_{ij}^* \delta\theta^i \delta \theta^j\right)^{1/2} = \RAdS\cosh^{-1}\left(\tfrac{p_x\.\mathcal{I}_{(x-y)}\.p_y}{|p_x||p_y|}\right)\,,
\end{aligned}\ee
with $|p|=\sqrt{-p^2}$. These formulas may be readily explained when the separation $(x{-}y)$ is purely in the time direction.
Then the impact parameter lives on the transverse hyperboloid $H_{d-1}$ and vanishes if $p_x$ differs from $p_y$ by a reflection $\mathcal{I}_{x-y}$.
(See figure \ref{fig:camera}: if the cannon shoots towards the right, the eye will see the signal arriving from left.)
Thus $b_{\rm bulk}$ is the chordal distance between points proportional to $p_x$ and $-\mathcal{I}_{(x{-}y)}\.p_y$.
The collision occurs at the halfway point in time, at a radius determined from the radial velocity. 
(In the regime relevant to \eqref{L flat}, the impact parameter is small, $b_{\rm bulk}\ll \RAdS$.)

In vacuum AdS, we can say more. The intersection between the future and past lightcone of $x$ and $y$ is geometrically an hyperboloid $H_{d-1}$,
and the chordal distance $b_{\rm bulk}$ in \eqref{sbulk} defines a natural impact parameter even when it is not small.
Exploiting symmetries, one can in fact calculate $\Pi_\Psi$ exactly in this case.  It turns out that the $u$ integral in \eqref{L} sources a shockwave supported on the null cone from $x$,
and the propagator \eqref{prop flat} becomes exactly that on the transverse hyperboloid:
\be
 \Pi_\Omega(x,p_x; y,p_y) = |p_x| |p_y| \Pi_{d-1}(\cosh(b_{\rm bulk}/\RAdS)) \label{prop vacuum}
\ee
with $\Pi_{\Delta}$ in \eqref{hyperbolic propagator}.
The calculation, presented in appendix \ref{app:shockwave},
is related to one in \cite{Cornalba:2006xk} by a conformal transformation.

\section{The AdS vacuum and conformal Regge theory}\label{sec:CRT}

The goal of this section is to better understand the camera by taking first images in the vacuum state of various conformal field theories.
The folded OPE limit is then equivalent to the well-studied Regge limit, which
will enable us to numerically study the correlator \eqref{camera} beyond Einstein's gravity.

For spacelike separated points, conformal correlators take the form
\be \label{G eucl}
 \< \Omega|\ \cO'(x_4)\ \cO'(x_3)\ \cO(x_2)\ \cO(x_1)\ |\Omega\> =
 \frac{1}{x_{12}^{2\Delta_\cO}x_{34}^{2\Delta_{\cO'}}} \cG(z,\zb)
\ee
where the nontrivial dependence factors through two cross-ratios defined by
\be
 \frac{x_{12}^2x_{34}^2}{x_{13}^2x_{24}^2}\equiv z\zb,\qquad
 \frac{x_{23}^2x_{14}^2}{x_{13}^2x_{24}^2}\equiv(1{-}z)(1{-}\zb). \label{cross-ratios}
\ee
For configurations in Euclidean space, $\zb$ is the complex conjugate of $z$
and $\cG$ is a single-valued function with no branch cut ambiguity.

Let us specialize to the kinematics in \eqref{OTOC position}.
With no loss of generality, in this section we fix $x=0$ and $y=e$  where $e$ is a unit timelike vector ($e^2=-1$),
and parametrize coordinates as 
\be
\vcenter{\hbox{
\begin{tikzpicture}
\draw (-1,0.3) -- (-1,0) -- (5,0) -- (5,-0.3) -- (-1,-0.3) -- (-1,-0.6) -- (5,-0.6) -- (5,-0.9) -- (-1,-0.9) -- (-1,-1.2);
\node [above] at (0,0) {$x_1=0$}; \node [above] at (4,0) {$x_4=e$};
\node [below] at (0.5,-0.9) {$x_2=\delta x$}; \node [below] at (4.5,-0.9) {$x_3=e+\delta y$};
\filldraw (0,0) circle (2pt); \filldraw (4,-0.3) circle (2pt); \filldraw (0.5,-0.6) circle (2pt); \filldraw (4.5,-0.9) circle (2pt);
\end{tikzpicture}}} \label{SK fourfold}
\ee
In the limit of small $\delta x$, $\delta y$, the cross-ratios \eqref{cross-ratios} are small and approximate to
\be
 z\zb \approx (\delta x_-)^2(\delta y_-)^2, \quad \tfrac12 (z+\zb) =
 -\delta x\. \delta y -2(e\.\delta x)(e\.\delta y) \equiv -\delta x\.\mathcal{I}_e\. \delta y \,.
 \label{zzb series}
\ee
Recall that $\mathcal{I}^{\mu\nu}_e=\eta^{\mu\nu}-2e^\mu e^\nu/e^2$ is geometrically a reflection in the $e$ direction.

Even though the numerical value of cross-ratios are small, the out-of-time correlator is not given
simply by \eqref{G eucl}: the nontrivial ordering is only reached after an analytic continuation.
This is reviewed in appendix \ref{app:path}, where it shown that the desired correlator \eqref{SK fourfold}
in a ``reference'' region $\delta x{\succ}0$, $\delta y{\prec}0$ (where $z,\zb$ are small, real and positive)
is reached after taking $z$ counter-clockwise around 1, holding $\zb>0$ fixed:
\be \label{OTOC CFT}
G_{3241}\equiv \<\Omega|\ \cO'(x_3)\ \cO(x_2)\ \cO'(x_4)\ \cO(x_1)\ |\Omega\> =
 \frac{e^{-i\pi(\Delta_\cO-\Delta_{\cO'})}}{(-x_{12}^2)^{\Delta_\cO}(-x_{34}^{2})^{\Delta_{\cO'}}} \cG(z,\zb)^{\circlearrowleft} \quad(\delta x{\succ}0,\delta y{\prec}0).
\ee
Other regions of $\delta x$, $\delta y$ (spacelike, past timelike, etc.) can then be reached
without leaving the region of small $z,\zb$, paying appropriate care to the $i0$ prescriptions in \eqref{zzb series}.
Examples are given in table \ref{regions}.
In particular, when $\delta x$ and $\delta y$ are both spacelike with
$\delta x \propto -{\cal I}_e\.\delta y$, the ``bulk-point limit'' of \cite{Maldacena:2015iua} is approached.

We briefly discuss the radar-like system \eqref{radar}, which fits in a Schwinger-Keldysh two-fold:
\be
\mbox{radar camera:}\qquad \vcenter{\hbox{
\begin{tikzpicture}
\draw (-2,0.3) -- (-2,0) -- (5,0) -- (5,-0.3) -- (-2,-0.3) -- (-2,-0.6);
\node [above] at (1.1,0) {$x_1=0$}; \node [above] at (-0.3,0) {$x_4=z$}; \node [above] at (4,0) {$x_3=y$};
\node [below] at (1.3,-0.3) {$x_2=\delta x$};
\filldraw (1.1,0) circle (2pt); \filldraw (-0.3,0) circle (2pt); \filldraw (4,0) circle (2pt); \filldraw (1.3,-0.3) circle (2pt);
\end{tikzpicture}}} \label{radar twofold}
\ee
There many kinematical regions one could explore with this system. In this paper
we will assume that $x$ and $z$ are spacelike separated (note that this is not apparent in the drawing),
and restrict attention to the case $\delta x\to 0$.
Physically this is a ``ballistic'' limit where the particle fired from $x$ has a high energy,
compared with the pulse sent from $z$ and recorded at $y$.
Cross-ratios vanish in this limit:
\be
 z\zb \to U_{y,z}^2 \delta x_-^2, \quad \tfrac12(z+\zb) \to
 U_{y,z}\.\delta x_- \quad\mbox{with}\quad U_{y,z}^\mu\equiv \frac{y^\mu}{y^2}-\frac{z^\mu}{z^2}\,.
\ee
In a reference region with $\delta x$ spacelike and $U_{y,z}\.\delta x>0$ ($U_{y,z}$ is always spacelike),
$z,\zb$ are small with a positive real part and the correlator is proportional to $\cG^{\circlearrowright}$.%
\footnote{The second term in the commutator \eqref{radar} yields \eqref{radar twofold} with $z$ inserted on the second branch.
This second term will be negligible when a positive energy is absorbed by the wavepacket at $y$, and we will ignore it here.
}
Thus, in the vacuum state of a conformal theory, the ballistic limit of the radar camera is equivalent
to the active camera up to a continuation sign and the replacement $-{\cal I}_e\.\delta y_- \mapsto U_{y,z}$.
(When the correlator becomes singular, such as near the bulk-point limit,
small imaginary parts from $z\mapsto z_+$ and $y\mapsto y_-$ should be added, which adds a negative shift $(U_{y,z})_-$.)

\subsection{Relating Regge correlators, detectors, and OTOCs}\label{ssec:relations}

The correlator $\cG(z,\zb)^{\circlearrowleft}$ is exactly the same one which controls the Regge limit (see for example \cite{Cornalba:2006xk,Costa:2012cb,Caron-Huot:2017vep}).
In this limit, a large relative boost is applied between two pairs of points.

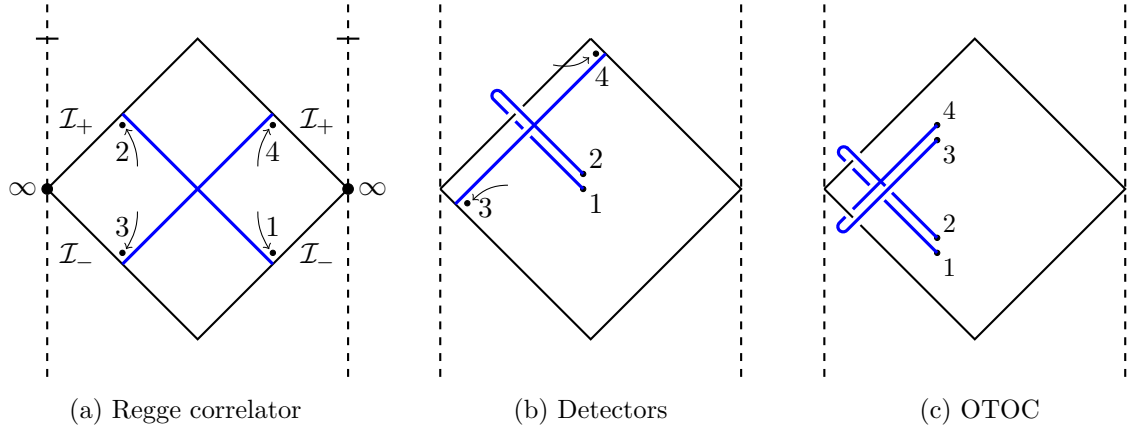
\begin{figure}[t]
\centering
\begin{subfigure}[b]{0.32\textwidth}\centering
\begin{tikzpicture}[baseline=(X.base)]
\draw[thick,dashed] (-2,-2.5) -- (-2,2.5) (2,-2.5) -- (2,2.5); \draw[thick] (-2.15,2) -- (-1.85,2) (2.15,2) -- (1.85,2);
\draw[thick] (-2,0) -- (0,2) -- (2,0) -- (0,-2) -- cycle;
\draw (1.6,0.9) node{$\scrI_+$} (-1.6,0.9) node{$\scrI_+$} (1.6,-0.9) node{$\scrI_-$} (-1.6,-0.9) node{$\scrI_-$};
\node[right] (X) at (2,0) {$\infty$}; \node[left] at (-2,0) {$\infty$};
\filldraw (2,0) circle (2pt); \filldraw (-2,0) circle (2pt);
\draw[very thick,blue] (-1,1) -- (1,-1); \draw[very thick,blue] (-1,-1) -- (1,1);
\filldraw (-1,-0.85) circle (1pt); \draw[<-] (-0.95,-0.8) [out=60,in=-90] to (-0.8,-0.3); \node at (-1,-0.5) {$3$};
\filldraw (1,-0.85) circle (1pt); \draw[<-] (0.95,-0.8) [out=120,in=-90] to (0.8,-0.3); \node at (1,-0.5) {$1$};
\filldraw (-1,0.85) circle (1pt); \draw[<-] (-0.95,0.8) [out=-60,in=90] to (-0.8,0.3); \node at (-1,0.5) {$2$};
\filldraw (1,0.85) circle (1pt); \draw[<-] (0.95,0.8) [out=-120,in=90] to (0.8,0.3); \node at (1,0.5) {$4$};
\end{tikzpicture}
\caption{Regge correlator}
\end{subfigure}
\hspace{0.02\textwidth}
\begin{subfigure}[b]{0.3\textwidth}\centering
\begin{tikzpicture}[baseline=(X.base)]
\draw[thick,dashed] (-2,-2.5) -- (-2,2.5) (2,-2.5) -- (2,2.5);
\draw[thick] (0,2) -- (2,0) -- (0,-2) -- (-2,0);
\filldraw (-1.64,-0.19) circle (1pt); \draw[<-] (-1.57,-0.15) [out=45,in=-180] to (-1.1,0.05); \node at (-1.4,-0.25) {$3$}; 
\filldraw (0.07,1.8) circle (1pt); \draw[<-] (-0.02,1.75) [out=-140,in=-10] to (-0.5,1.65); \node at (0.15,1.5) {$4$};
\draw[thick] (-2,0) -- (-1.0,1) (-0.9,1.1) --(0,2);
\draw[very thick,blue] (-1.8,-0.2) -- (0.2,1.8);
\filldraw (-0.1,0) circle (1pt) (-0.1,0.2) circle (1pt); 
\node[inner sep=2pt,below right] at (-0.1,0) {$1$}; \node[inner sep=2pt,above right] at (-0.1,0.2) {$2$};
\draw[very thick,blue] (-0.1,0)-- (-0.8,0.7) (-0.9,0.8)--(-1.0,0.9) (-1.1,1) -- (-1.3,1.2) arc (215:45:0.07071) (-1.2,1.3) -- (-0.1,0.2);
\end{tikzpicture}
\caption{Detectors}
\end{subfigure}
\hspace{0.02\textwidth}
\begin{subfigure}[b]{0.3\textwidth}\centering
\begin{tikzpicture}[baseline=(X.base)]
\draw[thick,dashed] (-2,-2.5) -- (-2,2.5) (2,-2.5) -- (2,2.5);
\draw[thick] (-2,0) -- (-1.62,0.38) (-1.52,0.48) -- (0,2) -- (2,0) -- (0,-2) -- (-1.52,-0.48) (-1.62,-0.38) -- (-2,0);
\node[right] (X) at (2,0) {};
\begin{scope}[shift={(-0.5,-0.35)}]
\filldraw (0,-0.5) circle (1pt) (0,-0.3) circle (1pt); 
\node[inner sep=2pt,below right] at (0,-0.5) {\small $1$}; \node[inner sep=2pt,above right] at (0,-0.3) {\small $2$};
\draw[very thick,blue] (0,-0.5) -- (-0.7,0.2) (-0.9,0.4) -- (-1.12,0.62) (-1.22,0.72) -- (-1.3,0.8) arc (215:45:0.07071) (-1.2,0.9) -- (-0.69,0.39) (-0.6,0.3) -- (0,-0.3);
\filldraw (0,1.2) circle (1pt) (0,1.0) circle (1pt); 
\node[inner sep=2pt,above right] at (0,1.2) {\small $4$}; \node[inner sep=2pt,below right] at (0,1.0) {\small $3$};
\draw[very thick,blue] (0,1.2)-- (-1.12,0.08) (-1.22,-0.02) -- (-1.3,-0.1) arc (135:315:0.07071) (-1.2,-0.2) -- (0,1); 
\end{scope}
\end{tikzpicture}
\caption{OTOC \label{fig:relationsC}}
\end{subfigure}
\caption{\label{fig:relations}
Geometric interpretation of the equivalence between three CFT correlators in the vacuum. The diamond is a Poincar\'e patch.
For ease of drawing, we display the Lorentzian cylinder in 1+1 spacetime dimension, with vertical dashed lines identified.
(More generally, the left- and right- halves of each pictures should be identified, with an $S^{d-2}$ sitting atop each point.)
(a) In the Regge limit, highly boosted operators approach null infinity and create the thick blue shockwaves.
(b) A shockwave at null infinity is an asymptotic measurement or ``detector''.
The picture is obtained from the first by translating northwest by a bit more than half a Poincar\'e patch, then folding.
(c) The out-of-time correlator is obtained by further translating the detector southwest by half a Poincar\'e patch, and folding again.
Layers follow the Schwinger-Keldysh folds in \eqref{SK fourfold} from bottom to top.
}
\end{figure}

An intuitive way to understand this relation is shown in figure \ref{fig:relations} in terms of the conformal theory on the Lorentzian cylinder $\mathbb{R}\times S^{d-1}$.
Our usual spacetime $R^{1,d-1}$ is embedded as a single Poincar\'e patch, which consists of the set of points spacelike-separated from a reference $\infty$.  The key point is that correlators that span multiple patches are
equivalent to correlators involving multiple timefolds on a single patch, up to a predictable overall phase (see \cite{Kravchuk:2018htv}).
The figure illustrates how the same configuration of four points on the cylinder can be interpreted in various ways,
depending on how one chooses to place them relative to an (arbitrary) Poincar\'e patch.

The dynamics of these limits are clearly similar: in all cases, the Landau singularities populate a complete null cone.
In the folded OPE of section 2, the two tips of the cone are at the same ``point'' but on different Schwinger-Keldysh timefolds.
In the Regge limit, rather, the two tips are at the ``same'' point at past and future null infinity, respectively.
The limit creates a shockwave on the $x^-=0$ null sheet, with excitations moving at different transverse positions $x_\perp\in R^{d-2}$;
this null sheet is nothing but (a stereographic projection) of the null cone from infinity.

The correlators dubbed ``detectors'' in \cite{Caron-Huot:2022eqs} combine these limits: operators created at the origin cross a shockwave at $\scrI_+$, and are then absorbed back near the origin on the second Schwinger-Keldysh branch.
The shock represents physically different ways to weight final states at infinity (for example, energy flux measurement),
while the two branches respectively compute the amplitude and its complex conjugate entering a cross-section.
In all cases, where lines intersect is where particles can collide.

The relation between the Regge limit and detectors are relatively well explored, since perhaps the work of \cite{Hofman:2008ar} on
conformal collider physics. In perturbative QCD, it is sometimes called timelike-spacelike correspondence, see for example \cite{Hatta:2008st,Caron-Huot:2015bja,Vladimirov:2016dll,Mueller:2018llt}.
The relation between OTOCs and the Regge limit was noted in the context of the bound on chaos \cite{Shenker:2014cwa,Maldacena:2015waa}. It will be further explored in this section.

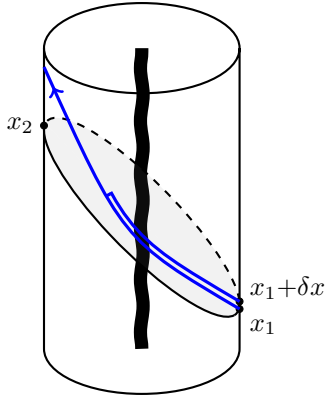
\begin{figure}[t]\centering
\begin{tikzpicture}[xscale=1.3]
\draw[thick] (-1,2.25) -- (-1,-1.75) arc (180:360:1 and 0.6) (1,-1.75) -- (1,2.25) arc (0:360:1 and 0.6);
\draw[line width=5pt,decorate,decoration={snake,amplitude=0.5,segment length=20}] (0,-1.75) -- (0,2.25);
\draw[thick,rotate=-54,fill=gray,fill opacity=0.1,shift=(0:1.61 and 0.4)] (0,0) arc(0:-170:1.61 and 0.4);
\draw[thick,dashed,rotate=-54,fill=gray,fill opacity=0.1,shift=(0:1.61 and 0.4)] (0,0) arc(0:190:1.61 and 0.4);
\filldraw (1,-1.22) ellipse(1pt and 1.3pt) (-1,1.22) ellipse(1pt and 1.3pt) (1,-1.12) ellipse(1pt and 1.3pt);
\node[below right] at (1,-1.22) {\small $x_1$}; \node[above right] at (1,-1.22) {\small $x_1{+}\delta x$};
\node[left] at (-1,1.22) {\small $x_2$};
\draw[very thick,blue,postaction=decorate,decoration={markings,mark=at position 0.93 with \arrow{>}}] (1,-1.22) .. controls (-0.2,-0.3) .. (-1,2);
\draw[very thick,blue,dash pattern=on 65pt off 200pt] (1,-1.12) .. controls (-0.2,-0.2) .. (-0.9,2.1);
\draw[very thick,blue] (-0.32,0.313) -- ++(-0.05,-0.035);
\end{tikzpicture}
\caption{In empty AdS, all lightrays from $x_1$ arrive at $x_2$ whether they pass through the boundary or bulk
(the interior of the cylinder).  When a macroscopic amount of energy (wavy black line) is added to the bulk,
bulk geodesics like the thick blue one experience a time delay and may not even return to the boundary,
thus washing out the Regge singularity at $(x_2-x_1)^2\to 0$.
In contrast, the folded OPE singularity between $x_1$ and $x_1+\delta x$ exists for any bulk state.
\label{fig:cyl}}
\end{figure}

We stress that these equivalences hold only around the vacuum. The limits become qualitatively distinct in excited states.
For example, if the bulk contains matter (or a black hole), geodesics that go through the bulk will no longer simultaneously arrive at the antipodal point due to Shapiro time delays. They may even never make it through.
Thus the Regge signal gets washed out when the bulk background geometry is far from AdS,
as depicted in figure \ref{fig:cyl}.  This phenomenon is known as saturation in high-energy scattering, or scrambling.

In contrast, the geodesics in the folded OPE limit perfectly follow each other on the different Schwinger-Keldysh branches,
no matter how far the background geometry is from the vacuum. We only need the probes to not significantly backreact on each other.
This can be ensured by keeping the center-of-mass energy below the Planck energy,
at the impact point. This state-independence of the folded OPE is crucial for our story.
It is why the cameras can image non-vacuum states.

\subsection{Conformal Regge Theory}

Conformal Regge Theory describes the Regge limit of correlators in terms of exchanges of Regge trajectories $J_i(\nu)$.
Assuming for notational simplicity that a single trajectory dominates, its contribution takes the form \cite{Costa:2012cb}
\be \label{CRT}
 \cG(z,\zb)^{\circlearrowleft} -1 \approx 2\pi i
 \int_0^\infty \frac{d\nu}{2\pi} \rho(\nu)\alpha(\nu)\ (z\zb)^{\frac{1-J(\nu)}{2}}
 \cP_{\frac{2-d}{2}+i\nu}\left(\frac{z+\zb}{2\sqrt{z\zb}}\right) \qquad (z,\zb\to 0).
\ee
The measure $\rho(\nu)$ and harmonic functions $\cP$ are generic kinematical objects defined in appendix \ref{app:harmonic}.
A specific formula for graviton exchange in AdS is obtained by inserting the
Regge trajectory $J=2$ and coefficient functions from eq.~(100) of \cite{Costa:2012cb}:
\be
\alpha_{\rm GR}(\nu) = -\frac{8\GN \RAdS^{1-d}}{\nu^2+\tfrac{d^2}{4}}
\frac{\gamma_{2\Delta_\cO+1}(\nu)\gamma_{2\Delta_{\cO'}+1}(\nu)}
{\Gamma(\Delta_\cO)\Gamma(\Delta_\cO-\frac{d-2}{2})\Gamma(\Delta_{\cO'})\Gamma(\Delta_{\cO'}-\frac{d-2}{2})}\ .
\label{alpha GR}
\ee
Here and below, $\Omega_d = \frac{2\pi^{d/2}}{\Gamma(d/2)}$ is the volume of the unit sphere $S^{d-1}$ and
\be \label{gamma}
 \gamma_a(\nu) \equiv \Gamma\big(\tfrac{a+1-d/2+i\nu}{2}\big)\Gamma\big(\tfrac{a+1-d/2-i\nu}{2}\big).
\ee
The expression \eqref{alpha GR} can be understood as follows: the quantum number $\nu$
is related to the scaling dimension of exchanged operators as $\Delta=\frac{d}{2}+i\nu$;
the Regge trajectory represents a $J$-plane singularity of OPE coefficients $c(\Delta,J)$ in the Lorentzian inversion formula \cite{Caron-Huot:2017vep,Caron-Huot:2020nem}.
The explicit poles in \eqref{alpha GR} thus correspond to exchange of a CFT operator with $\Delta=d$ (and its shadow at $\Delta=0$),
{\it ie.} the stress tensor, while the $\gamma$ factors have poles at double-twist exchanges $\Delta-J=2\Delta_\cO+2m$ for $m\geq 0$ and $J=2$.
Loosely speaking, eq.~\eqref{alpha GR} is the simplest function with all these poles and no other singularities.

The overall normalization is fixed by the residue on the stress tensor pole,
proportional to the inverse central charge $C_T^{-1}\sim \GN\ll 1$ as reviewed in appendix \ref{app:lorentzian}.
We used this constraint to write eq.~\eqref{alpha GR} in general $d$ (it was only explicitly quoted for $d=4$ in \cite{Costa:2012cb}).

It is significant that $\alpha_{\rm GR}(\nu)\sim e^{-\pi|\nu|}$ at large $\nu$.
This makes the correlator \eqref{OTOC CFT} a rather smooth and featureless function of the angle between $\delta x$ and $\delta y$
(if $\Delta_\cO,\Delta_{\cO'}$ are not parametrically large), in the reference kinematical region.
However, it ensures that the representation \eqref{CRT} converges even when analytically
continued to other regions. For example, for spacelike $\delta x,\delta y$, where $(z,\zb)\propto (e^{i\varphi},e^{-i\varphi})$,
we get $\cP_{\frac{2-d}{2}+i\nu}(\cos\varphi) \sim e^{\pm\varphi \nu}$.
In the bulk point limit $\varphi\to\pi$, the integral becomes only distributionally convergent.

Thus the fast decay of $\alpha$ enables to continue \eqref{CRT} (a priori defined only for $z,\zb>0$) to spacelike regions,
which are much better at probing bulk physics (see also \cite{Afkhami-Jeddi:2016ntf,Afkhami-Jeddi:2017rmx}). To our knowledge, it has not been established in an abstract CFT, only in examples.

To access those more explicitly, we rewrite the result \eqref{OTOC CFT}
in a form that eschews cross-ratios and separates the dependence on $\delta x$ and $\delta y$.
This can be achieved by integrating over angles (parametrized by a null direction $n^\mu$) using the identity \eqref{harmonic}:
\be\begin{aligned} \label{CRT nice} 
 G_{3241}\big|_{\rm CRT} &\approx \frac{1}{(\delta x_-^2)^{\Delta_\cO}(\delta y_-^2)^{\Delta_{\cO'}}}
 \ +2\pi i \int_0^\infty \frac{d\nu}{2\pi}\rho(\nu) \alpha(\nu) \frac{1}{(\delta x_-^2)^{\frac{2\Delta_{\cO}+J(\nu)-1}{2}}(\delta y_-^2)^{\frac{2\Delta_{\cO'}+J(\nu)-1}{2}}}
\\ &\qquad\qquad \times
\frac{1}{\Omega_{d-1}}\int_{S^{d-2}} \frac{d^{d-2}n}{(-n\.\hat{\delta x}_-)^{\frac{d-2}{2}+i\nu}(-n\.\mathcal{I}_e\.\hat{\delta y}_-)^{\frac{d-2}{2}-i\nu}}.
\end{aligned}\ee
Assuming the mentioned decay of $\alpha(\nu)$, eq.~\eqref{CRT nice} can be safely continued from the reference region $\delta x{\succ}0$, $\delta y{\prec}0$, where the parentheses in the second line are positive.

It is now straightforward to integrate \eqref{CRT nice} against wavepackets
since we can transform with respect to $\delta x$ and $\delta y$ separately.
The dependence on $n$ and $p^2$ is then determined by symmetries.
Inserting \eqref{nice transform}-\eqref{nice transform 1} (twice) into \eqref{CRT nice} and reverting the $n$ integral \eqref{harmonic},
one thus obtains a concise formula for the contribution of a Regge trajectory to \eqref{camera}:
\be \label{CRT mom}
 G(0,p,L;e,p',L')\big|_{\rm CRT}-1 \approx -i\int_0^\infty \frac{d\nu}{2\pi}\rho(\nu) \tilde\alpha(\nu) \left(|p||p'|\right)^{J(\nu)-1} \cP_{\frac{2-d}{2}+i\nu}(\hat{p}\.\mathcal{I}_e\.\hat{p}') e^{-\frac{\nu^2}{2}\sopt^2}
\ee
where
\begin{align}
\label{at from a}
 \tilde{\alpha}(\nu) &= -\alpha(\nu)\frac{2\pi}{4^{J(\nu)-1}}
 \frac{\Gamma(\Delta_\cO)\Gamma(\Delta-\frac{d-2}{2})}{\gamma_{2\Delta_\cO+J(\nu)-1}(\nu)}
 \frac{\Gamma(\Delta_{\cO'})\Gamma(\Delta'-\frac{d-2}{2})}{\gamma_{2\Delta_{\cO'}+J(\nu)-1}(\nu)}, \\
 \sopt^2 &= \frac{1}{L^2|p|^2} + \frac{1}{L'^2|p'|^2} \,. \label{sigma opt}
\end{align}
This is essentially the ``impact parameter amplitude'' \cite{Cornalba:2006xm,Kulaxizi:2017ixa,Meltzer:2017rtf}, with an extra Gaussian factor accounting for the optical resolution discussed below \eqref{shooting angle uncertainty}.
Our normalization of $\tilde\alpha$ was chosen to simplify the relation to flat space scattering amplitudes, see \eqref{a from M} below.

The main feature of \eqref{at from a} is that most factors in the gravity expression \eqref{alpha GR} neatly cancel,
in particular all the double-trace poles, leaving only the graviton exchange:
\be
 \tilde{\alpha}_{\rm GR}(\nu) = \frac{4\pi \GN \RAdS^{1-d}}{\nu^2+\tfrac{d^2}{4}}\,. \label{at GR}
\ee
For gravity, the $\nu$-integral in \eqref{CRT mom} can then be done analytically as it is merely
a fancy way of writing a propagator on the transverse hyperboloid $H_{d-1}$.
Specifically, as $L\to\infty$ one finds simply the function $\Pi_{d-1}(\hat{p}\.\mathcal{I}_e\.\hat{p}')$ in \eqref{hyperbolic propagator}, in precise agreement with section \ref{ssec:limits}.

Conceptually, this works because the $\nu$-representation can be used already at the level
of the OPE \eqref{grav OPE}.  The lightray $L_{x,\theta_x}$ is a function of an angle $\theta\in H_{d-1}$,
and a complete basis for such functions can be labelled by a null vector $n$ and a momentum $\nu$, such that the following
transform is invertible (see \eqref{fourier direct}):
\be
 L_{x,\theta_x}[h_{uu}] = \int\limits_0^\infty\frac{d\nu}{2\pi} \rho(\nu) \int_{S^{d-2}} \frac{d^{d-2}n}{(-n\.\theta)^{\frac{d-2}{2}-i\nu}}
 L_{x;n,\nu}[h_{uu}]. \label{Lnu}
\ee
Because the OPE \eqref{grav OPE} is Lorentz covariant, its transform must be proportional to $L_{x,\theta_x}$
and we find indeed:
\be\begin{aligned} \label{OPE with Lnu}
 \cO_2(\delta x)\ \cO_1(0)
& \sim 
\< \cO_2(\delta x)\ \cO_1(0)\> \\
&+
\frac{\tilde{C}_{\Delta_\cO}}{2(-\delta x_-^2)^{\Delta_\cO+\frac12}}
\int\limits_0^\infty\frac{d\nu}{2\pi} \rho(\nu) \int_{S^{d-2}} \frac{d^{d-2}n\ L_{x;n,\nu}[h_{uu}] }
{(-n\.\widehat{\delta x}_-)^{\frac{d-2}{2}-i\nu}}\times
  \frac{\pi^{\frac{d-2}{2}}\gamma_{2\Delta_\cO+1}(\nu)}{2\Gamma(2\Delta_\cO+1)}\,.
\end{aligned}\ee
Intuitively, the operator $L_{x;n,\nu}$ creates a bulk shockwave with momentum $\nu$ in the transverse $H_{d-1}$,
and the last factor is the transform of the bulk-to-boundary propagator.
We see from the $\gamma$ factor, which decays rapidly at large $\nu$,
that the scalar's bulk-to-boundary propagator (when $\delta x$ is timelike) has a smoothing effect which
obstructs access to large momenta.
The Fourier transform, which commutes with the harmonic decomposition, cancels this $\gamma$ factor, as above.
This is why it nicely localizes to a bulk geodesic.

All this was, in fact, discussed in the classic papers \cite{Cornalba:2006xk,Cornalba:2006xm}.
The only novelty here is that conformal symmetry plays no role:
we are only using the Lorentz symmetry of the tangent space near $x$.
This makes the folded OPE \eqref{OPE with Lnu} also apply in nontrivial bulk geometries far from the AdS vacuum.

Instead of Fourier-transforming \eqref{OPE with Lnu}, for our purposes one could equivalently
convolve it with any kernel which approximately cancels $\gamma$.
The wavepackets with finite support $|\nu|<\nu_{\rm max} \approx \Delta_{\rm gap}^{\rm higher-spin}$ used in \cite{Caron-Huot:2021enk} 
for example, enable to do everything without leaving positive-timelike $\delta x$.

\subsection{Relation to bulk $S$-matrix} \label{ssec:flat}

To plot the singular behavior of correlators as realistically as possible, below we will account for stringy corrections to pure gravity.
This can be readily done by relating the OTOC to a local scattering process in the bulk.

Intuitively, such a relation exists in a holographic theory in the approximation where one neglects the interactions between the bulk particles created by $\cO'(x_4)\cO(x_1)|\Omega\>$,
up to near the point marked as a  ``star'' in figure \ref{fig:relations}.   Then we have two ``asymptotic'' excitations hitting each other in a locally flat region.
This approximation is certainly valid at large-$N$. More generally, it captures the local bulk interactions which we expect to be responsible for small-angle singularities.

The bulk two-particle state is characterized by center-of-mass energy and impact parameter given in \eqref{sbulk}, namely:
\be \label{sbulk1}
s_{\rm bulk} = \frac{|p||p'|}{\RAdS^2}, \qquad \frac{b_{\rm bulk}}{\RAdS} = \cosh^{-1}\left(\frac{p\.\mathcal{I}_e\.p'}{|p||p'|}\right)=\mbox{small}\,.
\ee
After the collision, its products freely evolve back to the boundary of AdS.
Notice from figure \ref{fig:relationsC} that the scattering occurs backward in time since it is on the second timefold.
Its amplitude is thus given by the complex conjugate of the $2\to 2$ flat space $S$-matrix: $S^\dagger\approx 1-i\cM^\dagger$.

Let us first discuss the intended regime of operation of the camera: $1/|p| \ll L\ll |x-y|$, where
the wavepackets have well-defined position and shooting angles.
The fact that the shooting positions are well-separated ($L\ll |x-y|$)
implies that the bulk scattering angle is small.
(Otherwise, the particle fired from $x\pm O(L)$ would not return near $x$ after scattering and reverse time evolution.)
This is why shooting angles control the bulk impact parameter in \eqref{sbulk1}, as opposed to a bulk scattering angle.
The eikonal graviton exchange formula in \eqref{L flat} can thus be replaced, for a generic scattering process,
by the transform of $1-i\cM^\dagger$ to impact parameter:
\be
 G(x,p; y,p') \approx 1 - i\int \frac{d^{d-1}q_\perp}{(2\pi)^{d-1}} e^{iq_{\perp}\.b_{\rm bulk}}
 \frac{\cM^\dagger(s_{\rm bulk},t=-q_\perp^2)}{2s_{\rm bulk}} \qquad (1/|p| \ll L\ll |x-y|). \label{G from Mflat}
\ee
Here $b_{\rm bulk}$ in the exponent is any $(d-1)$-vector with the magnitude determined in \eqref{sbulk1}.
For tree-level gravity, where $\cM(s,t)\approx \frac{8\pi\GN s^2}{-t}$, this precisely agrees with \eqref{L flat}.

A similar reasoning led to eq.~(3.11) of ref.~\cite{Maldacena:2015iua} in the bulk point limit.
However, our setups differ in important ways.
First, we only see small-angle scattering with a fixed impact parameter, rather than fixed angle.
Second, here our $2\to 2$ scattering is localized around a unique bulk point, in any spacetime dimension,
as opposed to a $(d-2)$-dimensional manifold.
Third, the center-of-mass energy at the collision is fixed, rather than integrated over.
These differences are caused by our use of out-of-time-order correlators as well as to our use of wavepackets.

To compare with \eqref{CRT mom}, we simply need the fact that at large $\nu$ and small impact parameter (and $\nu b$ fixed),
the harmonic transform reduces to a transverse Fourier transform:
\be
\int_0^\infty \frac{d\nu}{2\pi}\rho(\nu)  \cP_{\frac{2-d}{2}+i\nu}(\cosh(b/\RAdS))F(\nu)
 \quad \Rightarrow\quad \RAdS^{d-1}\int \frac{d^{d-1}p_\perp}{(2\pi)^{d-1}} e^{ip_\perp{\cdot}b} F(\RAdS |p_\perp|).
\ee
Equating \eqref{G from Mflat} with \eqref{CRT mom} then gives the desired approximate relation,
which is most simply stated in the case that $\cM$ is controlled by a single Regge trajectory:
\be
\boxed{\frac{\cM(s,t)}{2s}\approx (\RAdS^2s)^{j(t)-1} \cM'(t) \quad\Rightarrow\quad \tilde{\alpha}(\nu) \approx \cM'_j(t)^*\big|_{t=-\nu^2/\RAdS^2}
\qquad (\nu\gg 1).}
\label{a from M}
\ee
The Regge trajectory $J(\nu) \approx j(\frac{-\nu^2}{\RAdS^2})$ is obtained by the same map.
This is exemplified below for the Veneziano-Shapiro amplitude.

The finite-size of wavepackets modifies this discussion through the uncertainty \eqref{shooting angle uncertainty} in the shooting angle,
which amounts to a Gaussian convolution in the impact parameter $b_{\rm bulk}$.  This is precisely accounted for by the Gaussian factor in \eqref{CRT mom}.
Note that a Gaussian uncertainly $\sim L$ in the shooting origin should \emph{not} be added.
This effect is absent because in \eqref{camera} we only integrate over the relative displacement $\delta x$
between creation and annihilation operators; the effect would be present if we had instead inserted independent wavepackets for each.

Let us briefly comment on the radar-like camera.
Integrating $\delta x$ against a wavepacket in the correlator \eqref{radar twofold},
the $\gamma_{\Delta_\cO+1}$ factor gets canceled and the analog of \eqref{CRT mom} is of the form
\be
 G_{\rm radar}(x,p_x;y,z) \sim \int d\nu \rho(\nu) \frac{\alpha(\nu)}{\gamma_{2\Delta_\cO+1}(\nu)}
 \cP(\hat{p_x}\.(\hat{U_{y,z}})_-)\,.
\ee
The argument of the harmonic function is a peculiar dot product of a timelike and spacelike vector.
The angle $\cosh^{-1}(\hat{p_x}\.(\hat{U_{y,z}})_-)$ is imaginary and approches $i\pi/2$ when the dot product vanishes.
This effectively cancels the second $\gamma$ factor at large $\nu$ and creates a singularity (or peak) when
\be
 p_x\cdot \left( \frac{y}{y^2} - \frac{z}{z^2}\right) =0. \label{radar singularity}
\ee
Mathematically this is similar to the bulk point singularity of the correlator in the small angle regime.
The condition \eqref{radar singularity} means that the future lightcone of $z$ and the past lightcone of $y$
intersect the geodesic $(0,p_x)$ at the same point.
This is precisely the expected signal from the radar camera \eqref{radar}, which was intended to record
at $y$ the reflection of a pulse sent from $z$.
There is a signal when $z$ and $y$ satisfy \eqref{radar singularity}
even before we Fourier transform with respect to $y$ (or $z$). The transform
would reveal more information about the angle of arrival.
(The holographic radar is further illustrated in figure \ref{fig:BTZ radar} in the next section.)

It may seem surprising that the pulse arrives and leaves at independent angles.
A good analogy is Compton scattering where a soft photon hits an ultrarelativistic electron sideways.
In the lab frame, the scattered photon can leave at any angle, and will simply experience a correlated Doppler-like energy shift
(related here to the ratio of $\frac{\partial}{\partial y}(p_x\. U_{y,z})$ and $\frac{\partial}{\partial z}(p_x\. U_{y,z})$).
From the center-of-mass frame perspective, the deflection is however
small and the Mandelstam invariants generically satisfy $-t/s\ll 1$.
The microscopic interaction is thus dominated by $t$-channel graviton exchange.  This is why
the eikonal approximation of the last section (for the $(0,p_x)$ geodesic) makes sense for $p_x$ sufficiently large.

\subsection{First images: $\mathcal{N}=4$ SYM at weak and strong coupling and 3D Ising model}\label{ssec:first images}

We are now ready to experiment with the AdS camera in various theories, not only pure AdS gravity.
We numerically study $\mathcal{N}=4$ super Yang-Mills (sYM) at strong and weak coupling,
as well as the critical 3D Ising model.

In the sYM model, we consider the OTOC of two half-BPS operators
in the stress-tensor multiplet: $\cO\sim {\rm Re}{\rm Tr}[XX]$ and $\cO'\sim {\rm Re}{\rm Tr}[YY]$,
where $X,Y$ are two complex scalar fields.
At large $\lambda$, sYM is dual to string theory where the string length is
related to the 't Hooft coupling $\lambda$ as
\be
\alpha' \RAdS^2= \frac{1}{\sqrt{\lambda}}\,.
\ee
As $\lambda\to\infty$, states on the leading trajectory are long rotating strings \cite{Gubser:2002tv}.
As long as $\nu\ll \sqrt{\lambda}$, the string remains short compared with the AdS radius, and therefore looks like one in flat space.
This means that the trajectory is approximately linear over a broad range extending to $\nu\sim\sqrt{\lambda}$, $J\sim \sqrt{\lambda}$:
\be
 J^{\mathcal{N}=4}_{\rm strong}(\nu) \approx 2 - \frac{1}{2\sqrt{\lambda}}(\nu^2+\tfrac{d^2}{4}) +
 O(\nu^2/\lambda) \qquad (\nu^2\ll \lambda). \label{N=4 strong j}
\ee
The corresponding amplitude can be obtained from the flat space Virasoro-Shapiro amplitude,
which we only need in the Regge limit:
\be\begin{aligned}
 \mathcal{M}^{\rm flat}_{VS}(s,t) &= \frac{8\pi \GN s^2}{-t}
 \frac{\Gamma(1-\tfrac14\alpha's)\Gamma(1-\tfrac14\alpha't)\Gamma(1-\tfrac14\alpha'u)}
 {\Gamma(1+\tfrac14\alpha's)\Gamma(1+\tfrac14\alpha't)\Gamma(1+\tfrac14\alpha'u)}
 \\& \quad \to  \frac{8\pi \GN s^2}{-t} \left( \frac{\alpha' s}{4}\right)^{\tfrac12\alpha't} e^{-i\frac14\pi\alpha't}
\frac{\Gamma(1-\tfrac14\alpha't)}{\Gamma(1+\tfrac14\alpha't)}.
\end{aligned}\ee
Using \eqref{a from M}, this provides a high-momentum completion of the gravity result \eqref{at GR}:
\be
 \tilde\alpha^{\mathcal{N}=4}_{\rm strong}(\nu) \approx 
\frac{2\pi^2}{N_c^2} \frac{1}{\nu^2+\tfrac{d^2}{4}}
 (-4i\sqrt{\lambda})^{\frac12\frac{\nu^2}{\sqrt{\lambda}}}
\frac{\Gamma(1+\tfrac14\frac{\nu^2}{\sqrt{\lambda}})}{\Gamma(1-\tfrac14\frac{\nu^2}{\sqrt{\lambda}})}.
\label{N=4 strong a}
\ee
This expression is now valid for any $\nu\ll \sqrt{\lambda}$, including small $\nu$.\footnote{We believe it should be equivalent
to eqs.~(106) and (112) of \cite{Costa:2012cb} but we have not compared in detail.
}
Note that this expression should only be used when $\nu^2 \lesssim s_{\rm bulk}\RAdS^2$.
In practice, we satisfy this constraint by terminating the $\nu$-integral in \eqref{CRT mom} at $\nu \sim \sopt^{-1}\times {\rm few}$, 
where the factor $e^{-\nu^2\sopt^2/2}$ already makes the integrand negligible.

At weak coupling, we borrow the following formulas from \cite{Costa:2012cb} (see also \cite{Cornalba:2008qf,Balitsky:2009yp}):
\be\begin{aligned}
J^{\mathcal{N}=4}_{\rm weak}(\nu) &= 1 + \frac{\lambda}{4\pi^2}\left(2\psi(1)-\psi\left(\tfrac{1+i\nu}{2}\right)-\psi\left(\tfrac{1-i\nu}{2}\right)\right) + O(\lambda^2),
\\
\tilde{\alpha}^{\mathcal{N}=4}_{\rm weak}(\nu) &= \frac{2i\lambda^2}{N_c^2} \frac{\tanh\tfrac{\pi\nu}{2}}{\nu(\nu^2+1)^2}+O(\lambda^3/N_c^2). \label{a weak}
\end{aligned}\ee

\begin{figure}[t]
\centering
\be\vcenter{\hbox{\includegraphics[width=12cm]{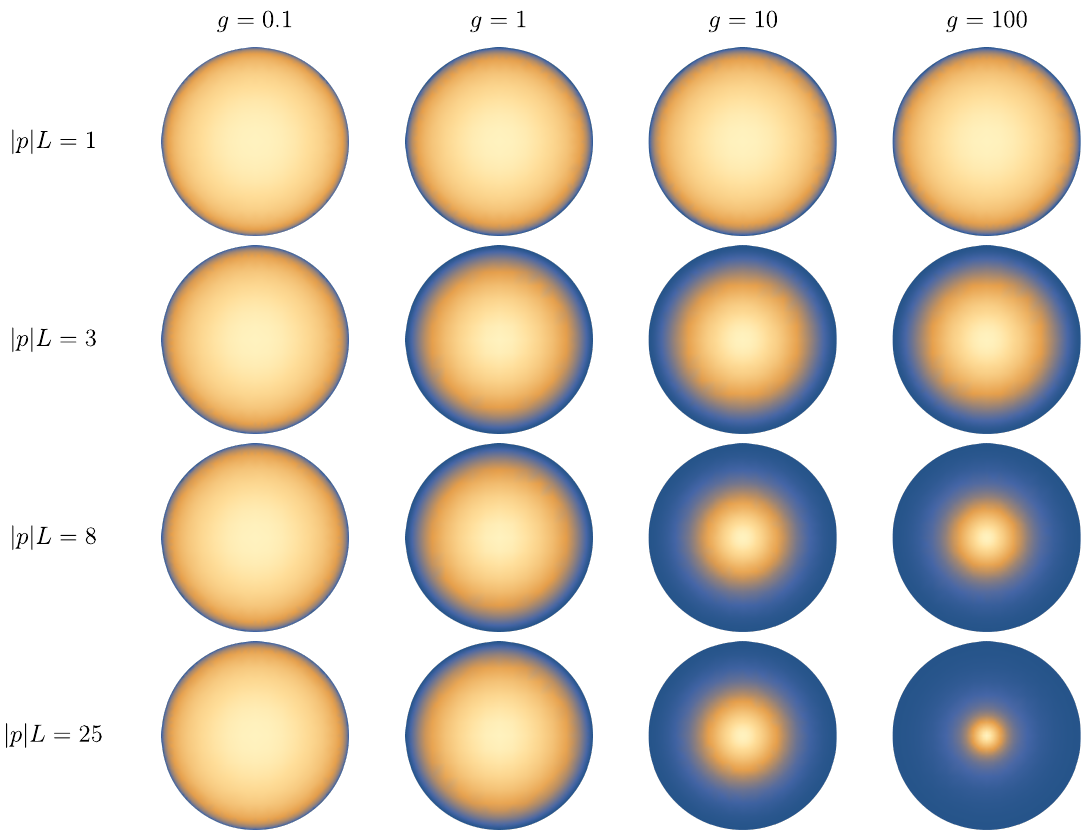}}}\qquad
\vcenter{\hbox{\includegraphics[width=8mm]{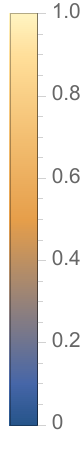}}}\nonumber\ee
\caption{Absolute value of the holographic camera signal \eqref{camera} in planar $\mathcal{N}=4$ sYM theory,
for various values of optical parameter $|p|L$ and coupling constant $g$.
Colors show the amplitude for receiving angles $(\tfrac{p_y^1}{p_y^0},\tfrac{p_y^2}{p_y^0})$
inside the unit disk, normalized by peak value.
A narrow peak requires good optics \emph{and} strong coupling.
Above the diagonal, the camera's angular resolution is limited by poor optics; below the diagonal,
by dynamics; along the diagonal, these limitations are roughly commensurate.
\label{fig:sYM}}
\end{figure}

In figure \ref{fig:sYM} we show the numerical integration of the right-hand-side of \eqref{CRT mom} in these models,
where we fix the projectile's momentum $p_x^\mu$ to be purely in the time direction and we display the amplitude of the signal
for different angles of $p_y^\mu$.  A sharp peak is interpreted as a signature of local bulk dynamics.
The two most important parameters are the
optical quality $|p_x|L_x=|p_y|L_y =|p|L\sim 1/\sqrt{\rm eff}$ and coupling $g^2\equiv \frac{\lambda}{16\pi^2}$.
We used the weak-coupling formulas for $g=0.1$, and strong-coupling for other values.
(These approximations may not be great at intermediate $g$.  We expect the trends and
qualitative features to be robust, though.)

Evidently, one sees that if $\sopt\sim 1/|p|L$ is not small, the camera has poor optics and any putative bulk object will seem blurred.
This is the familiar uncertainty relation: by observing the wave in a small region we can't know its momentum (ie. angle of arrival) accurately.  This effect can be mitigated by increasing the energies $p_x,p_y$ or by making the wavepackets wider,
within the constraint that $L_x,L_y\ll |x-y|$.

The next cause of blur is dynamical.
At strong coupling, this can be understood from the transverse broadening described by the Veneziano-Shapiro amplitude:
$\Delta b^2 \sim \alpha'\log s$.
A crude approximation to the observed signal is a Gaussian which adds optical and dynamical effects in quadrature:
\be
 \sigma_{\rm eff}^2 \approx \frac{1}{|p_x|^2L_x^2}+\frac{1}{|p_y|^2L_y^2} + \frac{\log \frac{|p_x||p_y||x-y|^2}{4\pi g}}{4\pi g}\,.
\label{sigma eff}
\ee
We see that even if one uses energies above the string scale, and ideal optics,
the camera can never resolve bulk impact parameters shorter than the string scale.
(The last term is only meaningful when the bulk center of mass energy exceeds the string scale, so that the logarithm is positive.  For lower energies, this term is in any case negligible compared with the others since $L_x,L_y\ll |x-y|$.)

\begin{figure}[t]
\centering
\includegraphics[width=10cm]{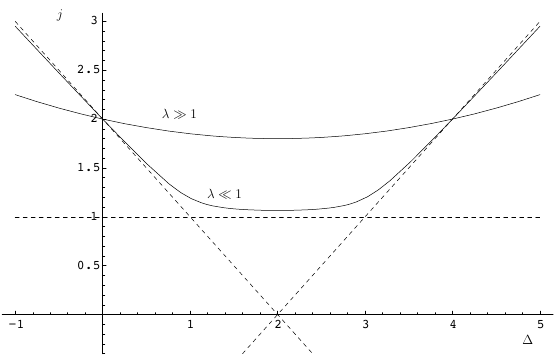}
\caption{Qualitative features of the leading Regge trajectory in planar gauge theory in $d=4$ (from \cite{Brower:2006ea}).
All trajectories pass through the point $(\Delta,J)=(4,2)$ and are symmetrical under the shadow reflection $\Delta\mapsto d-\Delta$.
At weak coupling, the trajectory is approximatively flat in the center  but bends rapidly near the intersection
with the twist 2 trajectories $\Delta-J\approx 2$.  At strong coupling, the curvature is small and inversely proportional to
the scaling dimension $\Delta_{\rm gap}^{\rm higher-spin}$ of the first single-trace spin-4 operator, through which the curve must pass.
The horizontal axis is related to the transverse momentum $\nu$ of bulk excitations as $\Delta=\frac{d}{2}+i\nu$.
\label{fig:BPST}}
\end{figure}

The large difference between weak and strong coupling is perhaps best explained from the sketch of Regge trajectories
shown in figure \ref{fig:BPST}.
The last factor in \eqref{sigma eff} is essentially the curvature of the leading (single-trace) Regge trajectory.
Good resolution is possible at strong coupling because the trajectory is nearly flat.
In general the curvature is controlled by the higher-spin gap which gives the lower bound
$\sigma_{\rm eff} \gtrsim 1/\Delta_{\rm gap}^{\rm higher-spin}$.
At weak coupling, the high curvature of the leading trajectory obstructs access to large transverse momenta; no amount of optical wizardry would bring ``bulk objects'' into focus.\footnote{
At very small coupling $g\to 0$,
one might object that the BFKL trajectory is approximately flat $J(\nu)\approx 1$ along the integration contour $\nu=$real
in \eqref{CRT mom}.  However, the comparatively rapid decay
$\tilde{\alpha}^{\mathcal{N}=4}_{\rm weak}(\nu)\sim  \nu^{-5}$ 
still makes the image blurry, and is in any case incompatible with an interpretation in terms of exchange of a finite set of bulk particles, which requires $\tilde\alpha\sim\nu^{-2}$.  The rapid decay of $\tilde\alpha$ for real $\nu$ is only possible because it has
complex poles at $\Delta-J\approx 2,4,6,\ldots$ and these exist at weak coupling
precisely because the trajectory is not flat for real $\Delta=\frac{d}{2}+i\nu$.}

\begin{figure}[t]
\centering
\begin{subfigure}[b]{0.6\textwidth}
\includegraphics[width=\textwidth]{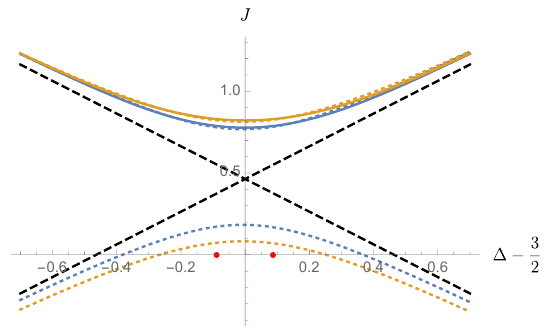}
\caption{\label{fig:Ising traj}}
\end{subfigure}
\hspace{0.1\textwidth}
\begin{subfigure}[b]{0.28\textwidth}
\raisebox{5mm}{\includegraphics[width=\textwidth]{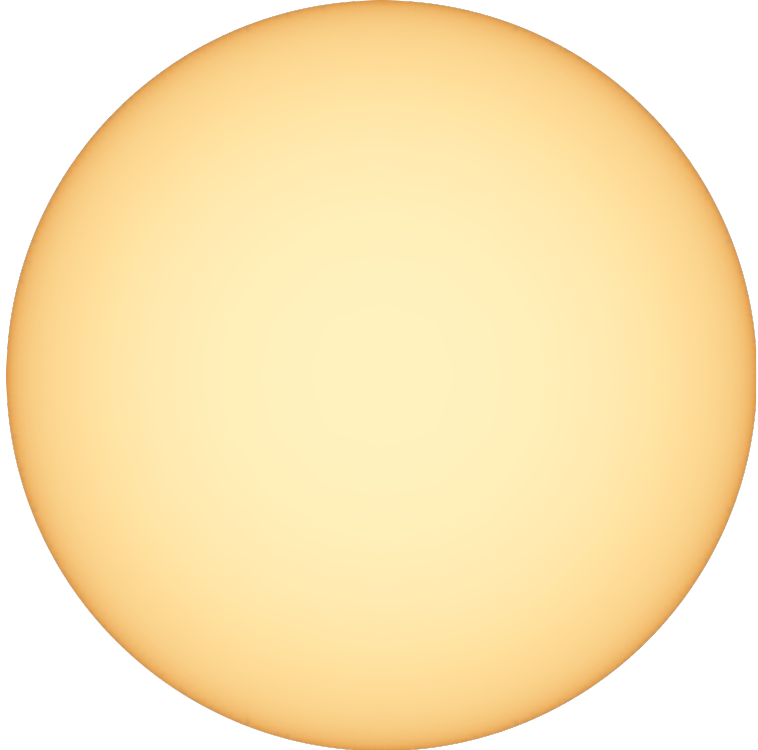}}
\caption{\label{fig:Ising mom}}
\end{subfigure}
\caption{(a) Estimate of the leading Regge trajectory in the critical 3D Ising Model (from \cite{Caron-Huot:2020ouj}),
showing its two branches, the orange (outermost) curve being the preferred one.
(b) Angular dependence of the camera signal, similar to figure \ref{fig:sYM}.
The optical parameter ($|p|L=8$ was used) has little influence on the plot.
\label{fig:Ising}}
\end{figure}

Let us now turn to the 3D Ising model.
Its leading trajectory is well understood for integer spins, from various analytic and numerical methods \cite{Alday:2015ota,Simmons-Duffin:2016wlq,Su:2022xnj}.
Less is known about its analytic continuation to the principal series, where $\Delta=\tfrac32+i\nu$ with real $\nu$.
Ref.~\cite{Caron-Huot:2020ouj} gave numerical estimates of the trajectory near the origin, with intercept $j_*\approx 0.8$, as reproduced in figure \ref{fig:Ising traj} (see \cite{Liu:2020tpf} for a related discussion of the O(N) model).
Here we will also need functional forms for the trajectory and coefficient functions.
We do not know it rigorously but we will attempt here a reasonable approximation
using the $\ep$-expansion of the Wilson-Fisher theory.

There are in fact two nearby trajectories. In the free theory these are the leading-twist trajectory and its shadow, which intersect at $\nu=0$. At small but finite coupling they mix, and the intersection is resolved, as detailed recently \cite{Caron-Huot:2022eqs}:
\be
 J^{\rm W-F}_\pm(\nu) \approx \tfrac{\ep}{2} \pm \sqrt{\nu_0^2-\nu^2} +O(\ep^2) ,\qquad \nu_0=\tfrac{\sqrt{2}}{3}\ep\,.
\label{Ising j}
\ee
Here $\ep=4-d$ and we take $\nu\sim \ep$.
Of course, one can cannot really trust this formula when $\ep=1$.
For modelling purposes, we will simply set $\ep\to 0.8$ in \eqref{Ising j},
which we find to be reasonably close to the numerical curve in figure \ref{fig:Ising traj}.

Ref.~\cite{Caron-Huot:2022eqs} gave operator representations which in principle enable to calculate the residue $\alpha(\nu)$.
We take a shortcut. From the Lorentzian inversion formula, $\alpha$ is the residue of a coefficient function $c(\Delta,J)$ with poles at \eqref{Ising j}. Using known OPE coefficients along the leading trajectory away from the intercept,
we can deduce the fist few orders in the $\ep$-expansion near the intercept, as detailed in \eqref{alpha Ising}.
The Regge limit of correlators may then be obtained by plugging into \eqref{CRT}.
Further simplifications arise because the trajectories have a large slope away from the intercept.
The steepest descent method then reduces the total contribution to a discontinuity of $\alpha$, as discussed in the same appendix:
\be\label{CRT Ising}
 \cG(z,\zb)^{\circlearrowleft}_{\rm W-F}-1 \approx
\oint \frac{\pi\ d\nu}{\sqrt{\nu^2-\nu_0^2}}
\frac{(z\zb)^{\frac12-\frac{\ep}{4} -\frac{i}{2}\sqrt{\nu^2-\nu_0^2}}}{\frac{\ep}{2}+i\sqrt{\nu^2-\nu_0^2}} \rho(\nu)\cP_{\frac{2-d}{2}+i\nu}\left(\frac{z+\zb}{2\sqrt{z\zb}}\right) + O((z\zb)^{1+O(\e)})
\ee
where the contour encircles the square-root cut counter-clockwise.

As a nontrivial sanity check, we can study \eqref{CRT Ising} as $\ep\to 0$.
The cut becomes infinitesimal and can be computed exactly.
The dependence on cross-ratio is simply proportional to the function $\cP_{-1}$,
while the explicit factors (and $\rho(\nu)\approx \frac{\nu^2}{\pi}\sim \ep^2$) control the $\ep$-scaling. We find:
\be
 \cG(z,\zb)^{\circlearrowleft}_{\rm W-F}-1 \approx \frac{2\pi i\ep}{3}\frac{z\zb}{z-\zb}\log\frac{z}{\zb} + O(\ep^2)\,. \label{limit W-F}
\ee
This can be compared with the connected diagram at $O(\lambda)$, which can be calculated exactly as
\be
 \cG(z,\zb)_{\rm W-F}\Big|_{\lambda} = -\frac{\lambda}{16\pi^2} \frac{z\zb}{z-\zb}
 \left[2{\rm Li}_2(z)-2{\rm Li}_2(\zb) + \log(z\zb)\log\frac{1-z}{1-\zb}\right].
\ee
After the analytic continuation to $\cG^\circlearrowleft$, the square bracket becomes $-2\pi i\log\frac{z}{\zb}$ in the Regge limit.
In the critical dimension ($\frac{\lambda}{16\pi^2} \to \frac{\ep}{3}$) this agrees precisely with \eqref{limit W-F}.
This calculation confirms that the operator $L[\phi^2]$ in \eqref{phi4 OPE} is not independent at small but finite coupling
and can be re-expressed as a short Regge cut involving double-twist operators.

Our proposed model for the wavepacket-transformed correlator
in the 3D Ising model is to take the integral \eqref{CRT Ising} at finite $\ep$ and
multiply by the factors in \eqref{at from a} evaluated as $\ep\to 0$. This gives the analog of \eqref{CRT mom}:
\be \label{Ising mom}
 G(0,p,L;e,p',L')\big|_{\rm Ising}-1 \approx
\oint \frac{36\pi\ d\nu}{\sqrt{\nu^2-\nu_0^2}}
\frac{(|p||p'|/4)^{-1+\frac{\ep}{2}+i\sqrt{\nu^2-\nu_0^2}}}{\frac{\ep}{2}+i\sqrt{\nu^2-\nu_0^2}}
\rho(\nu)\cP_{\frac{-1}{2}+i\nu}(\hat{p}\.\mathcal{I}_e\.\hat{p}') e^{-\frac{\nu^2}{2}\sopt^2}
\ee
Note that in the 3D Ising model, the signal's magnitude l gets smaller with increasing energy.
In figure \ref{fig:Ising mom} we display the numerical integration of \eqref{Ising mom},
with $\rho$ and $\cP$ evaluated at $d=3$, and $\ep=0.8$ in all other factors, as a function of the angle between $p$ and $p'$.
We observe a very smooth signal as a function of these angles:
the camera shows no hint of bulk point-like particles in the critical Ising model.

While there is admittedly ambiguity in the above model  (we only used the first term in the $\ep$-expansion),
we believe that the conclusion is robust.
The leading Regge trajectory is simply too curved to pump large transverse momenta into putative bulk particles.

It is generally believed that local bulk dynamics (at sub-AdS distances)
requires the dual CFT to have a large central charge $c_T\gg 1$ and a large (single-trace)
higher-spin gap $\Delta_{\rm gap}^{\rm higher-spin}\gg 1$ \cite{Heemskerk:2009pn}.
We see here that these criteria are precisely those needed for the holographic camera to produce sharp images.
The large central charge is needed to make the overall coefficient  on the right of \eqref{CRT mom} small,
so that we can access high energies before reaching saturation (ie. before $s_{\rm bulk}$ approaches the Planck scale).
The large gap is synonymous with the leading (single-trace) Regge trajectory being nearly flat.

\section{More images: CFT$_2$ thermal state and BTZ black hole}\label{sec:BTZ}

Before highlighting the qualities of the camera in a nontrivial state,
let us summarize some of its limitations we have seen so far:
\begin{enumerate}
\item If $pL$ is not large, the shooting angle will not be precise ($\Delta \theta \sim 1/(pL)$)
\item If the wavelength $1/p$ is not small compared with typical bulk structures,
waves of different frequencies will disperse
\item If bulk center-of-mass energies overcome the Planck scale $\sim 1/C_T$,
we can no longer ignore higher-order terms in \eqref{grav OPE} and scrambling may occur
\item Momentum transfers larger compared with the higher-spin scale $\Delta_{\rm gap}^{\rm higher-spin}$
will be damped by \eqref{sigma eff}.
\end{enumerate}
The first two are standard limits of geometrical optics: these would apply to any
camera-like device. They are generally alleviated by using shorter wavelengths.
The last two are dynamical limits which limit one's access to short wavelengths,
thus fundamentally limiting the focusing power of holographic cameras.

Consider a (spatially infinite) thermal state in a two-dimensional CFT.  The Euclidean CFT is defined on a spacetime with the topology of a cylinder $S^1\times R$ where the circle has circumference $1/T$.
We work above the temperature of the Hawking-Page transition so that
$1/T$ is smaller than the circumference of a (possible compact) spacial circle.
In two dimensions, this geometry can be conformally mapped to the plane, which allows to write thermal correlators in terms of vacuum ones \cite{Cardy:1984rp}.
The same applies to a general boosted state, which is conventionally described with left- and right-moving temperatures.

\def\boost{{\varphi}}

If the rest temperature is $T=\sqrt{T_LT_R}$, the conformal transformation is
\be
  z_i = \exp\left(2\pi T_R(x_i^1-x_i^0)\right),\qquad  \zb_i =\exp\left(2\pi T_L(x_i^1+x_i^0)\right)\,,
\ee
where $T=\sqrt{T_L T_R}$ is the temperature in the rest frame.
We are interested in the OTOC configuration similar to \eqref{SK fourfold}:
\be
 x_{1,2}^\mu = \mp \tfrac12 |\delta x| (\cosh\boost_x,\sinh \boost_x), \qquad
 x_{3,4}^\mu = (\Delta t,y) \mp \tfrac12 |\delta y| (\cosh\boost_y,\sinh \boost_y),
\ee
where the cross ratios go to, as $\delta x,\delta y\to 0$:
\be
 \sqrt{z\zb} \to \frac{\pi^2 T_RT_L\ \delta x\delta y}{\sinh(\pi T_R(\Delta t-y)\sinh(\pi T_L(\Delta t+y))},\quad
 \sqrt{\frac{z}{\zb}} \to e^{-\boost_x-\boost_y} \frac{T_R \sinh(\pi T_L(\Delta t+y))}{T_L \sinh(\pi T_R(\Delta t-y))}\,.
\ee
The correlation functions are then given by the vacuum correlator $\cG^{\circlearrowleft}(z,\zb)$ in \eqref{CRT}.
Since the dependence of cross-ratios on the small vectors $\delta x^\mu$ is similar to \eqref{cross-ratios}, the wavepacket-transformed correlator can be obtained as in \eqref{CRT mom}.
Parametrizing the momenta as $p_i^\mu=|p_i|(\cosh \boost_i,\sinh \boost_i)$ and specializing the transform to 2d, we have\footnote{
The same result may apply to a generic heavy energy eigenstate, following \cite{Fitzpatrick:2015zha}.}
\be \label{G BTZ}
 G(0,p_x,L_x; y,p_y,L_y)_{T_L,T_R}-1 \approx
 -i\int_{-\infty}^\infty \frac{d\nu}{2\pi}\tilde\alpha(\nu) \left(|p_x||p_y|\mathcal{I}(\Delta t,y)\right)^{J(\nu)-1}  e^{i\nu(\boost_x+\boost_y+\Delta\boost)} e^{-\frac{\nu^2}{2}\sopt^2}\,,
\ee
with $\tilde\alpha(\nu)$ and $\sopt$ as in \eqref{CRT mom} and
\be
\mathcal{I}(\Delta t,y)=\frac{\sinh(\pi T_R(\Delta t-y)\sinh(\pi T_L(\Delta t+y))}{\pi^2 T_R T_L}, \quad\Delta \boost=\log \frac{T_L \sinh(\pi T_R(\Delta t-y))}{T_R \sinh(\pi T_L(\Delta t+y))}\,. \label{I and boost}
\ee
These will be interpreted shortly in terms of the bulk geometry of the BTZ black hole dual to the thermal state in the CFT:
the intensity ${\cal I}$ is related to the blueshift factor at the center-of-mass energy at the collision and other factors;
the offset $\Delta \boost$ means that a null geodesic fired from the origin with boost angle $\boost_x$ intersects a null geodesic fired backward from $(\Delta t,y)$ with $\boost_y=-(\boost_x+\Delta \boost)$.

To enhance contrast, in all plots of this section
we take two derivatives of \eqref{G BTZ} with respect to $\theta_y$. This multiplies the integrand by $-\nu^2$.

In ``movie'' \ref{movie:BTZ} we display \eqref{G BTZ} as a function of receiving angle $\boost_y$ for various times $\Delta t$,
for a fixed geodesic initially fired at a square angle into the bulk: $\boost_x=0$. 
Note that while the bulk is ($2{+}1$)-dimensional, the image on a given time slice is only one-dimensional, similar to how a conventional camera records
a codimension-one projection of a scene, resolving the angle-of-arrival of photons but not depth.
Depth information can however be recovered in various ways, for example by
using two cameras displaced by a small offset $\Delta y$ (called ``left eye'' and ``right eye'' in the figure).
The following parameters were used:
\be
\quad 2\pi T_R=1,\quad 2\pi T_L=\tfrac12,\quad
 L_x=L_y=\frac{1}{100},\quad  |p_x|L_x=|p_y|L_y=100, \quad \Delta y =0.1. \label{BTZ params}
\ee
Since stringy effects were analyzed in the preceding section, here we simply set $\alpha'=0$ (and $c_T\gg 1$).
The correlator does not dependent on the scaling dimensions $\Delta_\cO$ and $\Delta_{\cO'}$ of
external operators, in the considered regime of geometrical optics.

\begin{movie}[t]
\centering
\raisebox{7ex}{\begin{minipage}[t]{0.11\textwidth}\raggedleft left eye\\ right eye\end{minipage}}
\hspace{-0.06\textwidth} \includegraphics[width=0.9\textwidth]{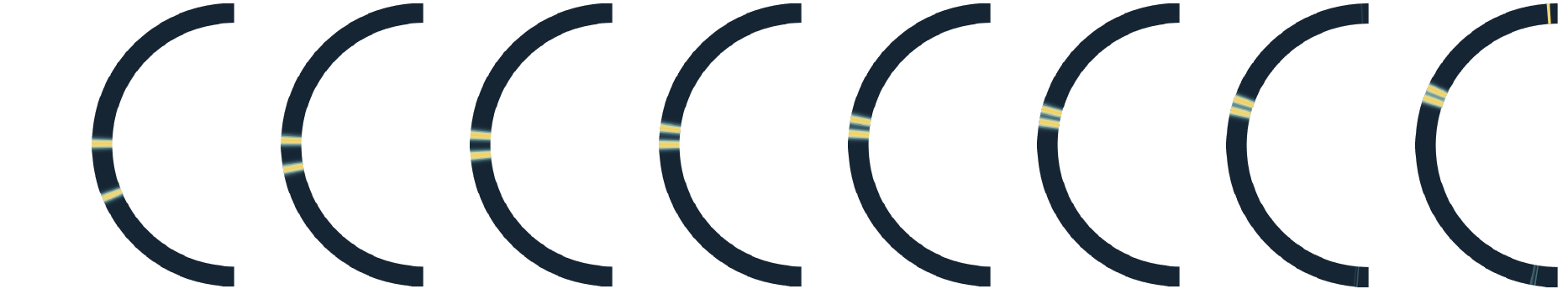}
\caption{A particle, initially fired at a right angle from the boundary at $t=0$, moves in the geometry of the BTZ black hole.
Each slide shows the intensity detected by the active camera as a function of the return angle $\boost_y$.
The shown times range over $t=0.5\ldots4$ in increments of $0.5$, with the temperature and other parameters as in \eqref{BTZ params}.
To add depth information, the images from two cameras displaced by $\Delta y=0.1$ have been superposed.
\label{movie:BTZ}}
\end{movie}

Several features can be observed. First, the image from just the left eye (at $y=0$)
shows that the particle initially at the origin starts drifting towards the right.
This happens because of the plasma's velocity.
Second, the angular separation between the two eyes' images decreases with time.
Intuitively, this happens as the particle goes deeper into the bulk,
the separation being inversely proportional to the radial coordinate, by stereography.
The fact that the angular separation stabilizes at a finite value at late times indicates that the radial motion stops.
Here this happens as the particle approaches a causal horizon in the bulk. This will be confirmed below by the signal intensity.
Finally, sidebands appear at $\Delta t\gtrsim 3$
because, in fact, we defined the theory on a periodic circle $y\simeq y+3$; past this time, the receiving geodesic
 $Y(v)$ can travel once around the circle.\footnote{In the geometrical optics approximation, we obtain the full signal simply by summing over images.}

\begin{figure}[t]
\centering
\begin{subfigure}[b]{0.4\textwidth}\centering
\includegraphics[width=\textwidth]{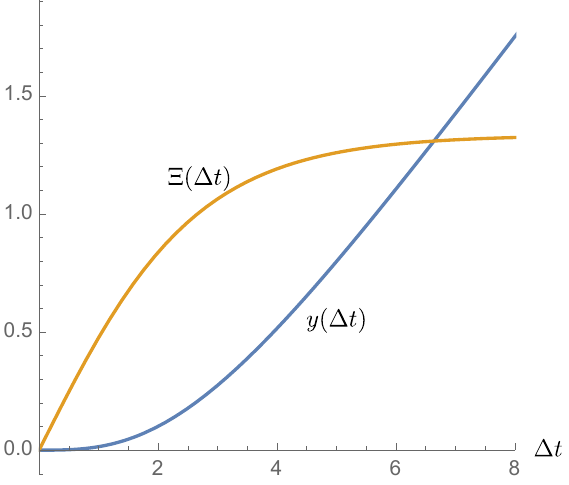}
\end{subfigure}
\hspace{0.15\textwidth}
\begin{subfigure}[b]{0.4\textwidth}\centering
\includegraphics[width=\textwidth]{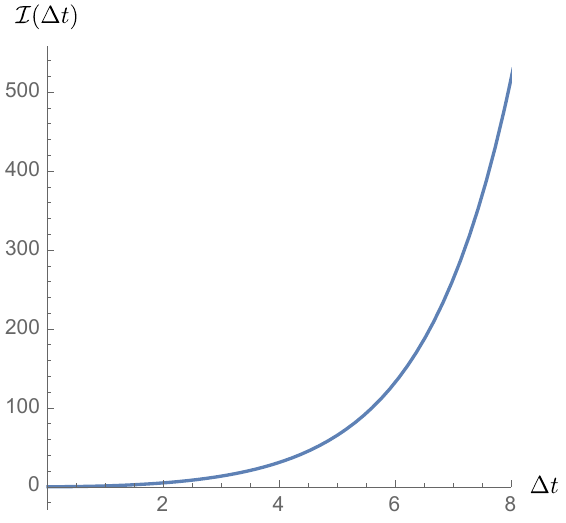}
\end{subfigure}
\caption{Three functions which characterize the bulk geodesic in figure \ref{movie:BTZ}: the horizontal displacement $y(\Delta t)\big|_{\boost_y=0}$ needed for the signal to come back at a right angle; the parallax $\Xi(\Delta t)$ in \eqref{Xi};
the signal intensity ${\cal I}(\Delta t)$ from \eqref{G BTZ}.  Parameters again as in \eqref{BTZ params}.
\label{fig:BTZ}}
\end{figure}

As a natural alternative to measuring the angle $\boost_y(\Delta t)\big|_{y=0}$, one could orient the receiver at a right angle
and scan over lateral displacements, measuring the peak's position $y(\Delta t)\big|_{\boost_y=0}$.
This data will be somewhat simpler to interpret below.
To get a sense of depth, it is natural again to use two nearby receivers.
``Parallax'' $\Xi$ can be defined as the ratio of lateral movement to angular change:
\be
 \Xi(\Delta t) \equiv -\frac{d\Delta y}{d\boost_y} \Big|_{\boost_y=0}\,.  \label{Xi}
\ee
Thus, using only (near) right-angle geodesics, we can measure three functions of time:
$\Delta y(\Delta t)$, $\Xi(\Delta t)$, and the intensity ${\cal I}(\Delta t)$.
The former can be determined numerically from \eqref{I and boost} by solving for $\Delta\boost=0$,
after which the other two follow readily.  This procedure was used to make the plots in figure \ref{fig:BTZ}.

These plots again show readily interpretable features.
The parallax $\Xi(\Delta t)$ saturates at late times as the particle reaches the horizon, as just described.
The linear rise in $y(\Delta t)$ quantifies the drift velocity at the horizon.
At early times, the intensity rises quadratically: ${\cal I}\propto (\Delta t)^2$,
indicating scale-invariant dynamics, ie. that the bulk geometry is asymptotically AdS.
The intensity later becomes exponential ${\cal I}\sim e^{2\pi T \Delta t}$,
which is the smoking gun of a causal horizon \cite{Shenker:2013pqa}.\footnote{
The exponential growth will stop after a scrambling time $t\sim \frac{1}{2\pi T}\log c_T$.
By that time, the exterior metric has already been essentially fully reconstructed.}

The fact that ${\cal I}$ grows with time may be counter-intuitive. It is because excitations, both from $x$ and $y$,
blueshift as they go deeper into the bulk. The locally measured center-of-mass energy at the collision thus increases with time.
This is a peculiar feature of the active camera.
The same would not happen for the radar camera, where
the incoming pulse indeed experiences the expected Doppler redshift as it bounces off the receding target.

The three measured functions in figure \ref{fig:BTZ} essentially give the bulk geometry.
To see this explicitly, we make a general ansatz for an asymptotically-AdS metric invariant under $x$ and $t$ translations:
\be
 ds^2_{\rm ansatz} = \frac{\RAdS^2}{r^2} \left[ -A(r) dt^2 + \frac{dr^2}{B(r)} + (dx-C(r)dt)^2\right]\,, \label{ds2 ansatz}
\ee
where $A(r),B(r)\to 1$ and $C(r)\to 0$ at the $r=0$ boundary, but are otherwise arbitrary functions.
Using eq.~\eqref{L flat} we can predict the signal in terms of the metric and its null geodesics.
Integrating the geodesic equation (this is easy since momenta corresponding to spacetime translations are conserved),
we find for small angles
\begin{subequations} \label{metric ansatz ABC}
\begin{align}
\frac{\Delta t}{2} &= \int_0^{r_*} \frac{dr}{\sqrt{A(r)B(r)}}\,,
\hspace{-20mm}& \frac{y}{2}&=\int_0^{r_*} \frac{dr\ C(r)}{\sqrt{A(r)B(r)}}\,,
\\
\mathcal{I}(\Delta t) &\propto \frac{r_*}{A(r_*)} \Xi(\Delta t)\,,
& \Xi&=\int_0^{r_*} dr \sqrt{\frac{A(r)}{B(r)}}\,.
\end{align}
\end{subequations}
The first equation determines the collision depth $r_*(\Delta t)$ as a function of time, so the other three equations
essentially define $A(r_*)$, $B(r_*)$, $C(r_*)$.
(In spacetime dimension $d$, the intensity is proportional to a factor of $\Xi^{3-d}$, see \eqref{L flat}.)
Taking various derivatives yields the following more useful relations:
\be
C(r_*(\Delta t))=\frac{dy(\Delta t)}{d\Delta t}, \qquad A(r_*(\Delta t))=2\frac{d\Xi(\Delta t)}{d\Delta t}\,,
\qquad r_*(\Delta t) \propto  \frac{\mathcal{I}(\Delta t)}{\Xi(\Delta t)^{3-d}} \frac{d\Xi(\Delta t)}{d\Delta t}\,.
\label{recipe}
\ee
For future reference, we have written the last relation in general spacetime dimension. The proportionality
constant is fixed by requiring that $r_*\to \frac{\Delta t}{2}$ at short times; physically, the constant defines the AdS radius.
Note that all quantities on the right-hand-sides are just derivatives of the numerical data in figure \ref{fig:BTZ}.
Thus from data one directly finds the metric functions $C$ and $A$ \emph{and} the bulk coordinate $r$ in the gauge \eqref{ds2 ansatz},
and therefore the functions $C(r)$ and $A(r)$!
The function $B(r)$ follows readily using either $\frac{\Delta t}{dr_*}$ or $\frac{d\Xi}{dr_*}$ from \eqref{metric ansatz ABC}.

For the correlator in \eqref{G BTZ}, the resulting functions agree precisely with the Ba\~{n}ados-Teitelboim-Zanelli metric \cite{Banados:1992wn,Banados:1992gq}:
\be
 A(r)=B(r) = (1-4\pi^2r^2T^2)(1-4\pi^2r^2T^2\ep^2), \qquad C(r)=4\pi^2 r^2 T^2\e
\ee
where $T=\frac{T_R+T_L}{2}$ and $\e=\frac{T_R-T_L}{T_R+T_L}$.
We have tested this numerically but did not attempt to prove it analytically (the equations are implicit and give $A$, $B$, $C$ and $r$ as functions of $\Delta t$).

What have we learnt?  Of course, there is nothing particularly remarkable about recovering the BTZ geometry from CFT correlators:
the metric is essentially determined by symmetry (it is locally AdS$_3$), just like the correlator \eqref{G BTZ} we started with.
The interesting thing about the recipe in \eqref{recipe} is that it makes no reference to Einstein's equations in the bulk:
it is just a method to extract a metric from correlation functions, and is valid whenever
geometrical optics applies in the bulk (so that we can unambiguously measure $y(\Delta t)$ and $\Xi(\Delta t)$ from the positions of
sharp peaks).  The specific recipe \eqref{recipe} applies to any situation with translation invariance, but the idea is more general.
One could write down an arbitrary metric (not necessarily satisfying any bulk equations of motion),
compute a sufficient number of OTOCs in this metric, and a colleague in principle could recover the metric from this data.

Note that we are not claiming that any metric can be ``easily'' reconstructed in this way.
The general problem, in the absence of symmetries, may be somewhat analogous to reconstructing the mass distribution of a galaxy from
gravitational lensing data or other kinds of images. Getting the images is just one step.

\subsection{More on translation-invariant states in higher dimensions, and radar}

The simple reconstruction \eqref{recipe} applies to translation-invariant and rotationally invariant states
in any spacetime dimension. It enables to write the bulk Einstein's equations directly in terms of boundary observables.
Working in the plasma's rest frame for simplicity, so that $y(\Delta t)=0=C(r)$, we find
\be \label{Einstein}
\mbox{Einstein's equations:}\ \left\{ \begin{array}{l}
 \frac{A(r)}{B(r)} =\mbox{Constant}(=1),\\
 (r\partial_r-d)(B(r)-1)=0,
  \end{array}\right.
\!\!\quad  \Leftrightarrow\quad
\left\{ \begin{array}{l}
\mathcal{I}(\Delta t) \Xi(\Delta t)^{d-4} \frac{d\Xi(\Delta t)}{d\Delta t} = \mbox{Constant},\\
\frac{1}{\Xi(\Delta t)^d} \left( 2\frac{d\Xi(\Delta t)}{d\Delta t} -1\right) = \mbox{Constant}\,.
\end{array}\right.
\ee
We verified that this is solved by the AdS-Schwarzschild metric ($A(r)=B(r)=1-(r/r_0)^d$).

\begin{figure}[t]
\centering
\includegraphics[width=0.35\textwidth]{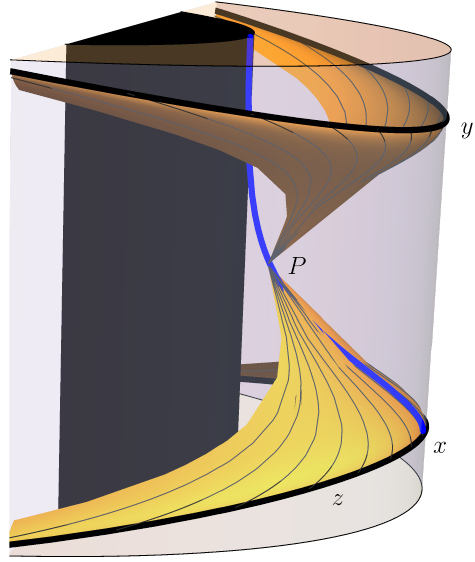}
\caption{
Geometry of the holographic radar:
a pulse sent from a point $z$ reflects off an object (following the thick blue geodesic) at a point $P$ and is later recorded at $y$.
This is shown here in the geometry surrounding a BTZ black hole (the inner cylinder),
while all measurements are done within the finite-temperature system at the outer boundary of the cylinder.
The yellow surface displays null geodesics from $P$ that reach the boundary.
Any points $z$ and $y$ at the intersection of this surface and the boundary would probe the same bulk point $P$.
\label{fig:BTZ radar}}
\end{figure}

It is very surprising to find the bulk dynamics expressed by a differential equation on \emph{boundary} observables.
This is possible here because the parallax $\Xi(\Delta t)$ is a quantitatively accurate radial coordinate,
not simply a qualitative measure of size or renormalization scale.%
\footnote{Entanglement surfaces and Euclidean geodesics can also accurately resolve bulk coordinates,
see for example \cite{Kusuki:2019hcg,Jokela:2020auu}.}
It is an important open question to interpret eqs.~\eqref{Einstein} from the boundary perspective.

In figure \ref{fig:BTZ radar} we describe the working of the radar camera in the BTZ black hole geometry,
with $T_L=T_R=1.1/(2\pi)$ and $x$-period $2\pi$, ie. just above the Hawking-Page temperature.
After Fourier transforming with respect to $\delta x$ (which singles out a projectile geodesic $(x,p_x)$),
the correlator displays a singularity at the conformally transformed version of \eqref{radar singularity}.

The bulk picture is that the forward lightcone from the pulse $z$ intersects the geodesic $(x,p_x)$ at a bulk point $P$.
Holding this data fixed, the reflected pulse can be observed at any point $y$ along the future lightcone of $P$.
Conversely, fixing $(x,p_x)$ and $y$, we would observe a singularity at any $z$ on the past lightcone of $P$.
The microscopic process is graviton exchange in the bulk as discussed below \eqref{radar singularity}.
(The points $x$ and $z$ shouldn't be too close, to avoid signal loss by Doppler redshift.)
Note that the projectile does not need to be recovered after the experiment, since its position is not observed beyond the point $P$.

The same construction works in any spacetime dimension and only requires four-point correlators.
Any bulk point $P$ whose past and future lightcone reach the boundary in some open sets
can be probed in this way.  Fourier transforming with respect to small variations $\delta y$ (or $\delta z$) would reveal additional information such as the parallax described above.

\section{Summary and discussion}\label{sec:conclusions}

In this paper we considered four-point correlation functions
as a tool to image point-like objects in the putative bulk dual of a field theory.
The correlators are decoded by a standard Fourier transform.
The images come out blurred in theories that do not satisfy the large $N$, large $\Delta_{\rm gap}^{\rm higher-spin}$ criterion,
as exemplified in figure \ref{fig:sYM}.

When geometrical optics applies in the bulk,
an arbitrary bulk point $P$ (causally reachable from the boundary) can be labelled by a triplet ($x$, $\theta_x$, $y$),
where $x$ and $y$ are points in the boundary spacetime and $\theta_x$ is a timelike tangent vector around $x$
(proportional to the momentum injected near $x$).
The corresponding bulk point lies at the intersection of a null geodesic labelled by $(x,p_x)$ and the lightcone from $y$,
see figure~\ref{fig:BTZ radar}.  More precisely, a bulk point may be viewed as an equivalence class of such triplets.

The construction is technically similar to the bulk point and light-cone cuts
\cite{Maldacena:2015iua,Engelhardt:2016wgb}, with two novel ingredients: Fourier transforms
and operator orderings. These tweaks enable to single out a unique bulk geodesic $(x,\theta_x)$ using just two points,
by Fourier transforming the folded OPE limit $\cO(x+\delta x)(\cdots)\cO(x)$ with respect to $\delta x$.
There are two options for using this in a four-point correlator, as depicted in figure \ref{fig:cameras}:
one can make a ``radar camera'' by using two extra points $y$, $z$ to send and receive a pulse,
or one can make an ``active camera'' by doing the same for a second geodesic $(y,\theta_y)$.
In both cases, the signal comes from a singularity or peak caused by a bulk scattering process
when geodesics nearly intersect.

The fact that the correlator data \emph{accurately} specifies a bulk point is vividly exemplified by eq.~\eqref{Einstein}:
in some situations, the bulk's Einstein's equations is precisely equivalent to a simple differential equation on the four-point correlator data.
It will be important to understand the meaning of this equation from the boundary perspective.

The active camera may be  of mostly theoretical interest since
it requires an out-of-time-order correlator that may be difficult or impossible to measure experimentally,
if one had access to a candidate holographic quantum system.
The radar camera, on the other hand, is a perfectly conventional expectation value
(described by the usual Schwinger-Keldysh two-fold as opposed to a four-fold).

The cameras do not have good analogs in two-dimensional models of gravity:
peaks at short impact parameters are a crucial part of the signal.
This requires the bulk to have at least 3 spacetime dimensions.

Much, evidently, remains to be understood.

A basic question is whether similar correlators can probe other bulk fields beyond the Einstein frame metric.
A natural candidate is to generalize the OPE \eqref{grav OPE} to products which vanish at leading power,
such as a field $\cO$ and the stress tensor in the presence of a background $\phi^\Psi$
(schematically and omitting indices, we have not tested such formulas):
\be
 \cO_2(x+\delta x)T^{\mu\nu}_1(x) \stackrel{?}{\sim} \int_{H_{d-1}} \frac{d^{d-1}\theta}{(-\theta\.\delta x)^{\Delta_\cO + d}},
 L_{x,\theta}[h \partial \phi^\Psi].
\ee
Again one would Fourier transform in $\delta x$ to focus on a single geodesic $\theta_x$,
and try to single out a point on it using small-impact-parameter singularities caused by graviton exchange with a second
geodesic from $y$. Related Regge OPEs with spin have been discussed in \cite{Cornalba:2008qf,Afkhami-Jeddi:2016ntf,Meltzer:2017rtf}.
This would naturally define (if it works) bulk fields in a boundary-anchored coordinate system $(x,p_x,y)$.
From this perspective, the extrapolate dictionary $\phi(x,r)\sim r^{\Delta_\cO}\cO(x)$ as $r\to 0$ would be a simple
consequence of the usual local OPE since $r\to 0$ can be approached by taking all four points to be nearby.

This type of description is highly redundant: a bulk point $P$ can be labelled by many triples $(x,\theta_x,y)$. 
It is unclear whether this is related to the highly entangled, quantum error-correcting nature
of the vacuum in holographic theories \cite{Almheiri:2014lwa}.
From that perspective, it is also generally believed that observations in a boundary region $R$ can probe
the full entanglement wedge of that region \cite{Jafferis:2015del}.
It is an important question whether, with suitable add-ons, a camera could image this generally larger region.

We assumed that $1/N$ and $1/\lambda$ corrections at small bulk impact parameters are controlled by flat space physics,
and effectively treated them as a nuisance.  It could be interesting to understand them better.

The camera signals have an interesting interpretation in large-$N$ algebras.
Consider a theory with central charge $\sim N_c^2$ where the OPE can be expanded in single-traces, double-traces, etc.
In a classical state $|\Psi\>$ with large expectation values, $\cO_i = \<\cO_i\>_\Psi + \delta \cO_i$,
such that all multi-traces need to be resummed, one finds an OPE for fluctuations $\delta \cO_i$ that becomes
effectively state-dependent.
It seems reasonable to expand the commutator between two operators in this regime as integrals over causal diamonds (see \cite{Czech:2016xec,deBoer:2016pqk} for related ``OPE blocks'')
\be\begin{aligned}
\, [\delta\cO_i(x),\delta\cO_j(y)] =& A_{ij}(x,y) + \frac{1}{N_c}\sum_k \int_{\Diamond} d^dz\ B_{ijk}(x,y;z) \cO_k
\\&\ + \frac12 \frac{1}{N_c^2} \sum_{k,l} \int_{\Diamond\times\Diamond} d^dz d^dz'\ C_{ijkl}(x,y;z,z') \cO_k(z)\cO_l(z') +\ldots \label{effective OPE}
\end{aligned}\ee
The identity coefficients $A_{ij}$ define a generalized free field algebra which nicely encodes some bulk properties \cite{Leutheusser:2021frk,Chandrasekaran:2022eqq}.
Under suitable conditions, where one can ignore single- and higher- traces,
the out-of-time-order correlator \eqref{camera} is essentially the double-trace coefficient $C_{ijkl}(x,y;z,z')$
in the limit  $z\to x$ and $z'\to y$ where the corners of the diamond are approached, up to a convolution against two-point functions.
What we saw, thus, is that the double-trace OPE coefficients near corners faithfully images bulk scattering processes near the ``waist'' of the bulk causal diamond.
This is reminiscent of \cite{Faulkner:2017vdd}, although we didn't use modular Hamiltonians.

The support in the second line of \eqref{effective OPE} has an interesting implication: it makes the Fourier transform analytic
in upper-half-plane energy planes for $p_x^\mu$ and $p_y^\mu$.
This is crossing symmetry of the bulk scattering amplitude!
Changing the sign of $p$ relates bulk particles to bulk antiparticles.
This property is somewhat mysterious from the correlator perspective.
Clarifying it could help adapt recent bootstrap constraints in vacuum AdS, which use dispersive sum rules
to control bulk equations of motion and bound their deviation from Einstein's theory
by powers of $1/\Delta_{\rm gap}^{\rm higher-spin}$ \cite{Caron-Huot:2021enk,Caron-Huot:2022ugt,Caron-Huot:2022jli}.

Given Einstein's equation around vacuum AdS,
a fundamental question in understanding the bulk
is why the same local equations of motion should be satisfied around any point in a nontrivial bulk geometry,
ie. the equivalence principle.
Morally, from the bulk perspective, this is related by consistency of interactions involving
massless (at least, compared with the AdS radius) spin-two particle \cite{Weinberg:1964ew}.
This suggests that bootstrapping stress tensor correlators could be particularly fruitful.

Four operators is just the right number for bootstrap.
It is clearly of interest to nonperturbatively bootstrap camera correlators in non-vacuum states.
Note that a bulk metric does not always exist, but correlators always do.
Potentially, this could extend or replace Einstein's equations in situations where they do not apply.

These questions have analogs in asymptotically flat spacetimes,  where a classical bulk geometry may also be probed using camera-like
four-point scattering amplitudes in its background.

Shockwaves at future null infinity ($\mathcal{I}_+$) also deserve further exploration.
Matrix elements of the bulk light-ray operators \eqref{Lnu} at infinity are related to $({\rm energy})^{J-1}$ flux
measured in a region of small angular size $\theta$, around the direction $n$ on the celestial sphere \cite{Caron-Huot:2022eqs}
\be
 L_{\infty;n,\nu} \propto \lim_{\theta\to 0} \frac{1}{\theta^{\frac{d}{2}+i\nu-J(\nu)}} \< E_\theta^{J(\nu)-1} \>_{\Psi}\,.
\ee
Experimentally, if one could add up the signals with different momenta $\nu$ and angles $n$,
in such a way as to compute the integral \eqref{OPE with Lnu}, one would be measuring the bulk average null energy (ANEC).
In conformal collider physics this could reveal that what appears from energy flux measurements
to be a spherical shell in the boundary \cite{Hofman:2008ar}, is in fact a point particle in the bulk.
QCD might not be holographic in the sense required here (at least not at high energies),
but it could still be interesting to study $({\rm energy})^{J-1}$ fluxes with complex $J$
in a collider context, perhaps similarly to the nucleon-energy correlators in \cite{Liu:2022wop}.

\section*{Acknowledgements}	

The author thanks Joydeep Chakravarty, Netta Engelhardt,
Jonah Kudler-Flam, Yue-Zhou Li and Juan Maldacena for discussions and comments.
This work was supported in parts by the Simons Collaboration on the Nonperturbative Bootstrap,
the Simons Fellowships in Theoretical Physics, and the Canada Research Chair program.

\begin{appendix}

\section{Formulas for Fourier, eikonal, and harmonic integrals}

Here we record a number of handy formulas used in the main text.
We start with the Fourier transform of a power law in Euclidean signature:
\be
 \int \frac{d^d p_E}{(2\pi)^d} e^{ip\.x} (p^2)^{\Delta-\frac{d}{2}} =  \frac{P_\Delta}{(x^2)^\Delta},
 \qquad P_\Delta \equiv \frac{4^\Delta\Gamma(\Delta)}{(4\pi)^{\frac{d}{2}}\Gamma(\tfrac{d}{2}-\Delta)}. \label{PDelta}
\ee
Since we work with mostly plus metric, we only indicate the signature with a subscript on the measure: $i d^dp_E=d^dp$.
The Wick rotation then gives the time-ordered correlator
\be
 -i\int \frac{d^d p}{(2\pi)^d} e^{ip\.x} (p^2-i0)^{\Delta-\frac{d}{2}} =  \frac{P_\Delta}{(x^2+i0)^\Delta}.
\ee
The Wightman function in momentum space is related to the discontinuity across the positive-frequency cut of the above
(this is because the time-ordered and Wightman functions coincide for positive times). Let us record the inverse transform:
\be \label{Fourier Wightman}
 \int d^d x \frac{e^{-ip\.x}}{(x^2_-)^\Delta} = \theta(p^0)\theta(-p^2)(-p^2)^{\Delta-\frac{d}{2}}(P_\Delta^>)^{-1},
 \qquad P_\Delta^> \equiv \frac{1}{2\pi}
 \frac{4^{\Delta}\Gamma(\Delta)\Gamma(\Delta-\tfrac{d-2}{2})}{(4\pi)^{\frac{d}{2}}}.
\ee

We now describe $v$ integrals like \eqref{scalar integral near cone}, which occur in position-space eikonal calculations:
\be
I_{\Delta_1,\Delta_2} \equiv \int\limits_{-\infty}^{+\infty} \frac{dv}{(-2v+i0)^{\Delta_1}(-2(v+\delta+i0))^{\Delta_2}}. \label{I12}
\ee
The integrand is real and positive when $v$ is sufficiently negative (corresponding physically to spacelike propagators),
and the $+i0$ indicates the correct analytic continuation to other regions. 
Thus, the first propagator has a branch cut at $v\geq 0$ where its phase is $e^{-i\pi\Delta_1}$ since
the contour runs below the cut; the second propagator has its cut above the contour.
We evaluate the integral by wrapping the contour around the $v=0$ cut:
\begin{align}
\label{I12 first step}
 I_{\Delta_1,\Delta_2} &=(e^{-i\pi \Delta_1}-e^{i\pi \Delta_1})\int_0^\infty \frac{dv}{(2v)^{\Delta_1}} \frac{e^{i\pi \Delta_2}}{(2(v+\delta+i0))^{\Delta_2}}
\\ 
&=\frac{\Gamma(\Delta_1+\Delta_2-1)}{\Gamma(\Delta_1)\Gamma(\Delta_2)}
\frac{-i\pi\ e^{i\pi\Delta_2}}{(2\delta+i0)^{\Delta_1+\Delta_2-1}}.  \label{I12 result}
\end{align}
The denominator is real when $\delta>0$, and is defined elsewhere by analytic continuation.

Two comments are in order. First, the original integral \eqref{I12} has a symmetry under $\Delta_1\leftrightarrow \Delta_2$, $\delta\mapsto-\delta$ and complex conjugation.  This symmetry survives in the final answer since
\be
 \frac{-i\pi\ e^{i\pi\Delta_2}}{(2\delta+i0)^{\Delta_1+\Delta_2-1}}=  \frac{+i\pi\ e^{-i\pi\Delta_1}}{(-2\delta-i0)^{\Delta_1+\Delta_2-1}}\,.
\ee
Second, to get from \eqref{I12} to \eqref{I12 first step} we have dropped arcs at infinity.
This is only valid for sufficiently large or generic $\Delta_1$ and $\Delta_2$.

\subsection{Harmonic analysis} \label{app:harmonic}

A complete basis of functions for a scalar field on $H_{d-1}$ is labelled by a null vector $n^\mu \in {\rm S}^{d-2}$ and a momentum $\nu$,
with eigenfunction $\sim 1/(-n\.\theta)^{\frac{d-2}{2}+i\nu}$.
Suppose $\theta$ is a unit timelike vector ($\theta^2=-1$). 
The harmonic transform is the following Fourier-like transform and its inverse:
\begin{align}
 \widehat{f}(n,\nu) &\equiv \frac{1}{\Omega_{d-1}} \int_{H_{d-1}} \frac{d^{d-1}\theta}{(-n\.\theta)^{\frac{d-2}{2}+i\nu}}f(\theta), \label{fourier direct} \\
 f(\theta) &= \int\limits_0^\infty\frac{d\nu}{2\pi} \rho(\nu) \int_{S^{d-2}} \frac{d^{d-2}n}{(-n\.\theta)^{\frac{d-2}{2}-i\nu}}
 \widehat{f}(n,\nu), \label{fourier inverse}
\end{align}
with $\Omega_d =2\pi^{\frac{d}{2}}/\Gamma(\tfrac{d}{2})$.
The measure is (note that $\rho(\nu)$ here differs from $\rho(\nu)$ in \cite{Caron-Huot:2021enk} by $\frac{2^{d-2}}{\Omega_{d-1}}$):
\be
 \rho(\nu)=\frac{2^{d-2}}{\Omega_{d-1}}
 q_{\frac{2-d}{2}+i\nu}q_{\frac{2-d}{2}-i\nu} \quad\mbox{where}\quad q_J\equiv \frac{(d-2)_J}{(\tfrac{d-2}{2})_J}\,. \label{rho}
\ee
Integrate over all angles $n$ gives, for $x,y$ two unit vectors, the harmonic functions
\be \label{harmonic}
\frac{1}{\Omega_{d-1}}\int \frac{d^{d-2}n}
{(-n\.x)^{\frac{d-2}{2}+i\nu}(-n\.y)^{\frac{d-2}{2}-i\nu}} =\cP_{\frac{2-d}{2}+i\nu}\left(-x.y\right)
\ee
where
\be
\cP_J(\eta) = {}_2F_1\big({-}J,d{-}2+J,\tfrac{d-1}{2},\tfrac{1-\eta}{2}\big)\,.
\ee
For integer $J$, this function is proportional to a spherical harmonic (Gegenbauer polynomial), whence our notation.
For more background and spinning generalizations,  see \cite{Karateev:2018oml}.

Since the Laplacian on $H_{d-1}$ is multiplication by minus $\nu^2+(\tfrac{d-2}{2})^2$,
harmonic representations of Green's functions are readily obtained.
For example, Poisson's equation
\be
 \left(\square_{{\rm H}_{d-1}} - \Delta(\Delta+2-d)\right) \Pi_\Delta(-x\.y) = -\delta^{d-1}(x,y)
\ee
is inverted by the following transform
\begin{align}
\Pi_\Delta(\eta)&\equiv
\int_0^\infty \frac{d\nu}{2\pi}\rho(\nu)\frac{1}{\nu^2+\big(\Delta-\tfrac{d- 2}{2}\big)^2} \cP_{\frac{2-d}{2}+i\nu}(\eta)
\\ &= 
\frac{C^{(d-2)}_\Delta}{(2\eta)^{\Delta}} {}_2F_1\big(\tfrac{\Delta}{2},\tfrac{\Delta+1}{2},\Delta+\tfrac{4-d}{2},\tfrac{1}{\eta^2}\big), \qquad C_\Delta^{(d-2)}= \frac{\Gamma(\Delta)}{2\pi^{\tfrac{d-2}{2}}\Gamma(\Delta+2-\tfrac{d}{2})}.
\label{hyperbolic propagator}
\end{align}
The constant is a dimension-shifted version of $C_\Delta$ in \eqref{bulktobdy}.
This is used in the main text to write the graviton propagator on the transverse hyperboloid as $\Pi_{d-1}(\eta)$.

Finally, the harmonic representation plays nicely with other Lorentz-covariant operations such as ordinary Fourier transforms:
the $d$-dimensional Fourier transform of an eigenfunction is an eigenfunction.
Below \eqref{CRT nice} we used the following identity, involving the wavefunction \eqref{psi} in the plane-wave limit $\psi_{p,L=\infty}$,
with the integral running over all $d$-vector $x$ (which can be either spacelike or timelike).
It can be derived by combining the denominators using Schwinger parameters and using \eqref{Fourier Wightman}:
\be
\int  \frac{d^dx\ \psi_{p,L=\infty}(x)}{(x_-^2)^{\frac{2\Delta+J-1}2}(-n\.\hat{x}_-)^{\frac{d-2}{2}+i\nu}} =
\frac{1}{2^{J-1}} \frac{(-p^2)^{\frac{J-1}{2}}}{(-n\.\hat{p}_-)^{\frac{d-2}{2}+i\nu}}\frac{\Gamma(\Delta)\Gamma(\Delta-\frac{d-2}{2})}{\gamma_{2\Delta+J-1}(\nu)}, \label{nice transform}
\ee
with $\gamma$ in \eqref{gamma}. As in the main text,
we have assumed that $p$ is positive-frequency timelike and defined $(-n\.\hat{x}_-)^q$ by analytic continuation from the positive-timelike region, where the parenthesis is positive.
Recall that hatted vectors are normalized: $\hat{p}=p^\mu/\sqrt{-p^2}$.

To account for the finite width of wavepackets, we convolve \eqref{nice transform} with a Gaussian of width $\Delta\theta= 1/(|p|L)$ in the orientation of $p$, as mentioned below \eqref{AdS OPE p}.
We are only interested in situations where this width is small.  The convolution then has essentially no effect unless is large
$\nu\sim 1/\Delta\theta$.
In the rest frame of $p$, say, one can then approximate $(-n\.\hat{p}_-)^{\frac{d-2}{2}+i\nu}\sim e^{-i \nu  \vec{n}\.\vec{p}/p^0}$
so the convolution simply gives a Gaussian in $\nu$.  Since our wavepackets are covariant, this result holds in any frame:
\be
\int  \frac{d^dx\ \psi_{p,L}(x)}{(x_-^2)^{\frac{2\Delta+J-1}2}(-n\.\hat{x}_-)^{\frac{d-2}{2}+i\nu}} \approx \mbox{eq.~\eqref{nice transform}} \times e^{-\frac{\nu^2}{2|p|^2L^2}}\,.
\label{nice transform 1}
\ee

\section{Graviton propagator and shockwave geometry} \label{app:shockwave}

The bulk retarded graviton propagator for fluctuations $g_{ab}=g_{ab}^{\Psi}+\kappa h_{ab}$ around a background $g_{ab}^{\Psi}$
satisfies Einstein's linearized equations with source:
\begin{align}
 &\< h^{cd}(X) \left(D^2  h_{ab} + D_aD_b h_e^e-D_e(D_a h_b^e+D_b h_a^e) + \Lambda\frac{d-3}{d-1}h_{ab}\right)(Y)\>_{\Psi,\rm ret}
\\
 &\qquad = -\left(\tfrac12 (\delta^c_a \delta^d_b+\delta^d_a \delta^c_b)-\tfrac{1}{d-1}g^{\Psi}_{ab}g^{\Psi,cd}\right)
 \frac{\delta^{d+1}(X,Y)}{\sqrt{-g^{\Psi}(X)}}\,, \label{linearized}
\end{align}
together with the boundary condition that it vanishes unless $X{\prec}Y$.
Here we describe what happens when we integrate this propagator over a null ray with affine parameter $u$.
Taking the metric in the form \eqref{ds2 cone},
the source on the right-hand-side of \eqref{linearized} simplifies to
\be
 \< \int du\ h_{uu}(X)\    \Big(\cdots\Big)(Y) \>_{\Psi,\rm ret} = -\delta_a^v\delta_b^v\frac{\delta(v)\delta^{d-1}(x_i,y_j)}{\sqrt{\det \tilde{g}_{ij}}}.
\label{linearized 1}
\ee
When the background geometry possesses a boost symmetry which fixes the $v=0$ null sheet,
we expect physically to find an Aichelburg-sexl type shockwave solutions supported on this sheet.
(For backgrounds that do not enjoy a boost symmetry, we do not know any simple way to solve \eqref{linearized 1}.)

Let us see this for the null cone from a point at the boundary of pure AdS$_{d+1}$.
Using the parametrization $x^\mu= (\frac{1}{-u}+v)\theta^\mu$, $r=\frac{1}{-u}-v$, with $\theta^2=-1$, we get the metric
\be
 ds^2_{{\rm AdS}_{d+1}} = \RAdS^2\left[\frac{-4du dv}{(1+uv)^2} + \frac{(1-uv)^2}{(1+uv)^2} d\theta^2\right]
\ee
with the null cone at $v=0$ and the AdS boundary at $v=-u^{-1}$.
(As in the main text, $u$ is an affine coordinate along geodesics which leave the boundary at $u=-\infty$.)
Eq.~\eqref{linearized 1} then admit the following shockwave solution:
\be
 h_{ab}(Y)dy^a dy^b = f(\theta_y)\delta(v)dvdv, \qquad \left(\square_{{\rm H}_{d-1}} +1-d\right) f(\theta_y) =
 -\delta^{d-1}(\theta_x,\theta_y)\,.  
\ee
This is solved by the propagator $f(\theta_y)=\Pi_{d-1}(-\theta_x\.\theta_y)$ in \eqref{hyperbolic propagator}.

From here it is a short step to obtain the doubly integrated correlator \eqref{L}, since the $\delta(v)$ just kills the $v$ integral.
Consider the case where the $v$ geodesics in \eqref{p' geodesic} reaches the
boundary at a point $e=(1,\vec{0})$ along direction $\theta_y$.
It intersects the forward lightcone of the origin at $(u,v,\theta)=(2e\.\theta_y,0,-{\cal I}_e\.\theta_y)$, with coordinate related
to the affine parameter as $dv = \frac14\RAdS^2dv_{\rm affine}$.
The integral \eqref{L} then gives, as quoted in \eqref{prop vacuum}:
\be \label{prop vacuum app}
 \Pi_{\Omega}(0,\theta_x; e,\theta_y) = \RAdS^{1-d}\Pi_{d-1}(\theta_x\.{\cal I}_e\.\theta_y).
\ee

\section{More on conformal Regge theory}

\subsection{Analytic continuation from Euclidean to OTO kinematics} \label{app:path}

\begin{figure}[t]
\centering
\begin{subfigure}[b]{0.35\textwidth}
\begin{tikzpicture}[thick,baseline=(X.base)]
\draw[->] (0,-0.5) -- (0,4);
\draw[->] (-0.5,0) -- (4,0);
\node[right] at (4,0) {$x$};
\node[above] at (0,4) {$t$};
\node[circle,fill=black,inner sep=0pt,minimum size=3pt,label=right:{$3$}] at (0,3) {};
\node[circle,fill=black,inner sep=0pt,minimum size=3pt,label=right:{$4$}] at (0,2.25) {};
\node[circle,fill=black,inner sep=0pt,minimum size=3pt,label=left:{$1$}] at (0,0.75) {};
\node[circle,fill=black,inner sep=0pt,minimum size=3pt,label=below left:{$2$}] (X) at (0,0) {};
\begin{scope}[gray,thick,decoration={markings,mark=at position 0.5 with {\arrow{>}}}]
	\node[circle,fill=black,inner sep=0pt,minimum size=3pt,label=below:{$3$}] at (3,0) {};
	\node[circle,fill=black,inner sep=0pt,minimum size=3pt,label=below:{$4$}] at (2.25,0) {};
	\node[circle,fill=black,inner sep=0pt,minimum size=3pt,label=below:{$1$}] at (0.75,0) {};
	\draw[postaction={decorate},dashed] (3,0) -- (0,3);
	\node[above right] at (1.45,1.45) {$a$};
	\draw[postaction={decorate}] (2.25,0) -- (0,2.25);
	\draw[postaction={decorate}] (0,2.25) -- (-0.75,3);
	\node[below left] at (1.2,1.2) {$b$};
	\draw[postaction={decorate},dashed] (-0.75,3) to [out=-90,in=180] (0,2.25); 
	\node[below left] at (-0.375,2.5) {$c$};
	\draw[postaction={decorate},dashed] (0.75,0) -- (0,0.75);
	\node[above right] at (0.33,0.33) {$c$};
\end{scope}
\draw[->] (2.5,2.5) -- (3,2);
\draw[->] (2.5,2.5) -- (3,3);
\node[below right,inner sep=0pt] at (3,2) {$z$};
\node[above right,inner sep=0pt] at (3,3) {$\zb$};
\end{tikzpicture}
\end{subfigure}
\hspace{0.15\textwidth}
\begin{subfigure}[b]{0.35\textwidth}
\begin{tikzpicture}[thick,baseline=(X.base)]
\draw (0,-0.5) -- (0,3.5);
\draw (-0.5,0) -- (4,0);
\node[circle,fill=black,inner sep=0pt,minimum size=3pt,label=below left:{$0$}] (X) at (0,0) {};
\node[circle,fill=black,inner sep=0pt,minimum size=3pt,label=below:{$1$}] at (3,0) {};
\node[circle,fill=black,inner sep=0pt,minimum size=3pt,label=above:{$\zb$}] at (0.5,0.2) {};
\node[gray,circle,fill=gray,inner sep=0pt,minimum size=3pt,label=below:{$z$}] at (0.5,-0.2) {};
\draw[dashed,->] (0.5,-0.2) to [out=-20,in=270] (3.5,0) to [out=90,in=20] (0.6,-0.15);
\draw (3,2.5) -- (3,2) -- (3.8,2);
\node[above right] at (3,2) {$z,\zb$};
\end{tikzpicture}
\end{subfigure}
\caption{Left: analytic continuation steps from  spacelike to timelike configuration in CFT vacuum, with all $\zb_i$'s kept fixed.
As discussed in the text, for the operator ordering $3241$ nothing happens to cross-ratios during steps $a$ and $c$. During
step $b$, $z$ winds counter-clockwise around $z=1$, as shown in the right panel.
The resulting correlator $\cG(z,\zb)^{\circlearrowleft}$ is analytic near $z{=}\zb$ so the relative positions of $z$ and $\zb$
needs not be specified.
\label{fig:path}}
\end{figure}
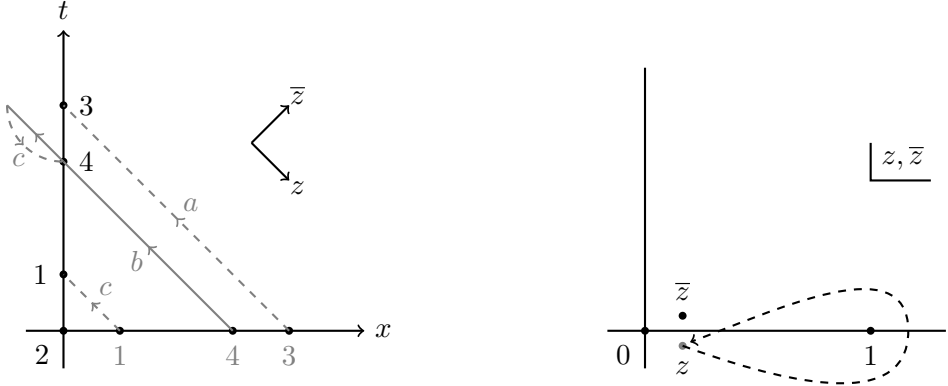

Here we derive the analytic continuation path from the Euclidean region, to the OTOC region with operator ordering
$3241$ and time ordering $2{\prec}1{\prec}4{\prec}3$.
(This will be equivalent to the ``reference region'' $1{\prec}2{\prec}3{\prec}4$ used in the main text.)
The discussion is similar to \cite{Cornalba:2006xk}.
A useful observation is that it suffices to restrict to $(1+1)$-dimensional kinematics, where $z_i=x_i{-}t_i$ and $\zb_i=x_i{+}t_i$,
and the cross-ratios are simply
\be
 z=\frac{z_{12}z_{34}}{z_{13}z_{24}},\qquad \zb=\frac{\zb_{12}\zb_{34}}{\zb_{13}\zb_{24}}
\ee
with $z_{ij}=z_i-z_j$.
We start from an Euclidean configuration where all $z$'s are real and $\zb_i=z_i$.
By starting from a suitable initial ordering $z_2{<}z_1{<}z_4{<}z_3$, we can keep the $\zb_i$'s constant
at all steps, as depicted in figure \ref{fig:path}.

In the first step, we drag operator $\cO'_3$ to the future region, from large positive $z_3$ to negative $z_3$.
Conformal correlators on the main sheet have branch points at $z=0,1$ (and $\infty$)
so we only need to find whether $z$ winds around 0 or 1. This is determined by looking at which factors of $z$ and $1-z$ change sign,
and the $i0$ prescription in each:
\be
\mbox{step a: } z= \frac{(z_1-z_2)(z_3-z_4{+}i0)^{\curvearrowleft}}{(z_4-z_2)(z_3-z_1{+}i0)^{\curvearrowleft}}\sim\mbox{no phase},\quad
1-z = \frac{(z_4-z_1)(z_3-z_2{+}i0)^{\curvearrowleft}}{(z_4-z_2)(z_3-z_1{+}i0)^{\curvearrowleft}}\sim\mbox{no phase} \,.
\ee
Thus nothing happened during this step.  The numerator and denominator both rotate in the same way because
$z_3$ has the largest $+i0$ of all operators. We could have added a large positive imaginary part to $z_3$ and then
the cross-ratio $z$ would have remained approximately constant during the whole step.

In the second step we move $z_4$ all the way to the left of $z_3$.
This is more interesting as not all the singularities are on the same side:
\be
\mbox{step b: } z= \frac{(z_1-z_2)(z_4-z_3{-}i0)^{\raisebox{\depth}{\scalebox{1}[-1]{$\curvearrowleft$}}}}
{(z_4-z_2{-}i0)^{\raisebox{\depth}{\scalebox{1}[-1]{$\curvearrowleft$}}}(z_1-z_3)}\sim\mbox{no phase},
\quad
1{-}z = \frac{(z_4-z_1{+}i0)^{\curvearrowleft}(z_2-z_3)}{(z_4-z_2{-}i0)^{\raisebox{\depth}{\scalebox{1}[-1]{$\curvearrowleft$}}}(z_1-z_3)}
\sim e^{2\pi i} \,.
\ee
Recall that the $i0$'s are easily read off from the ordering $3241$.
Thus $z$ went counter-clockwise around 1.  Finally, during step $c$ nothing happens as the
two factors in $(z_1-z_2-i0)(z_3-z_4+i0)$ wind in opposite ways.  Thus the reference region is reached by rotating $z$ around 1
as depicted on the right panel of figure \ref{fig:path}.

Other cases, such as $\delta x$ spacelike or ${\prec}0$, can then be reached without leaving the region of small $z,\zb$
simply by paying attention to the $i0$ prescriptions in \eqref{zzb series}. 
Table \ref{regions} summarizes the phases of the cross-ratios in most possible cases,
starting from the reference region in the top line, as well as the convergence properties of the $t$ and $u$-channel OPEs.
The $t$-channel OPE always converges. This is related to the fact that $\cO'(x_4)\cO(x_1)$ acts on the vacuum in the $3241$ operator ordering.
The $u$-channel OPE only converges in some regions, while the traditional $s$-channel OPE never converges for $cG(z,\zb)^{\circlearrowleft}$.

\begin{table}[t] \centering
\def\configpic#1#2#3#4{{\begin{tikzpicture} \filldraw (#1,0) circle (1pt);  \filldraw ({#2},-0.1) circle (1pt); \filldraw (#3,0) circle (1pt); \filldraw ({#4},-0.1) circle (1pt); \end{tikzpicture}}}
\begin{tabular}{c|c|c|c|c|c|c}  
$\delta x$ & $\delta y$ & schematics & $(z,\zb)$& $t$-channel & $u$-channel & description  \\\hline
${\prec}0$ & ${\succ}0$ & \configpic{0.1}{-0.1}{0.9}{1.1} & $({>}0,{>}0)$ & \cmark\cmark & \cmark & $t$-channel Regge \\\hline
${\succ}0$ & ${\prec}0$ & \configpic{-0.1}{0.1}{1.1}{0.9} & $({>}0,{>}0)$ & \cmark\cmark & \cmark & $t$-channel Regge  \\\hline
${\succ}0$ & ${\succ}0$ & \configpic{-0.1}{0.1}{0.9}{1.1} & ${\sim}(e^{i\pi},e^{i\pi})$ & \cmark & \cmark\cmark & $u$-channel Regge \\\hline
${\prec}0$ & ${\prec}0$ & \configpic{0.1}{-0.1}{1.1}{0.9} & ${\sim}(e^{-i\pi},e^{-i\pi})$ & \cmark & \xmark & no name \\\hline
${\sim}0$ & ${\sim}0$ & \configpic{0}{0}{1}{1} & ${\sim}(e^{-i\varphi},e^{i\varphi})$ & \cmark & \xmark & bulk point ($\varphi\to\pi$) \\\hline
${\sim}0$ & ${\succ}0$ & \configpic{0}{0}{0.9}{1.1} & $(e^{i\pi},{>}0)$ & \cmark & \xmark & no name \\\hline
${\sim}0$ & ${\prec}0$ & \configpic{0}{0}{1.1}{0.9} & $(e^{-i\pi},{>}0)$ & \cmark & \xmark & no name
\end{tabular}
\caption{Various kinematic regions classified by whether the small vectors $\delta x,\delta y$ are future timelike (${\succ}0$),
past timelike (${\prec}0$) or spacelike (${\sim}0$), 
showing the phase of small cross-ratios.
The schematics are meant to visually evoke positions along the timefold \eqref{SK fourfold}. 
The double check marks indicate exponential convergence of an OPE,
single check marks indicate marginal (distributional) convergence, and \xmark\ indicates that an OPE does not converge.
$\varphi$ is the angle between $\delta x$ and $-\mathcal{I}_e\.\delta y$.
\label{regions}
}
\end{table}

\subsection{Residues of partial wave amplitudes}\label{app:lorentzian}

Here we briefly review how residues of $\alpha(\nu)$ relate to local operators, matching with different conventions in the literature.
Our starting point are the functions $c^t(\Delta,J)$ and $c^u(\Delta,J)$, analytic in their two arguments,
defined by the Lorentzian inversion formula \cite{Caron-Huot:2017vep}. Their $\Delta$-poles at integer spin encode the operator spectrum and OPE coefficients:
\be
 {-}\!\Res\limits_{\Delta'=\Delta} \left[ c^t(\Delta',J) + (-1)^J c^u(\Delta',J)\right] = f_{12\cO_{\Delta,J}}f_{43\cO_{\Delta,J}}\,. \label{LIF residue}
\ee
There are also spurious poles at special values of $\Delta'$, see \cite{Caron-Huot:2017vep} for details.

The operator product expansion is traditionally written as a sum over spins of exchanged operator.  This sum ceases to converge when one analytically continue around a cross-channel,
as just discussed when computing $\cG^\circlearrowleft$ from the $s$-channel OPE. 
The $s$-channel OPE can nonetheless be extended into the second sheet using the Watson-Sommerfeld trick
of rewriting the sum as an integral.
This yields the exact representation of the analytically continued correlator in (3.24) of \cite{Caron-Huot:2020nem}.
Eq.~(3.26) there shows the result of expanding this formula as $z,\zb\to 0$ assuming that $c_t$ has discrete $j$-plane singularities (Regge trajectories):
\be\begin{aligned} \label{CRT exact}
e^{i\pi(a+b)} \cG^{\circlearrowleft}(z,\zb) - \cG(z,\zb) &= i \int_0^\infty \frac{d\nu}{2\pi} \sum_{j_n(\Delta)} \Res\limits_{J=j_n(\Delta)}\frac{ e^{i\pi J}c^t(\Delta,J)+c^u(\Delta,J)}{\sin(\pi J)} \frac{G^{(a,b)}_{1-J,1-\Delta}(z,\zb)}{\kappa^{(a,b)}_{\Delta+J}}
\\ &\phantom{=} -4i\sum_{J\geq 0} \int_0^\infty \frac{d\nu}{2\pi} \left[ c^t(\Delta,J)+(-1)^J c^u(\Delta,J)\right] s^{(-a,-b)}_{\Delta+J} F^{\prime(a,b)}_{\Delta,J}(z,\zb).
\end{aligned}\ee
This provides an all-order expansion around the Regge limit $z,\zb\to 0$. The $J$-residue is to be taken with
$\Delta=\frac{d}{2}+i\nu$ fixed. We refer to \cite{Caron-Huot:2020nem} for definitions of the various objects.
In general, $a=\frac{\Delta_2-\Delta_1}{2}$ and $b=\frac{\Delta_3-\Delta_4}{2}$ and in the formulas below we set those to zero.

In the main text we focused on the leading term, assuming that a single Regge trajectory dominates,
as in the classic conformal Regge theory paper \cite{Costa:2012cb}.
If one restricts to the leading term of blocks ($q_J$, $\rho(\nu)$ and $\cP_J(\eta)$ are defined around \eqref{rho}),
\be
 \lim_{z,\zb\to 0} G^{(a,b)}_{1-J,1-\Delta} (z,\zb)= (z\zb)^{\frac{1-J}{2}} \frac{\Omega_{d-1}}{2^{d-2}q_{\frac{2-d}{2}+i\nu}} \rho(\nu) \cP_{\frac{2-d}{2}+i\nu}\left(\tfrac{z+\zb}{2\sqrt{z\zb}}\right)\,, \label{Regge limit of block}
\ee
one sees that the first line of \eqref{CRT exact} takes precisely the form of \eqref{CRT} with
\be
 \alpha_n(\nu) = \frac{\Omega_{d-1}}{2^{d-1}\pi q_{\frac{2-d}{2}+i\nu}} \Res\limits_{J=j_n(\Delta)}\frac{e^{i\pi J} c^t(\Delta,J)+c^u(\Delta,J)}{\sin(\pi J)} \frac{1}{\kappa^{(a,b)}_{\Delta+J}}\,. \label{alpha from c}
\ee
Again, $\Delta=\tfrac{d}{2}+i\nu$.  For reference, $\kappa^{(0,0)}_\beta = \frac{\Gamma(\beta/2)^4}{2\pi^2\Gamma(\beta-1)\Gamma(\beta)}$.

The functions $c$ are supposed to have poles $\sim C_n(\Delta)/(J-j_n(\Delta))$ near Regge trajectories.
The $1/\sin(\pi J)$ factor in \eqref{alpha from c} then create poles in $\alpha_n(\nu)$ at $\Delta$ such that $j_n(\Delta)$ is integer, ie. local operators.
These poles are thus simply related to the OPE coefficients in \eqref{LIF residue}; we find:
\be \label{residue of alpha}
 \Res\limits_{\Delta'=\Delta} \alpha_n\big(\nu=-i(\Delta'{-}d/2)\big) =
 \frac{\Omega_{d-1}}{2^{d-1}\pi^2} \frac{1}{q_{\Delta+1-d}}\frac{1}{\kappa^{(a,b)}_{\Delta+J}}   f_{12\cO_{\Delta,J}} f_{43\cO_{\Delta,J}} \qquad (j_n(\Delta)=J\ \mbox{integer}).
\ee
This is essentially equivalent to eqs.~(49), (54) and (57) of \cite{Costa:2012cb}.  See also (5.21)-(5.22) of \cite{Kravchuk:2018htv}.

This can be used to check the residue of the stress tensor pole in \eqref{alpha GR}.
From Ward identities, the OPE coefficient for stress-tensor exchange is related to the normalization of the stress-tensor two-point function $C_T$ \cite{Dolan:2000ut},
itself related to Newton's constant \cite{Liu:1998bu}:
\be \label{CT}
 C_{\cO \cO T} = \frac{-\Delta}{\sqrt{C_T}} \frac{d}{d-1}, \qquad  C_T=\frac{\pi^{\frac{d-2}{2}}\Gamma(d+2)}{2 G_N^{(d+1)}\RAdS^{1-d} \Gamma(d/2)^3(d - 1)}\,,
\ee
in agreement with (103)-(104) of \cite{Costa:2012cb}.  For the $\mathcal{N}=4$ SYM theory with SU$(N_c)$ gauge group, $C_T=20\pi \RAdS^3/G_N^{(5)} = 40(N_c^2-1)$.
Finally, the OPE coefficients $C$ normalized as in \cite{Dolan:2000ut,Costa:2012cb} are related to $f$ above, which follow \cite{Caron-Huot:2017vep,Caron-Huot:2020nem},
by
\be
 f_{ij\cO_{\Delta,J}} =2^{-J/2} C_{ij\cO_{\Delta,J}}.
\ee

\subsection{Pomeron exchange in the Wilson-Fisher theory}
\def\hb{\overline{h}}

We discuss here the coefficient functions of the Wilson-Fisher theory in $d=4-\ep$ dimensions near the intercept.
We start with formulas from \cite{Henriksson:2018myn} for the leading trajectory for generic spin.
The anomalous dimension takes its most concise form in terms of a variable that we write here as
$\hb=\frac{\Delta+J}{2}=1+J+O(\ep)$:
\be
 \Delta(J)= 2\Df + J+\gamma(\hb), \quad \gamma(\hb) = -\frac{\ep^2}{9\hb(\hb-1)}
 - 2\ep^3 \frac{H(\hb{-}1) -\tfrac{11}{18}}{27\hb(\hb-1)} + O(\ep^4)\,. \label{Ising traj}
\ee
with $H(x)=\psi(1+x)-\psi(1)$ the (analytic continuation) of harmonic numbers.
Note that even though $\Delta$ appears here on both sides of this equation, the equation can be solved perturbatively in $\ep$
to recover the formulas from \cite{Henriksson:2018myn}.
The leading-trajectory residue takes a similarly concise form in terms of $\hb$,
\begin{align}
 c^{t}(\Delta,J) &= c^{u}(\Delta,J) = \frac{\Gamma(\hb)^2}{\Gamma(2\hb-1)}
\frac{\Gamma(\hb+\Df-1)}{\Gamma(\Df)^2\Gamma(\hb-\Df+1)}
\frac{r(\hb)}{2\Df+J+\gamma(\tfrac{\Delta+J}{2})-\Delta} + \mbox{pole-free}, \nonumber\\
r(\hb) &= 1 - \frac{\ep^2}{18\hb^2(\hb-1)^2} - \ep^3\left[ \frac{(H(\hb{-}1)-\tfrac{11}{18})}{27\hb^2(\hb-1)^2} - \frac{\pi^2}{162\hb(\hb-1)}\right]+ O(\ep^4)\,. \label{Ising residue}
\end{align}
More orders are available, but this will suffice for our purposes.

The above expansions are valid for generic $\hb$ but they break down when $\hb\to 1$ (where $\Delta\to 2+O(\ep)$ and $J\to O(\ep)$). As explained in \cite{Caron-Huot:2022eqs}, this singularity is caused by mixing between the trajectory \eqref{Ising traj} and its shadow (defined by the same formula with $\Delta\mapsto d{-}\Delta$).
We can get a smooth characteristic equation $\chi=0$ by multiplying \eqref{Ising traj} with its shadow and expanding in $\ep$ (see (2.27) of \cite{Caron-Huot:2022eqs}):
\begin{align}
\chi(\Delta,J) &\equiv -(\Delta-d/2)^2+(2\Df-d/2+J+\gamma)^2
\nonumber \\ &= -(\Delta-d/2)^2 + J^2-J \ep+\left(\frac{J}{27}+\frac{1}{4}-\frac{2}{9 (J+1)}\right) \ep^2
\nonumber \\ &\quad +\left(\frac{109 J^3+164 J^2+265 J-114}{2916 (J+1)^2} -\frac{4 H(J)}{27 (J+1)}\right)\ep^3 + O(\ep^4).
\label{Ising traj sym}
\end{align}
For generic $J$, \eqref{Ising traj sym} admits a solution $\Delta\approx J+2+\cO(\ep)$,
whose expansion in $\ep$ recovers the full information in \eqref{Ising traj}.
The advantage of the form \eqref{Ising traj sym} is that 
now all terms are regular at $J\to 0$. This is a nontrivial condition satisfied by $\gamma$.

Shadow symmetry is an exact property of the functions $c^t=c^u$ that come out of the Lorentzian inversion formula.
More precisely, the combination that enters $\alpha$ in \eqref{alpha from c} is symmetrical.
We can use this property to clarify the behavior near the intercept.
Since $c$ should have poles on the trajectory and its shadow, we expect to find a smooth result after factoring out $1/\chi$: 
\be
 \frac{c^t(\Delta,J)}{q_{\Delta+1-d} \kappa_{\Delta+J}^{(0,0)}} \simeq
\frac{4\pi^2}{\Gamma(\Df)^2} \frac{\Gamma(2h)}{\Gamma(h)\Gamma(h-\Df+1)}\frac{\Gamma(2\hb)}{\Gamma(\hb)\Gamma(\hb-\Df+1)}
\frac{\tilde{r}(J)}{\chi(\Delta,J)}\,,
\label{ctilde ansatz}
\ee
with $h=\frac{d-\Delta+J}{2}$. The equality should hold up to terms analytic near the leading trajectory and
for $J\gtrsim -1$ (the location of the next intersections). Note that the $\Gamma$-functions are regular near the intercept.
Comparing the fixed-$J$ residues of \eqref{ctilde ansatz} with \eqref{Ising residue} on the leading trajectory, we obtain that
\be
 \tilde{r}(J) = 1 - \ep + \ep^2\left( \frac{J^2+2J-5}{54(J+1)^2} +\frac{\pi^2}{12} + \frac{H(J)}{108}\right) +O(\ep^3)\,.
\ee 
The $O(\ep^3)$ term can actually be obtained from the data recorded above but we do not write it out here.
The crucial point is that all terms are regular near $J\to 0$.
In the main text we only use the leading term.
Now taking the fixed-$\Delta$ residue of \eqref{ctilde ansatz} on the trajectories
$J_\pm(\nu)\approx \frac{\ep}{2}\pm\sqrt{\frac{2}{9}\ep^2-\nu^2}$, following \eqref{alpha from c}, we find,
working to leading order in $J\sim \nu \sim \ep$:
\be
 \alpha_\pm^{\rm W-F}(\nu) =  \frac{\pm2\pi}{J_\pm(\nu)\sqrt{\nu_0^2-\nu^2}} + O(\ep^{-1}),\qquad
 \nu_0=\frac{\sqrt{2}}{3}\ep\,.
\label{alpha Ising}
\ee
As a simple check, $\alpha_-$ has a pole at $\nu=\frac{i\ep}{6}$ which is in agreement,
according to \eqref{residue of alpha}, with the OPE coefficient $f_{\sigma\sigma\e}^2=2+O(\ep)$.

Eq.~\eqref{alpha Ising} is an important result about the Regge limit of Ising correlators,
which follow from inserting the two trajectories into \eqref{CRT}.
Further simplifications are possible, however.
We will be careful with a subtlety in this model,
which can be seen from the more detailed analysis of \cite{Caron-Huot:2020nem},
who considered effects that are typically power-suppressed effects. In particular, a
``non-normalizable'' term in the harmonic decomposition (3.8) of \cite{Caron-Huot:2020nem},
coming from $\e$-exchange, where $\Delta_\e<\frac{d}{2}$, won't be negligible here as $\ep\to 0$.
Adding its continuation to \eqref{CRT exact}, the correct formula to leading power is:
\be\begin{aligned} \label{WF G 1}
\cG(z,\zb)^{\circlearrowleft}_{\rm W-F}-1 &\approx \frac{-if_{\sigma\sigma\e}^2}{\pi\kappa_{\Delta_\e}^{(0,0)}}
G_{1,1-\Delta_\e}(z,\zb)
\\ &\quad +
i\pi  \int_{-\infty}^{+\infty} \frac{d\nu}{2\pi} \rho(\nu)\cP_{\frac{2-d}{2}+i\nu}\left(\frac{z+\zb}{2\sqrt{z\zb}}\right)
 \sum_\pm\alpha_\pm^{W-F}(\nu)\ (z\zb)^{\frac{1-J_\pm(\nu)}{2}} + O(z\zb)\,.
\end{aligned}\ee
Both lines are of order $(z\zb)^{1/2+O(\ep)}$ in the Regge limit;
the block $F'$ on the second line of \eqref{CRT exact} contributes only at order $z\zb$ so we ignore it here.

The integration contour in \eqref{WF G 1} passes through the branch point of the square root,
but the integral is unambiguous since the two solutions are added.
For the discussion below, it will be convenient to displace the contour slightly below the real axis,
so that the two contributions can be separated.

The further simplification in the Regge limit of \eqref{WF G 1} comes from the fact that $\sim (z\zb)^{\frac{1-J}{2}}$
decreases rapidly with decreasing ${\rm Re}(J)$. The steepest descent method thus pushes the integral into that direction.
The contour for $J_+$ gets pushed up, while that for $J_-$ is pushed down.
The former picks up a contour integral around $[-\nu_0,\nu_0]$.
This is similar to the treatment of the fishnet theory in \cite{Caron-Huot:2020nem}.
Pushing further, both contours also pick up an extra residue at $J=0$ from the denominator of \eqref{alpha Ising}.
Nicely, through the identity \eqref{residue of alpha}, we find that these precisely cancel the first term in \eqref{WF G 1}.
This means that the Regge limit is controlled by just the short cut (the naive expectation), as
recorded in \eqref{CRT Ising} in the main text. The next singularities are at $J\approx -1$ and are power-suppressed.

\end{appendix}

\bibliographystyle{JHEP}
\bibliography{refs}

\providecommand{\href}[2]{#2}\begingroup\raggedright\begin{thebibliography}{10}

\bibitem{Maldacena:1997re}
J.~M. Maldacena, \emph{{The Large N limit of superconformal field theories and
  supergravity}}, \href{https://doi.org/10.1023/A:1026654312961}{\emph{Adv.
  Theor. Math. Phys.} {\bfseries 2} (1998) 231--252},
  [\href{https://arxiv.org/abs/hep-th/9711200}{{\ttfamily hep-th/9711200}}].

\bibitem{Heemskerk:2009pn}
I.~Heemskerk, J.~Penedones, J.~Polchinski and J.~Sully, \emph{{Holography from
  Conformal Field Theory}},
  \href{https://doi.org/10.1088/1126-6708/2009/10/079}{\emph{JHEP} {\bfseries
  10} (2009) 079}, [\href{https://arxiv.org/abs/0907.0151}{{\ttfamily
  0907.0151}}].

\bibitem{Hofman:2008ar}
D.~M. Hofman and J.~Maldacena, \emph{{Conformal collider physics: Energy and
  charge correlations}},
  \href{https://doi.org/10.1088/1126-6708/2008/05/012}{\emph{JHEP} {\bfseries
  05} (2008) 012}, [\href{https://arxiv.org/abs/0803.1467}{{\ttfamily
  0803.1467}}].

\bibitem{Arnold:2011qi}
P.~Arnold and D.~Vaman, \emph{{Jet quenching in hot strongly coupled gauge
  theories simplified}},
  \href{https://doi.org/10.1007/JHEP04(2011)027}{\emph{JHEP} {\bfseries 04}
  (2011) 027}, [\href{https://arxiv.org/abs/1101.2689}{{\ttfamily 1101.2689}}].

\bibitem{Hashimoto:2018okj}
K.~Hashimoto, S.~Kinoshita and K.~Murata, \emph{{Imaging black holes through
  the AdS/CFT correspondence}},
  \href{https://doi.org/10.1103/PhysRevD.101.066018}{\emph{Phys. Rev. D}
  {\bfseries 101} (2020) 066018},
  [\href{https://arxiv.org/abs/1811.12617}{{\ttfamily 1811.12617}}].

\bibitem{Hashimoto:2019jmw}
K.~Hashimoto, S.~Kinoshita and K.~Murata, \emph{{Einstein Rings in
  Holography}},
  \href{https://doi.org/10.1103/PhysRevLett.123.031602}{\emph{Phys. Rev. Lett.}
  {\bfseries 123} (2019) 031602},
  [\href{https://arxiv.org/abs/1906.09113}{{\ttfamily 1906.09113}}].

\bibitem{Kaku:2021xqp}
Y.~Kaku, K.~Murata and J.~Tsujimura, \emph{{Observing black holes through
  superconductors}}, \href{https://doi.org/10.1007/JHEP09(2021)138}{\emph{JHEP}
  {\bfseries 09} (2021) 138},
  [\href{https://arxiv.org/abs/2106.00304}{{\ttfamily 2106.00304}}].

\bibitem{Polchinski:1999yd}
J.~Polchinski, L.~Susskind and N.~Toumbas, \emph{{Negative energy,
  superluminosity and holography}},
  \href{https://doi.org/10.1103/PhysRevD.60.084006}{\emph{Phys. Rev. D}
  {\bfseries 60} (1999) 084006},
  [\href{https://arxiv.org/abs/hep-th/9903228}{{\ttfamily hep-th/9903228}}].

\bibitem{Shenker:2013pqa}
S.~H. Shenker and D.~Stanford, \emph{{Black holes and the butterfly effect}},
  \href{https://doi.org/10.1007/JHEP03(2014)067}{\emph{JHEP} {\bfseries 03}
  (2014) 067}, [\href{https://arxiv.org/abs/1306.0622}{{\ttfamily 1306.0622}}].

\bibitem{Maldacena:2015waa}
J.~Maldacena, S.~H. Shenker and D.~Stanford, \emph{{A bound on chaos}},
  \href{https://doi.org/10.1007/JHEP08(2016)106}{\emph{JHEP} {\bfseries 08}
  (2016) 106}, [\href{https://arxiv.org/abs/1503.01409}{{\ttfamily
  1503.01409}}].

\bibitem{Kajuri:2020vxf}
N.~Kajuri, \emph{{Lectures on Bulk Reconstruction}},
  \href{https://doi.org/10.21468/SciPostPhysLectNotes.22}{\emph{SciPost Phys.
  Lect. Notes} {\bfseries 22} (2021) 1},
  [\href{https://arxiv.org/abs/2003.00587}{{\ttfamily 2003.00587}}].

\bibitem{Hamilton:2006az}
A.~Hamilton, D.~N. Kabat, G.~Lifschytz and D.~A. Lowe, \emph{{Holographic
  representation of local bulk operators}},
  \href{https://doi.org/10.1103/PhysRevD.74.066009}{\emph{Phys. Rev. D}
  {\bfseries 74} (2006) 066009},
  [\href{https://arxiv.org/abs/hep-th/0606141}{{\ttfamily hep-th/0606141}}].

\bibitem{Hubeny:2012ry}
V.~E. Hubeny, \emph{{Extremal surfaces as bulk probes in AdS/CFT}},
  \href{https://doi.org/10.1007/JHEP07(2012)093}{\emph{JHEP} {\bfseries 07}
  (2012) 093}, [\href{https://arxiv.org/abs/1203.1044}{{\ttfamily 1203.1044}}].

\bibitem{Maldacena:2015iua}
J.~Maldacena, D.~Simmons-Duffin and A.~Zhiboedov, \emph{{Looking for a bulk
  point}}, \href{https://doi.org/10.1007/JHEP01(2017)013}{\emph{JHEP}
  {\bfseries 01} (2017) 013},
  [\href{https://arxiv.org/abs/1509.03612}{{\ttfamily 1509.03612}}].

\bibitem{Engelhardt:2016wgb}
N.~Engelhardt and G.~T. Horowitz, \emph{{Towards a Reconstruction of General
  Bulk Metrics}},
  \href{https://doi.org/10.1088/1361-6382/34/1/015004}{\emph{Class. Quant.
  Grav.} {\bfseries 34} (2017) 015004},
  [\href{https://arxiv.org/abs/1605.01070}{{\ttfamily 1605.01070}}].

\bibitem{Cornalba:2008qf}
L.~Cornalba, M.~S. Costa and J.~Penedones, \emph{{Eikonal Methods in AdS/CFT:
  BFKL Pomeron at Weak Coupling}},
  \href{https://doi.org/10.1088/1126-6708/2008/06/048}{\emph{JHEP} {\bfseries
  06} (2008) 048}, [\href{https://arxiv.org/abs/0801.3002}{{\ttfamily
  0801.3002}}].

\bibitem{Cornalba:2006xk}
L.~Cornalba, M.~S. Costa, J.~Penedones and R.~Schiappa, \emph{{Eikonal
  Approximation in AdS/CFT: From Shock Waves to Four-Point Functions}},
  \href{https://doi.org/10.1088/1126-6708/2007/08/019}{\emph{JHEP} {\bfseries
  08} (2007) 019}, [\href{https://arxiv.org/abs/hep-th/0611122}{{\ttfamily
  hep-th/0611122}}].

\bibitem{Balitsky:2001gj}
I.~Balitsky, \emph{{High-energy QCD and Wilson lines}},
  \href{https://arxiv.org/abs/hep-ph/0101042}{{\ttfamily hep-ph/0101042}}.

\bibitem{Kovchegov:2012mbw}
Y.~V. Kovchegov and E.~Levin, \emph{{Quantum chromodynamics at high energy}},
  vol.~33.
\newblock Cambridge University Press, 8, 2012,
  \href{https://doi.org/10.1017/CBO9781139022187}{10.1017/CBO9781139022187}.

\bibitem{Afkhami-Jeddi:2017rmx}
N.~Afkhami-Jeddi, T.~Hartman, S.~Kundu and A.~Tajdini, \emph{{Shockwaves from
  the Operator Product Expansion}},
  \href{https://doi.org/10.1007/JHEP03(2019)201}{\emph{JHEP} {\bfseries 03}
  (2019) 201}, [\href{https://arxiv.org/abs/1709.03597}{{\ttfamily
  1709.03597}}].

\bibitem{tHooft:1973wag}
G.~'t~Hooft and M.~J.~G. Veltman, \emph{{DIAGRAMMAR}},
  \href{https://doi.org/10.1007/978-1-4684-2826-1_5}{\emph{NATO Sci. Ser. B}
  {\bfseries 4} (1974) 177--322}.

\bibitem{Haehl:2017qfl}
F.~M. Haehl, R.~Loganayagam, P.~Narayan and M.~Rangamani, \emph{{Classification
  of out-of-time-order correlators}},
  \href{https://doi.org/10.21468/SciPostPhys.6.1.001}{\emph{SciPost Phys.}
  {\bfseries 6} (2019) 001},
  [\href{https://arxiv.org/abs/1701.02820}{{\ttfamily 1701.02820}}].

\bibitem{Alday:2010zy}
L.~F. Alday, B.~Eden, G.~P. Korchemsky, J.~Maldacena and E.~Sokatchev,
  \emph{{From correlation functions to Wilson loops}},
  \href{https://doi.org/10.1007/JHEP09(2011)123}{\emph{JHEP} {\bfseries 09}
  (2011) 123}, [\href{https://arxiv.org/abs/1007.3243}{{\ttfamily 1007.3243}}].

\bibitem{Mack:1976pa}
G.~Mack, \emph{{Convergence of Operator Product Expansions on the Vacuum in
  Conformal Invariant Quantum Field Theory}},
  \href{https://doi.org/10.1007/BF01609130}{\emph{Commun. Math. Phys.}
  {\bfseries 53} (1977) 155}.

\bibitem{Kravchuk:2018htv}
P.~Kravchuk and D.~Simmons-Duffin, \emph{{Light-ray operators in conformal
  field theory}}, \href{https://doi.org/10.1007/JHEP11(2018)102}{\emph{JHEP}
  {\bfseries 11} (2018) 102},
  [\href{https://arxiv.org/abs/1805.00098}{{\ttfamily 1805.00098}}].

\bibitem{Caron-Huot:2013fea}
S.~Caron-Huot, \emph{{When does the gluon reggeize?}},
  \href{https://doi.org/10.1007/JHEP05(2015)093}{\emph{JHEP} {\bfseries 05}
  (2015) 093}, [\href{https://arxiv.org/abs/1309.6521}{{\ttfamily 1309.6521}}].

\bibitem{Caron-Huot:2022eqs}
S.~Caron-Huot, M.~Kologlu, P.~Kravchuk, D.~Meltzer and D.~Simmons-Duffin,
  \emph{{Detectors in weakly-coupled field theories}},
  \href{https://arxiv.org/abs/2209.00008}{{\ttfamily 2209.00008}}.

\bibitem{Mueller:1994jq}
A.~H. Mueller and B.~Patel, \emph{{Single and double BFKL pomeron exchange and
  a dipole picture of high-energy hard processes}},
  \href{https://doi.org/10.1016/0550-3213(94)90284-4}{\emph{Nucl. Phys. B}
  {\bfseries 425} (1994) 471--488},
  [\href{https://arxiv.org/abs/hep-ph/9403256}{{\ttfamily hep-ph/9403256}}].

\bibitem{Balitsky:1995ub}
I.~Balitsky, \emph{{Operator expansion for high-energy scattering}},
  \href{https://doi.org/10.1016/0550-3213(95)00638-9}{\emph{Nucl. Phys. B}
  {\bfseries 463} (1996) 99--160},
  [\href{https://arxiv.org/abs/hep-ph/9509348}{{\ttfamily hep-ph/9509348}}].

\bibitem{Son:2002sd}
D.~T. Son and A.~O. Starinets, \emph{{Minkowski space correlators in AdS / CFT
  correspondence: Recipe and applications}},
  \href{https://doi.org/10.1088/1126-6708/2002/09/042}{\emph{JHEP} {\bfseries
  09} (2002) 042}, [\href{https://arxiv.org/abs/hep-th/0205051}{{\ttfamily
  hep-th/0205051}}].

\bibitem{Skenderis:2008dg}
K.~Skenderis and B.~C. van Rees, \emph{{Real-time gauge/gravity duality:
  Prescription, Renormalization and Examples}},
  \href{https://doi.org/10.1088/1126-6708/2009/05/085}{\emph{JHEP} {\bfseries
  05} (2009) 085}, [\href{https://arxiv.org/abs/0812.2909}{{\ttfamily
  0812.2909}}].

\bibitem{Glorioso:2018mmw}
P.~Glorioso, M.~Crossley and H.~Liu, \emph{{A prescription for holographic
  Schwinger-Keldysh contour in non-equilibrium systems}},
  \href{https://arxiv.org/abs/1812.08785}{{\ttfamily 1812.08785}}.

\bibitem{Chakrabarty:2019aeu}
B.~Chakrabarty, J.~Chakravarty, S.~Chaudhuri, C.~Jana, R.~Loganayagam and
  A.~Sivakumar, \emph{{Nonlinear Langevin dynamics via holography}},
  \href{https://doi.org/10.1007/JHEP01(2020)165}{\emph{JHEP} {\bfseries 01}
  (2020) 165}, [\href{https://arxiv.org/abs/1906.07762}{{\ttfamily
  1906.07762}}].

\bibitem{Witten:2019qhl}
E.~Witten, \emph{{Light Rays, Singularities, and All That}},
  \href{https://doi.org/10.1103/RevModPhys.92.045004}{\emph{Rev. Mod. Phys.}
  {\bfseries 92} (2020) 045004},
  [\href{https://arxiv.org/abs/1901.03928}{{\ttfamily 1901.03928}}].

\bibitem{Cornalba:2006xm}
L.~Cornalba, M.~S. Costa, J.~Penedones and R.~Schiappa, \emph{{Eikonal
  Approximation in AdS/CFT: Conformal Partial Waves and Finite N Four-Point
  Functions}},
  \href{https://doi.org/10.1016/j.nuclphysb.2007.01.007}{\emph{Nucl. Phys. B}
  {\bfseries 767} (2007) 327--351},
  [\href{https://arxiv.org/abs/hep-th/0611123}{{\ttfamily hep-th/0611123}}].

\bibitem{Meltzer:2017rtf}
D.~Meltzer and E.~Perlmutter, \emph{{Beyond $a = c$: gravitational couplings to
  matter and the stress tensor OPE}},
  \href{https://doi.org/10.1007/JHEP07(2018)157}{\emph{JHEP} {\bfseries 07}
  (2018) 157}, [\href{https://arxiv.org/abs/1712.04861}{{\ttfamily
  1712.04861}}].

\bibitem{Kulaxizi:2017ixa}
M.~Kulaxizi, A.~Parnachev and A.~Zhiboedov, \emph{{Bulk Phase Shift, CFT Regge
  Limit and Einstein Gravity}},
  \href{https://doi.org/10.1007/JHEP06(2018)121}{\emph{JHEP} {\bfseries 06}
  (2018) 121}, [\href{https://arxiv.org/abs/1705.02934}{{\ttfamily
  1705.02934}}].

\bibitem{Costa:2012cb}
M.~S. Costa, V.~Goncalves and J.~Penedones, \emph{{Conformal Regge theory}},
  \href{https://doi.org/10.1007/JHEP12(2012)091}{\emph{JHEP} {\bfseries 12}
  (2012) 091}, [\href{https://arxiv.org/abs/1209.4355}{{\ttfamily 1209.4355}}].

\bibitem{Caron-Huot:2017vep}
S.~Caron-Huot, \emph{{Analyticity in Spin in Conformal Theories}},
  \href{https://doi.org/10.1007/JHEP09(2017)078}{\emph{JHEP} {\bfseries 09}
  (2017) 078}, [\href{https://arxiv.org/abs/1703.00278}{{\ttfamily
  1703.00278}}].

\bibitem{Hatta:2008st}
Y.~Hatta, \emph{{Relating e+ e- annihilation to high energy scattering at weak
  and strong coupling}},
  \href{https://doi.org/10.1088/1126-6708/2008/11/057}{\emph{JHEP} {\bfseries
  11} (2008) 057}, [\href{https://arxiv.org/abs/0810.0889}{{\ttfamily
  0810.0889}}].

\bibitem{Caron-Huot:2015bja}
S.~Caron-Huot, \emph{{Resummation of non-global logarithms and the BFKL
  equation}}, \href{https://doi.org/10.1007/JHEP03(2018)036}{\emph{JHEP}
  {\bfseries 03} (2018) 036},
  [\href{https://arxiv.org/abs/1501.03754}{{\ttfamily 1501.03754}}].

\bibitem{Vladimirov:2016dll}
A.~A. Vladimirov, \emph{{Correspondence between Soft and Rapidity Anomalous
  Dimensions}},
  \href{https://doi.org/10.1103/PhysRevLett.118.062001}{\emph{Phys. Rev. Lett.}
  {\bfseries 118} (2017) 062001},
  [\href{https://arxiv.org/abs/1610.05791}{{\ttfamily 1610.05791}}].

\bibitem{Mueller:2018llt}
A.~H. Mueller, \emph{{Conformal spacelike-timelike correspondence in QCD}},
  \href{https://doi.org/10.1007/JHEP08(2018)139}{\emph{JHEP} {\bfseries 08}
  (2018) 139}, [\href{https://arxiv.org/abs/1804.07249}{{\ttfamily
  1804.07249}}].

\bibitem{Shenker:2014cwa}
S.~H. Shenker and D.~Stanford, \emph{{Stringy effects in scrambling}},
  \href{https://doi.org/10.1007/JHEP05(2015)132}{\emph{JHEP} {\bfseries 05}
  (2015) 132}, [\href{https://arxiv.org/abs/1412.6087}{{\ttfamily 1412.6087}}].

\bibitem{Caron-Huot:2020nem}
S.~Caron-Huot and J.~Sandor, \emph{{Conformal Regge Theory at Finite Boost}},
  \href{https://doi.org/10.1007/JHEP05(2021)059}{\emph{JHEP} {\bfseries 05}
  (2021) 059}, [\href{https://arxiv.org/abs/2008.11759}{{\ttfamily
  2008.11759}}].

\bibitem{Afkhami-Jeddi:2016ntf}
N.~Afkhami-Jeddi, T.~Hartman, S.~Kundu and A.~Tajdini, \emph{{Einstein gravity
  3-point functions from conformal field theory}},
  \href{https://doi.org/10.1007/JHEP12(2017)049}{\emph{JHEP} {\bfseries 12}
  (2017) 049}, [\href{https://arxiv.org/abs/1610.09378}{{\ttfamily
  1610.09378}}].

\bibitem{Caron-Huot:2021enk}
S.~Caron-Huot, D.~Mazac, L.~Rastelli and D.~Simmons-Duffin, \emph{{AdS bulk
  locality from sharp CFT bounds}},
  \href{https://doi.org/10.1007/JHEP11(2021)164}{\emph{JHEP} {\bfseries 11}
  (2021) 164}, [\href{https://arxiv.org/abs/2106.10274}{{\ttfamily
  2106.10274}}].

\bibitem{Gubser:2002tv}
S.~S. Gubser, I.~R. Klebanov and A.~M. Polyakov, \emph{{A Semiclassical limit
  of the gauge / string correspondence}},
  \href{https://doi.org/10.1016/S0550-3213(02)00373-5}{\emph{Nucl. Phys. B}
  {\bfseries 636} (2002) 99--114},
  [\href{https://arxiv.org/abs/hep-th/0204051}{{\ttfamily hep-th/0204051}}].

\bibitem{Balitsky:2009yp}
I.~Balitsky and G.~A. Chirilli, \emph{{High-energy amplitudes in N=4 SYM in the
  next-to-leading order}},
  \href{https://doi.org/10.1016/j.physletb.2010.02.084}{\emph{Phys. Lett. B}
  {\bfseries 687} (2010) 204--213},
  [\href{https://arxiv.org/abs/0911.5192}{{\ttfamily 0911.5192}}].

\bibitem{Brower:2006ea}
R.~C. Brower, J.~Polchinski, M.~J. Strassler and C.-I. Tan, \emph{{The Pomeron
  and gauge/string duality}},
  \href{https://doi.org/10.1088/1126-6708/2007/12/005}{\emph{JHEP} {\bfseries
  12} (2007) 005}, [\href{https://arxiv.org/abs/hep-th/0603115}{{\ttfamily
  hep-th/0603115}}].

\bibitem{Alday:2015ota}
L.~F. Alday and A.~Zhiboedov, \emph{{Conformal Bootstrap With Slightly Broken
  Higher Spin Symmetry}},
  \href{https://doi.org/10.1007/JHEP06(2016)091}{\emph{JHEP} {\bfseries 06}
  (2016) 091}, [\href{https://arxiv.org/abs/1506.04659}{{\ttfamily
  1506.04659}}].

\bibitem{Simmons-Duffin:2016wlq}
D.~Simmons-Duffin, \emph{{The Lightcone Bootstrap and the Spectrum of the 3d
  Ising CFT}}, \href{https://doi.org/10.1007/JHEP03(2017)086}{\emph{JHEP}
  {\bfseries 03} (2017) 086},
  [\href{https://arxiv.org/abs/1612.08471}{{\ttfamily 1612.08471}}].

\bibitem{Su:2022xnj}
N.~Su, \emph{{The Hybrid Bootstrap}},
  \href{https://arxiv.org/abs/2202.07607}{{\ttfamily 2202.07607}}.

\bibitem{Caron-Huot:2020ouj}
S.~Caron-Huot, Y.~Gobeil and Z.~Zahraee, \emph{{The leading trajectory in the
  2+1D Ising CFT}},  \href{https://arxiv.org/abs/2007.11647}{{\ttfamily
  2007.11647}}.

\bibitem{Liu:2020tpf}
J.~Liu, D.~Meltzer, D.~Poland and D.~Simmons-Duffin, \emph{{The Lorentzian
  inversion formula and the spectrum of the 3d O(2) CFT}},
  \href{https://doi.org/10.1007/JHEP09(2020)115}{\emph{JHEP} {\bfseries 09}
  (2020) 115}, [\href{https://arxiv.org/abs/2007.07914}{{\ttfamily
  2007.07914}}].

\bibitem{Cardy:1984rp}
J.~L. Cardy, \emph{{Conformal invariance and universality in finite-size
  scaling}}, {\emph{J. Phys. A} {\bfseries 17} (1984) L385--L387}.

\bibitem{Fitzpatrick:2015zha}
A.~L. Fitzpatrick, J.~Kaplan and M.~T. Walters, \emph{{Virasoro Conformal
  Blocks and Thermality from Classical Background Fields}},
  \href{https://doi.org/10.1007/JHEP11(2015)200}{\emph{JHEP} {\bfseries 11}
  (2015) 200}, [\href{https://arxiv.org/abs/1501.05315}{{\ttfamily
  1501.05315}}].

\bibitem{Banados:1992wn}
M.~Banados, C.~Teitelboim and J.~Zanelli, \emph{{The Black hole in
  three-dimensional space-time}},
  \href{https://doi.org/10.1103/PhysRevLett.69.1849}{\emph{Phys. Rev. Lett.}
  {\bfseries 69} (1992) 1849--1851},
  [\href{https://arxiv.org/abs/hep-th/9204099}{{\ttfamily hep-th/9204099}}].

\bibitem{Banados:1992gq}
M.~Banados, M.~Henneaux, C.~Teitelboim and J.~Zanelli, \emph{{Geometry of the
  (2+1) black hole}},
  \href{https://doi.org/10.1103/PhysRevD.48.1506}{\emph{Phys. Rev. D}
  {\bfseries 48} (1993) 1506--1525},
  [\href{https://arxiv.org/abs/gr-qc/9302012}{{\ttfamily gr-qc/9302012}}].

\bibitem{Kusuki:2019hcg}
Y.~Kusuki, Y.~Suzuki, T.~Takayanagi and K.~Umemoto, \emph{{Looking at Shadows
  of Entanglement Wedges}},
  \href{https://doi.org/10.1093/ptep/ptaa152}{\emph{PTEP} {\bfseries 2020}
  (2020) 11B105}, [\href{https://arxiv.org/abs/1912.08423}{{\ttfamily
  1912.08423}}].

\bibitem{Jokela:2020auu}
N.~Jokela and A.~P\"onni, \emph{{Towards precision holography}},
  \href{https://doi.org/10.1103/PhysRevD.103.026010}{\emph{Phys. Rev. D}
  {\bfseries 103} (2021) 026010},
  [\href{https://arxiv.org/abs/2007.00010}{{\ttfamily 2007.00010}}].

\bibitem{Almheiri:2014lwa}
A.~Almheiri, X.~Dong and D.~Harlow, \emph{{Bulk Locality and Quantum Error
  Correction in AdS/CFT}},
  \href{https://doi.org/10.1007/JHEP04(2015)163}{\emph{JHEP} {\bfseries 04}
  (2015) 163}, [\href{https://arxiv.org/abs/1411.7041}{{\ttfamily 1411.7041}}].

\bibitem{Jafferis:2015del}
D.~L. Jafferis, A.~Lewkowycz, J.~Maldacena and S.~J. Suh, \emph{{Relative
  entropy equals bulk relative entropy}},
  \href{https://doi.org/10.1007/JHEP06(2016)004}{\emph{JHEP} {\bfseries 06}
  (2016) 004}, [\href{https://arxiv.org/abs/1512.06431}{{\ttfamily
  1512.06431}}].

\bibitem{Czech:2016xec}
B.~Czech, L.~Lamprou, S.~McCandlish, B.~Mosk and J.~Sully, \emph{{A
  Stereoscopic Look into the Bulk}},
  \href{https://doi.org/10.1007/JHEP07(2016)129}{\emph{JHEP} {\bfseries 07}
  (2016) 129}, [\href{https://arxiv.org/abs/1604.03110}{{\ttfamily
  1604.03110}}].

\bibitem{deBoer:2016pqk}
J.~de~Boer, F.~M. Haehl, M.~P. Heller and R.~C. Myers, \emph{{Entanglement,
  holography and causal diamonds}},
  \href{https://doi.org/10.1007/JHEP08(2016)162}{\emph{JHEP} {\bfseries 08}
  (2016) 162}, [\href{https://arxiv.org/abs/1606.03307}{{\ttfamily
  1606.03307}}].

\bibitem{Leutheusser:2021frk}
S.~Leutheusser and H.~Liu, \emph{{Emergent times in holographic duality}},
  \href{https://arxiv.org/abs/2112.12156}{{\ttfamily 2112.12156}}.

\bibitem{Chandrasekaran:2022eqq}
V.~Chandrasekaran, G.~Penington and E.~Witten, \emph{{Large N algebras and
  generalized entropy}},  \href{https://arxiv.org/abs/2209.10454}{{\ttfamily
  2209.10454}}.

\bibitem{Faulkner:2017vdd}
T.~Faulkner and A.~Lewkowycz, \emph{{Bulk locality from modular flow}},
  \href{https://doi.org/10.1007/JHEP07(2017)151}{\emph{JHEP} {\bfseries 07}
  (2017) 151}, [\href{https://arxiv.org/abs/1704.05464}{{\ttfamily
  1704.05464}}].

\bibitem{Caron-Huot:2022ugt}
S.~Caron-Huot, Y.-Z. Li, J.~Parra-Martinez and D.~Simmons-Duffin,
  \emph{{Causality constraints on corrections to Einstein gravity}},
  \href{https://arxiv.org/abs/2201.06602}{{\ttfamily 2201.06602}}.

\bibitem{Caron-Huot:2022jli}
S.~Caron-Huot, Y.-Z. Li, J.~Parra-Martinez and D.~Simmons-Duffin,
  \emph{{Graviton partial waves and causality in higher dimensions}},
  \href{https://arxiv.org/abs/2205.01495}{{\ttfamily 2205.01495}}.

\bibitem{Weinberg:1964ew}
S.~Weinberg, \emph{{Photons and Gravitons in $S$-Matrix Theory: Derivation of
  Charge Conservation and Equality of Gravitational and Inertial Mass}},
  \href{https://doi.org/10.1103/PhysRev.135.B1049}{\emph{Phys. Rev.} {\bfseries
  135} (1964) B1049--B1056}.

\bibitem{Liu:2022wop}
X.~Liu and H.~X. Zhu, \emph{{The Nucleon Energy Correlators}},
  \href{https://arxiv.org/abs/2209.02080}{{\ttfamily 2209.02080}}.

\bibitem{Karateev:2018oml}
D.~Karateev, P.~Kravchuk and D.~Simmons-Duffin, \emph{{Harmonic Analysis and
  Mean Field Theory}},
  \href{https://doi.org/10.1007/JHEP10(2019)217}{\emph{JHEP} {\bfseries 10}
  (2019) 217}, [\href{https://arxiv.org/abs/1809.05111}{{\ttfamily
  1809.05111}}].

\bibitem{Dolan:2000ut}
F.~A. Dolan and H.~Osborn, \emph{{Conformal four point functions and the
  operator product expansion}},
  \href{https://doi.org/10.1016/S0550-3213(01)00013-X}{\emph{Nucl. Phys. B}
  {\bfseries 599} (2001) 459--496},
  [\href{https://arxiv.org/abs/hep-th/0011040}{{\ttfamily hep-th/0011040}}].

\bibitem{Liu:1998bu}
H.~Liu and A.~A. Tseytlin, \emph{{D = 4 superYang-Mills, D = 5 gauged
  supergravity, and D = 4 conformal supergravity}},
  \href{https://doi.org/10.1016/S0550-3213(98)00443-X}{\emph{Nucl. Phys. B}
  {\bfseries 533} (1998) 88--108},
  [\href{https://arxiv.org/abs/hep-th/9804083}{{\ttfamily hep-th/9804083}}].

\bibitem{Henriksson:2018myn}
J.~Henriksson and M.~Van~Loon, \emph{{Critical O(N) model to order $\epsilon^4$
  from analytic bootstrap}},
  \href{https://doi.org/10.1088/1751-8121/aaf1e2}{\emph{J. Phys. A} {\bfseries
  52} (2019) 025401}, [\href{https://arxiv.org/abs/1801.03512}{{\ttfamily
  1801.03512}}].

\end{thebibliography}\endgroup

\end{document}